\documentclass[sigplan,copyrightmode=0]{acmart}

\def\BibTeX{{\rm B\kern-.05em{\sc i\kern-.025em b}\kern-.08emT\kern-.1667em\lower.7ex\hbox{E}\kern-.125emX}}

\usepackage{wrapfig}
\usepackage{multirow}
\usepackage{balance}
\usepackage{xcolor}
\usepackage{hyperref}
\usepackage{url}

\hypersetup{pdfborder=0 0 0, colorlinks=true, citecolor=blue, urlcolor=blue, linkcolor=black}

\usepackage{pifont}

\newcommand{\smallcapital}{\fontsize{9pt}{10pt}\selectfont}

\newcommand{\skiourakisize}{\fontsize{8pt}{10pt}\selectfont}

\usepackage{verbatim}   

\usepackage{rotating}

\usepackage{arydshln}

\usepackage{placeins}

\begin{document}

%
\title{An Open-Source Benchmark Suite for Cloud and IoT Microservices}


\author{Yu Gan, Yanqi Zhang, Dailun Cheng, Ankitha Shetty, Priyal Rathi, Nayan Katarki, Ariana Bruno, Justin Hu, Brian Ritchken, Brendon Jackson, Kelvin Hu, Meghna Pancholi, Yuan He, Brett Clancy, Chris Colen, Fukang Wen, Catherine Leung, Siyuan Wang, Leon Zaruvinsky, Mateo Espinosa, Rick Lin, Zhongling Liu, Jake Padilla, and Christina Delimitrou}
\affiliation{%
  \institution{Cornell University}
  }
  \email{delimitrou@cornell.edu}

\begin{abstract}
	{Cloud services have recently started undergoing a major shift from monolithic applications, 
	to graphs of hundreds of loosely-coupled microservices. 
Microservices fundamentally change a lot of assumptions current cloud systems are designed with, and 
present both opportunities and challenges when optimizing 
for quality of service (QoS) and utilization. 

In this paper we explore the implications microservices have across the cloud system stack. 
We first present DeathStarBench, a novel, open-source benchmark suite built with microservices that is representative 
of large end-to-end services, modular and extensible. DeathStarBench includes
a social network, a media service, an e-commerce site, a banking system, and IoT applications for coordination control of UAV swarms. 
We then use DeathStarBench to study the architectural characteristics of microservices, 
their implications in networking and operating systems, their challenges with respect to cluster management, 
and their trade-offs in terms of application design and programming frameworks. 
Finally, we explore the tail at scale effects of microservices in real deployments with hundreds of users, 
and highlight the increased pressure they put on performance predictability. 
}

\end{abstract}

%
%



%

\maketitle

\section{Introduction}


Large-scale datacenters host an increasing number of popular online cloud services that span all aspects of human endeavor. 
Many of these applications are \textit{interactive}, \textit{latency-critical} services that must meet strict performance (throughput and tail latency),
and availability constraints, while also handling frequent software updates~\cite{tailatscale,BarrosoBook,Meisner11,Lo14,Lo15,Delimitrou13,Delimitrou14,Delimitrou15,Kasture16,sirius,Delimitrou16,Delimitrou17,Delimitrou13d,Delimitrou13e,Delimitrou14b}.
The effort to satisfy these often contradicting requirements has pushed datacenter applications on the verge of a major design shift,
from complex \textit{monolithic} services that encompass the entire application functionality in a single binary,
to graphs with tens or hundreds of single-purpose, loosely-coupled \textit{microservices}. 
This shift is becoming increasingly pervasive with large cloud providers, such as Amazon, Twitter, Netflix, Apple, and EBay
having already adopted the microservices application model~\cite{Cockroft15,Cockroft16,twitter_decomposing,Gan18}, and 
Netflix reporting more than 200 unique microservices in their ecosystem, as of the end of 2016~\cite{Cockroft15,Cockroft16}. 

\begin{wrapfigure}[15]{l}{0.26\textwidth}
	\centering
	\includegraphics[scale=0.28, trim=0 0.2cm 0 5.4cm, clip=true]{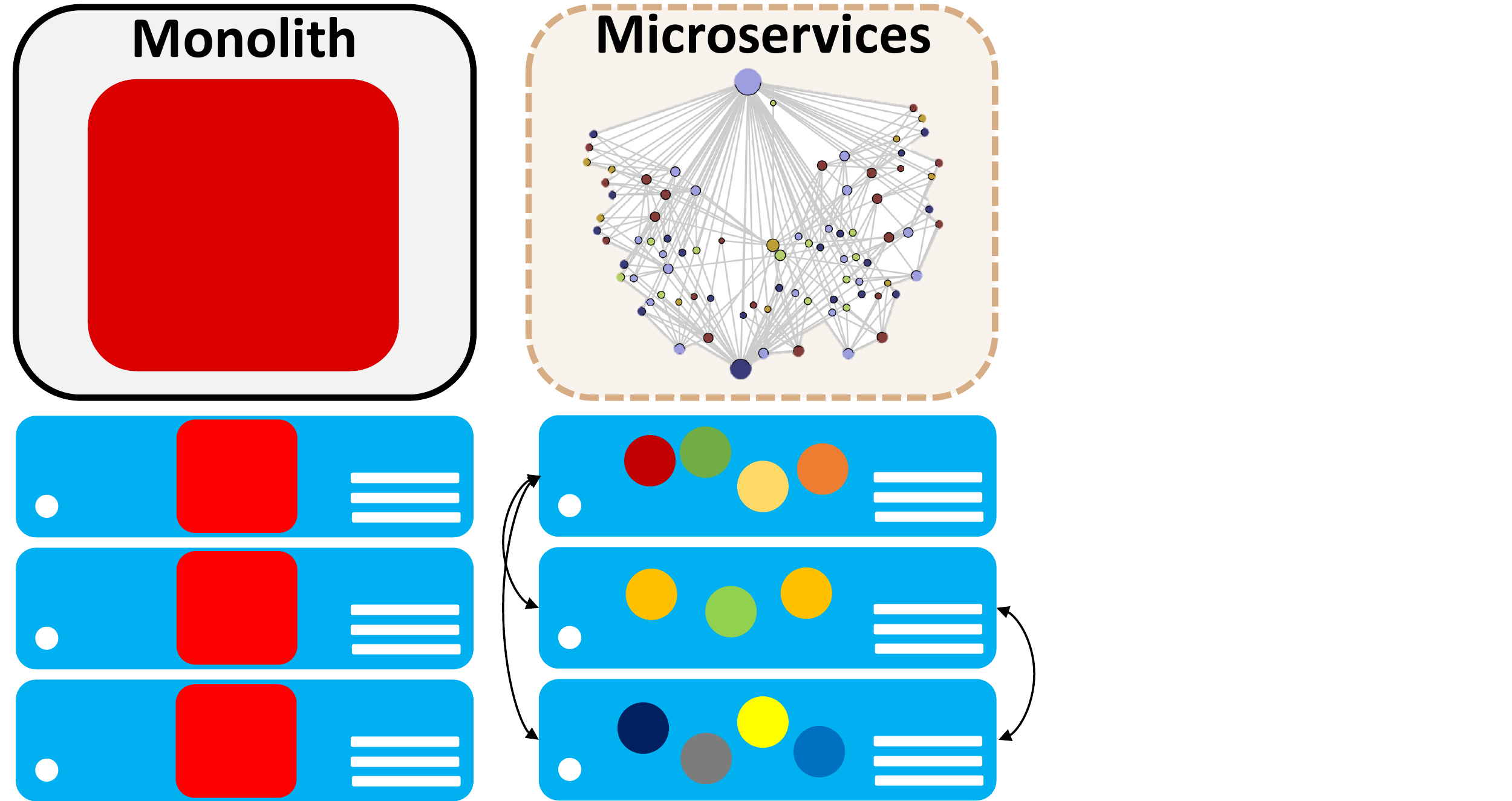}
	\caption{\label{fig:motivation} Differences in the deployment of monoliths and microservices. }
\end{wrapfigure}

The increasing popularity of microservices is justified by several reasons. 
First, they promote composable software design, simplifying and accelerating development,
with each microservice being responsible for a small subset of the application's functionality. The richer the functionality of cloud services becomes, the more 
the modular design of microservices helps manage system complexity. 
They similarly facilitate deploying, scaling, and updating individual microservices independently, avoiding long development cycles, and improving elasticity. 
Fig.~\ref{fig:motivation} shows the deployment differences between a traditional monolithic service, and an application built with microservices. 
While the entire monolith is scaled out on multiple servers, 
microservices allow individual components of the end-to-end application 
to be elastically scaled, with microservices of
complementary resources bin-packed on the same physical server.
Even though modularity in cloud services was already part of 
the Service-Oriented Architecture (SOA) design approach~\cite{soa}, 
the fine granularity of microservices, and their independent deployment 
create hardware and software challenges different from those in traditional SOA workloads.

Second, microservices 
enable programming language and framework heterogeneity, with each tier 
developed in the most suitable language, 
only requiring a common API for microservices
to communicate with each other; typically over remote procedure calls (RPC)~\cite{thrift,finagle,grpc} or a RESTful API.
In contrast, \textit{monoliths} limit the languages used for development, and make frequent updates cumbersome and error-prone. 

Finally, microservices simplify correctness and performance debugging, as bugs can be isolated
in specific tiers, unlike monoliths, where resolving bugs often involves troubleshooting the entire service. This makes them additionally 
applicable to internet-of-things (IoT) applications, that often host mission-critical computation, which puts more pressure on correctness verification~\cite{flinn,Hasan15}. 

Despite their advantages, microservices represent a significant departure from the way cloud services are traditionally designed, and have broad implications 
ranging from cloud management and programming frameworks, to operating systems and datacenter hardware design. 

In this paper we explore the implications microservices have across the cloud system stack, from hardware all the way to application design,  
using a suite of new end-to-end and representative applications built with tens of microservices each. 
The DeathStarBench suite~\footnote{Named after the DeathStar graphs that visualize dependencies between microservices~\cite{Cockroft15,Cockroft16}. } includes six 
end-to-end services that cover a wide spectrum of popular cloud and edge services: a \textit{social network}, a \textit{media service} (movie reviewing, renting, streaming), 
an \textit{e-commerce site}, a \textit{secure banking system}, and \textit{Swarm}; an IoT service for coordination control of drone swarms, with and without a cloud backend. 

\begin{wrapfigure}[12]{r}{0.26\textwidth}
	\vspace{-0.08in}
	\centering
	\includegraphics[scale=0.46, trim=0 0.2cm 0 12.2cm, clip=true]{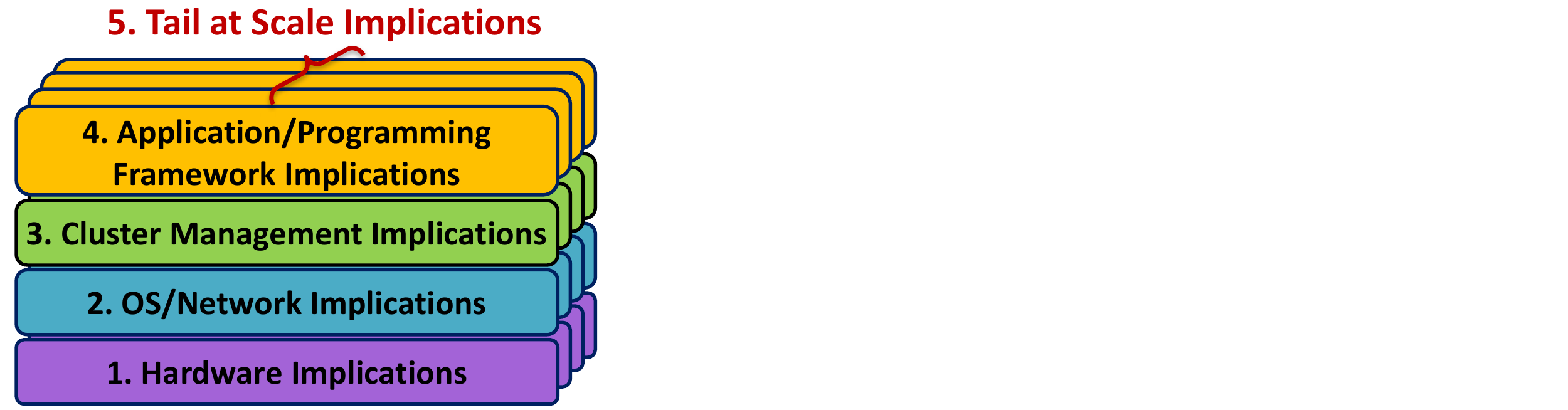}
	\caption{\label{fig:study} {Exploring the implications of microservices across the system stack.} }
\end{wrapfigure}
Each service includes tens of microservices in different languages and programming models, including node.js, 
Python, C/C++, Java, Javascript, Scala, and Go, and leverages open-source applications, such as NGINX~\cite{nginx}, 
memcached~\cite{memcached}, MongoDB~\cite{mongodb}, Cylon~\cite{cylon}, and Xapian~\cite{Kasture16}. 
To create the end-to-end services, we built custom {\smallcapital RPC} and {\smallcapital REST}ful APIs using popular open-source frameworks like Apache Thrift~\cite{thrift}, and gRPC~\cite{grpc}. 
Finally, to track how user requests progress through microservices, 
we have developed a lightweight and transparent to the user distributed tracing system, similar to Dapper~\cite{dapper} and Zipkin~\cite{zipkin}  
that tracks requests at {\smallcapital RPC} granularity, associates {\smallcapital RPC}s belonging to the same end-to-end request, and records traces 
in a centralized database. We study both traffic generated by real users of the services, 
and synthetic loads generated by open-loop workload generators. 

We use these services to study the implications of microservices spanning the system stack, as seen in Fig.~\ref{fig:study}. 
First, we quantify how effective current \textit{datacenter architectures} are at running microservices, 
as well as how datacenter hardware needs to change to 
better accommodate their performance and resource requirements (Section~\ref{sec:architecture}). 
This includes analyzing the cycle breakdown in modern servers, examining whether big 
or small cores are preferable~\cite{Reddi10,Reddi11,hoelzle,Gupta13,amdahls18,Gan18,Chen17}, 
determining the pressure microservices put on instruction caches~\cite{Cloudsuite12,Kaynak13}, and 
exploring the potential they have for hardware acceleration~\cite{catapult,catapult2,brainwave,tpu,firestone18}. We show that despite 
the small amount of computation per microservice, the latency requirements of each individual tier 
are much stricter than for typical applications, putting more pressure on 
predictably high single-thread performance. 

\begin{wrapfigure}[15]{r}{0.28\textwidth}
	\centering
	\begin{tabular}{cc}
		\includegraphics[scale=0.14, viewport = 135 0 555 380]{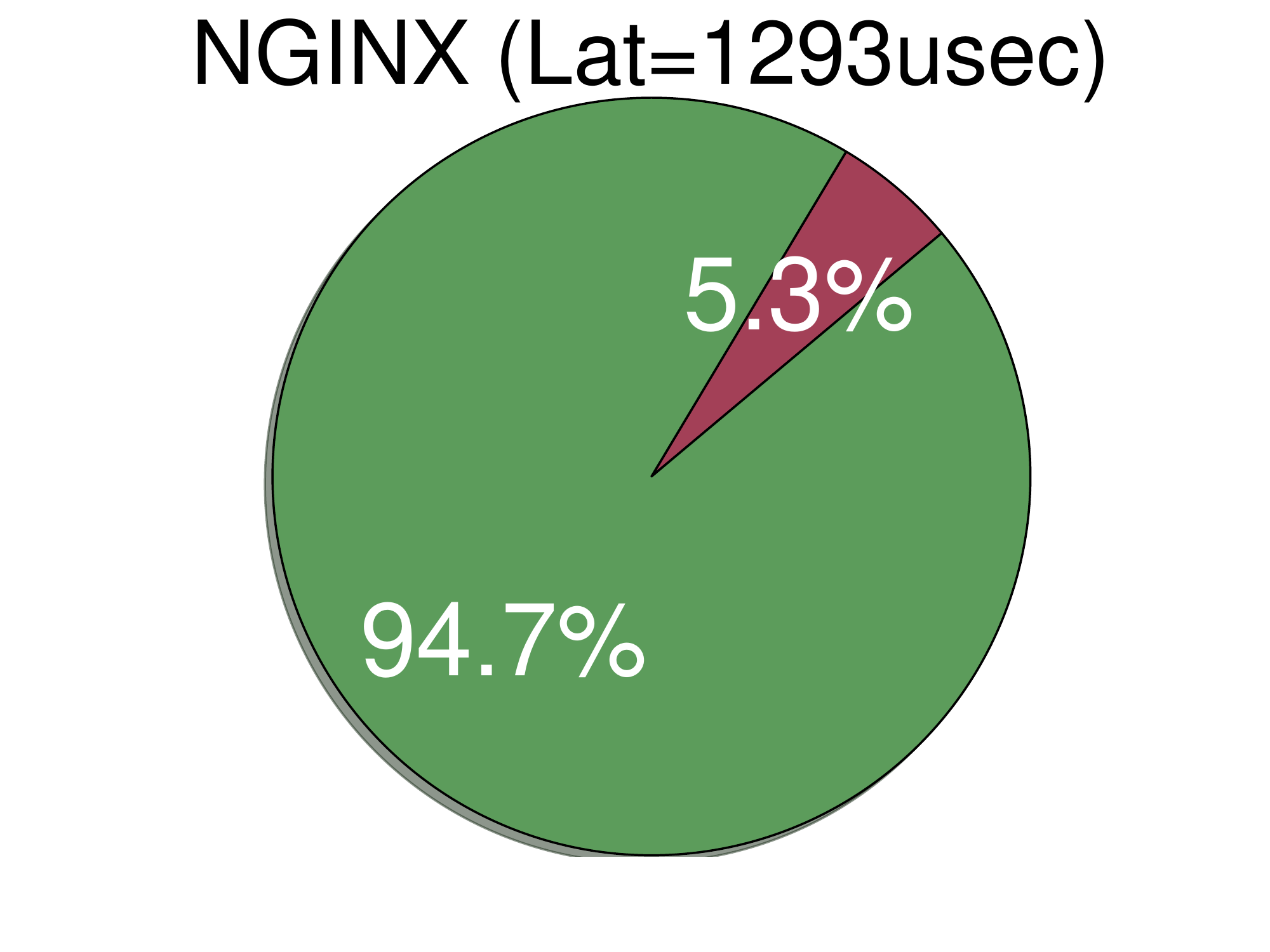} &
		\includegraphics[scale=0.14, viewport = 115 0 555 380]{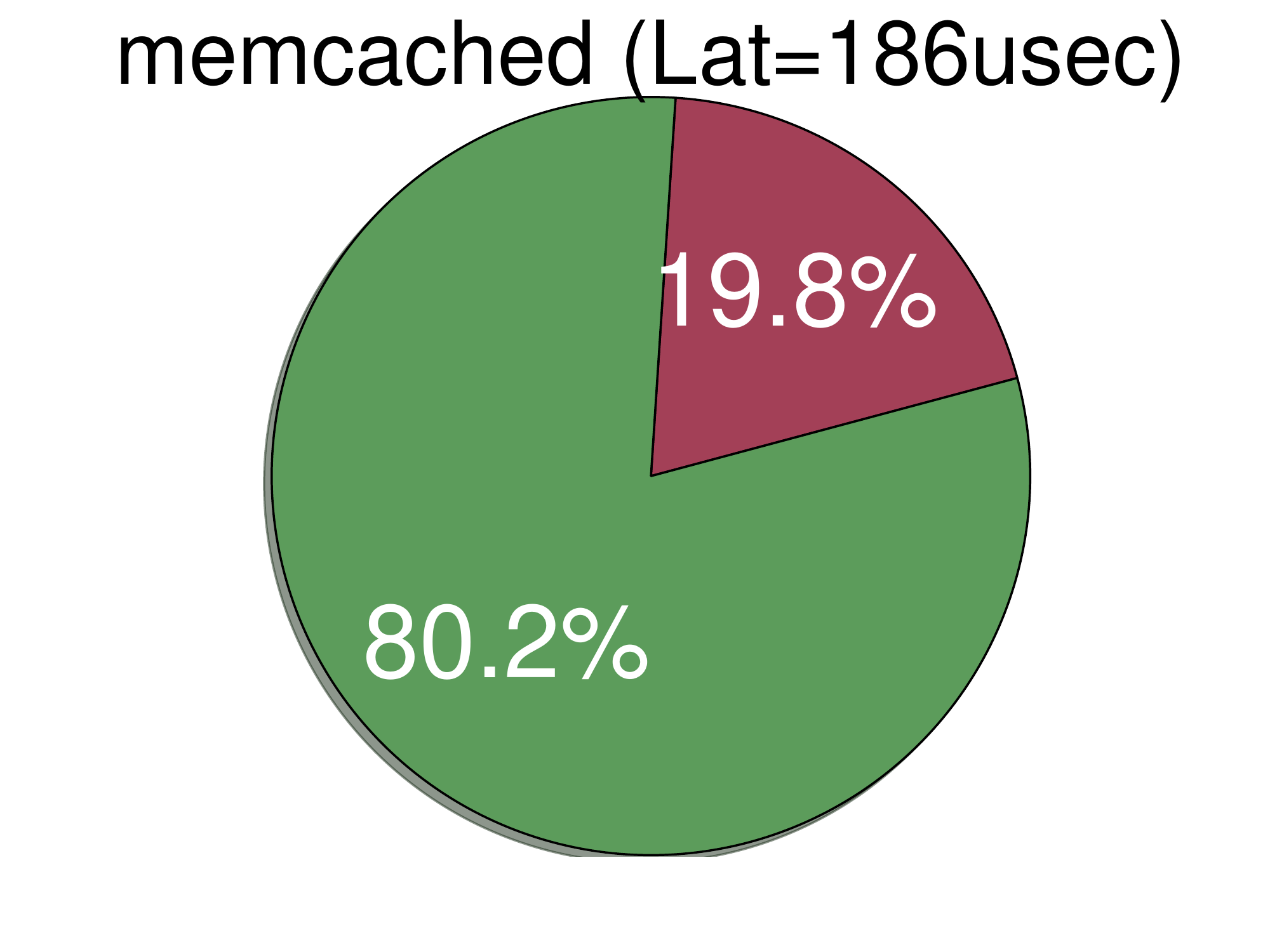} \\
		\includegraphics[scale=0.14, viewport = 135 60 555 420]{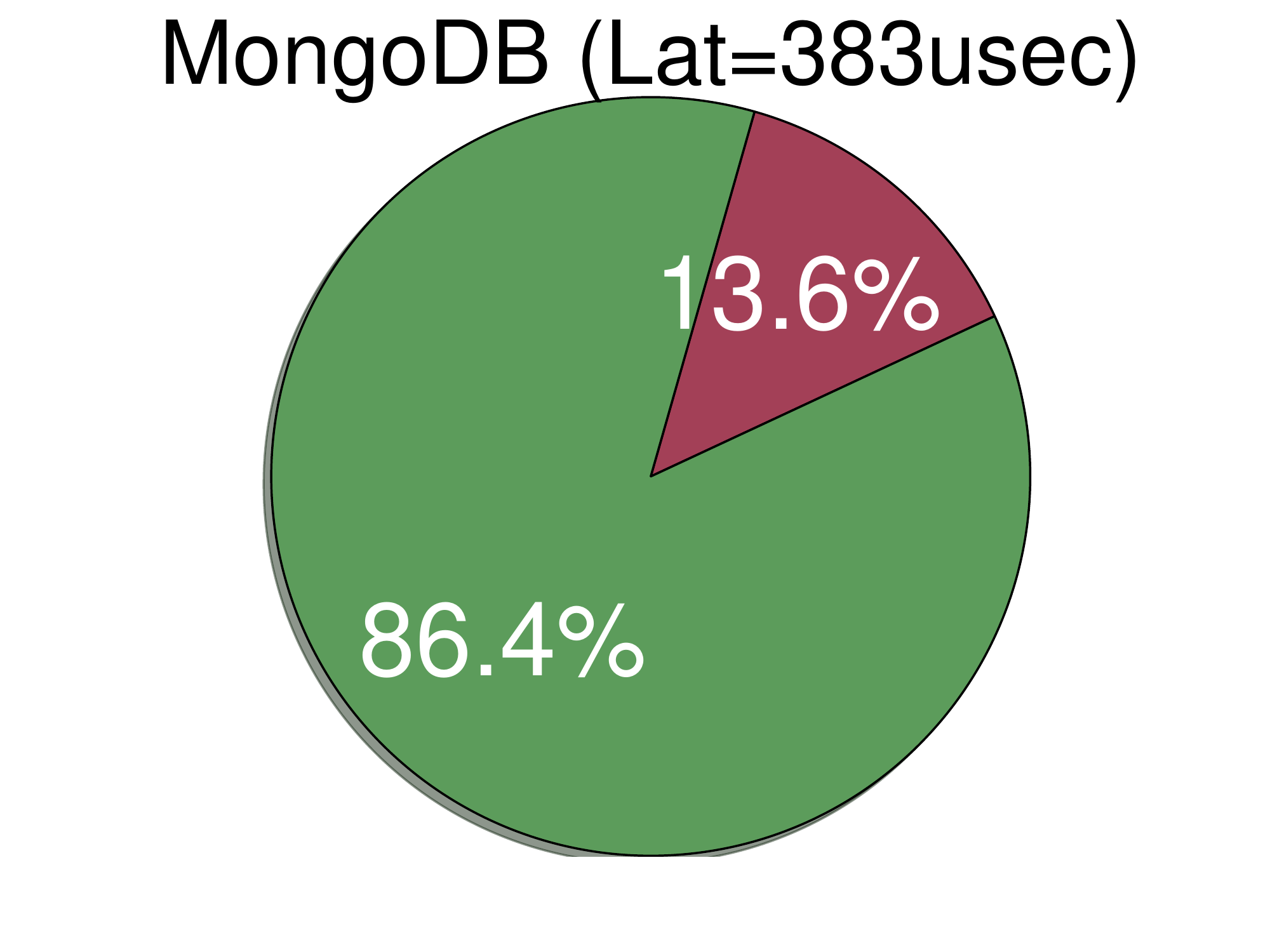} &
	\includegraphics[scale=0.14, viewport = 105 60 555 420]{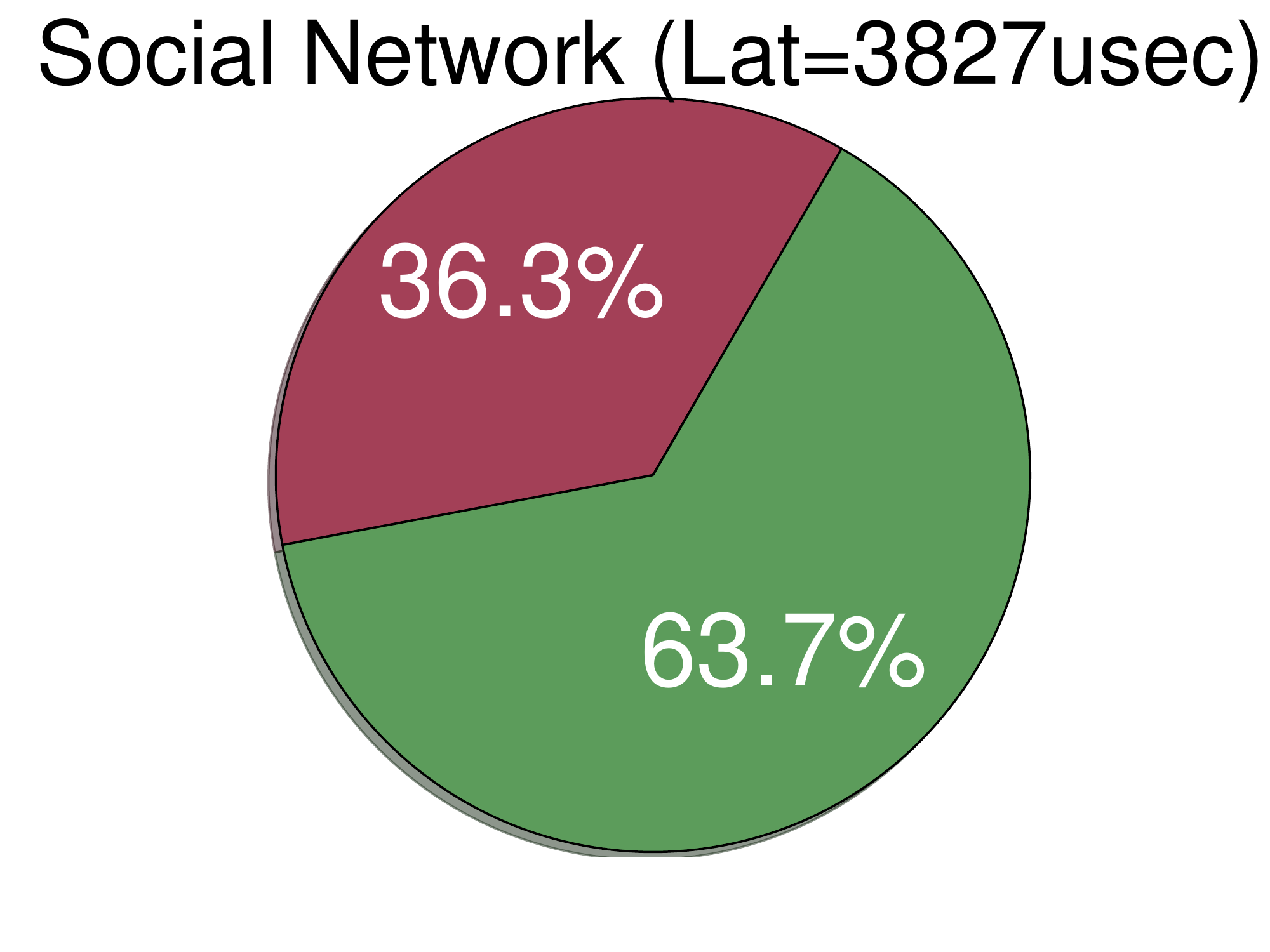}
\end{tabular}
\caption{\label{fig:micro_vs_mono} Network (red) vs. application processing (green) for monoliths and microservices. }
\end{wrapfigure}

Second, we quantify the \textit{networking and operating system} implications of microservices. Specifically we show that, similarly to traditional cloud applications, microservices spend 
a large fraction of time in the kernel. Unlike monolithic services though, microservices spend much more time sending and processing network requests over {\smallcapital RPC}s or other {\smallcapital REST} {\smallcapital API}s. 
Fig.~\ref{fig:micro_vs_mono} shows the breakdown of execution time to network (red) and application processing (green) for three monolithic services (NGINX, memcached, MongoDB) and the end-to-end Social Network application. 
While for the single-tier services only a small amount of time goes towards network processing, when using microservices, this time increases to 36.3\% of total execution time, causing the system's 
resource bottlenecks to change drastically. In Section~\ref{sec:os_network} we show that offloading {\smallcapital RPC} processing to an {\smallcapital FPGA} tightly-coupled with the host server, 
can improve network performance by 10-60$\times$. 

Third, microservices significantly complicate \textit{cluster management}. Even though the cluster manager can scale out individual microservices on-demand instead of the entire monolith, 
dependencies between microservices introduce backpressure effects and cascading QoS violations that quickly propagate through the system, making performance unpredictable. 
Existing cluster managers that optimize for performance and/or 
utilization~\cite{Delimitrou13,Delimitrou16,Mars13a,Mars13b,Nathuji07,Nathuji10,Delimitrou14, Borg, Lin11, Delimitrou15,Ousterhout13,Lo14,Lo15,Mesos11, Omega13} 
are not expressive enough to account for the impact each pair-wise dependency has on end-to-end performance. In Section~\ref{sec:management}, we show that mismanaging even a single such dependency dramatically hurts 
tail latency, e.g., by 10.4$\times$ for the Social Network, and requires long periods for the system to recover, compared to the corresponding monolithic service. We also show that 
traditional autoscaling mechanisms, present in many cloud infrastructures, fall short of addressing QoS violations caused by dependencies between microservices.

Fourth, in Section~\ref{sec:application}, we identify microservices creating bottlenecks in the end-to-end service's critical path, 
quantify the performance trade-offs between {\smallcapital RPC} and {\smallcapital REST}ful {\smallcapital API}s, and explore the performance 
and cost implications of running microservices on \textit{serverless} programming frameworks. 

Finally, given that performance issues in the cloud often only emerge at large scale~\cite{tailatscale}, in Section~\ref{sec:tail_at_scale} 
we use real application deployments with hundreds of users to 
show that tail-at-scale effects become more pronounced in microservices compared to monolithic applications, as 
a single poorly-configured microservice, or slow server can degrade end-to-end latency by several orders of magnitude. 


As microservices continue to evolve, it is essential for datacenter hardware, operating and networking systems, cluster managers, and programming frameworks to also evolve with them, 
to ensure that their prevalence does not come at a performance and/or efficiency loss. 
DeathStarBench is currently used in several academic and industrial institutions with applications in serverless compute, hardware acceleration, and runtime management. 
We hope that open-sourcing it to a wider audience will encourage more research in this emerging field. 


\begin{table*}
\skiourakisize
\centering
\begin{tabular}{ccccccc}
\hline
\multirow{2}{*}{\bf Service} & \multicolumn{1}{c}{\bf Total New} & {\bf{Comm.}} & \multicolumn{2}{c}{\bf{LoCs for RPC/REST}} & {\bf{Unique}} & {\bf{Per-language LoC breakdown}} \\
			     & {\bf{LoCs}} & {\bf{Protocol}} & {\bf{Handwritten}} & {\bf{Autogen}} & {\bf{Microservices}} & {\bf{(end-to-end service)}} \\		
\hline
{\bf{Social}} & \multirow{2}{*}{\texttt{15,198}} & \multirow{2}{*}{\bf RPC} & \multirow{2}{*}{\texttt{9,286}} & \multirow{2}{*}{\texttt{52,863}} & \multirow{2}{*}{\texttt{36}} & {34\% C, 23\% C++, 18\% Java, 7\% node.js, } \\
{\bf{Network}} & & & & & & {6\% Python, 5\% Scala, 3\% PHP, 2\% Javascript, 2\% Go} \\ 
\hdashline[0.5pt/2.5pt]
{\bf{Movie}} & \multirow{2}{*}{\texttt{12,155}} & \multirow{2}{*}{\bf RPC} & \multirow{2}{*}{\texttt{9,853}} & \multirow{2}{*}{\texttt{48,001}} & \multirow{2}{*}{\texttt{38}} & {30\% C, 21\% C++, 20\% Java, 10\% PHP, } \\
{\bf{Reviewing}} & & & & & & {8\% Scala, 5\% node.js, 3\% Python, 3\% Javascript} \\
\hdashline[0.5pt/2.5pt]
{\bf{E-commerce}} & \multirow{2}{*}{\texttt{16,194}} & {\bf REST} & {\texttt{4,798}} & {\texttt{-}} & \multirow{2}{*}{\texttt{41}} & {21\% Java, 16\% C++, 15\% C, 14\% Go, 10\% Javascript, } \\
{\bf{Website}} & &{\bf{RPC}} & {\texttt{2,658}}&\texttt{12,085} & & {7\% node.js, 5\% Scala, 4\% HTML, 3\% Ruby} \\
\hdashline[0.5pt/2.5pt]
{\bf{Banking}} & \multirow{2}{*}{\texttt{13,876}} & \multirow{2}{*}{\bf{RPC}} & \multirow{2}{*}{\texttt{4,757}} & \multirow{2}{*}{\texttt{31,156}} & \multirow{2}{*}{\texttt{34}} & {29\% C, 25\% Javascript, 16\% Java, } \\
{\bf{System}} & & & & & & {16\% node.js, 11\% C++, 3\% Python} \\
\hdashline[0.5pt/2.5pt]
{\bf{Swarm}} & \multirow{2}{*}{\texttt{11,283}} & {\bf{REST}} & \texttt{2,610} & \texttt{-} & \multirow{2}{*}{\texttt{25}} & {\footnotesize 36\% C, 19\% Java, 16\% Javascript, } \\
{\bf{Cloud}} & &{\bf{RPC}} &\texttt{4,614} &\texttt{21,574} & & {\footnotesize 14\% node.js, 13\% C++, 2\% Python}\\ 
\hdashline[0.5pt/2.5pt]
{\bf{Swarm}} & \multirow{2}{*}{\texttt{13,876}} & \multirow{2}{*}{\bf{REST}} & \multirow{2}{*}{\texttt{4,757}} & \multirow{2}{*}\texttt{-} & \multirow{2}{*}{\texttt{21}} & {\footnotesize 29\% C, 25\% Javascript, 16\% Java, } \\
{\bf{Edge}} & & & & & & {\footnotesize 16\% node.js, 11\% C++, 3\% Python} \\
\hline
\end{tabular}
\caption{\label{loc_stats} {Characteristics and code composition of each end-to-end microservices-based application. } } 
\end{table*}
















\section{Related Work}
\label{sec:RelatedWork}

Cloud applications have attracted a lot of attention over the past decade, with several benchmark suites 
being released both from academia and industry~\cite{Cloudsuite12, sirius, Kasture16, Reddi15, bigdatabench}. 
Cloudsuite for example, includes both batch and interactive services, such as memcached, and has been used to study the architectural implications of cloud benchmarks~\cite{Cloudsuite12}. 
Similarly, TailBench aggregates a set of interactive benchmarks, from web servers and databases to speech recognition and machine translation systems and proposes a new methodology 
to analyze their performance~\cite{Kasture16}. Sirius also focuses on intelligent personal assistant workloads, such as voice to text translation, 
and has been used to study the acceleration potential for interactive ML applications~\cite{sirius}.

A limitation of these benchmark suites is that they focus on single-tier applications, or at most services with two or three tiers, which drastically deviates from the 
way cloud services are deployed today. For example, even applications like websearch, which is a classic multi-tier workload, 
are configured as independent leaf nodes, which does not capture correlations across tiers. As we show in Sections~\ref{sec:architecture}-\ref{sec:application} studying the effects of microservices 
using existing benchmarks leads to fundamentally different conclusions altogether. 

The emergence of microservices has prompted recent work to study their characteristics and requirements~\cite{Sriraman18,Zhou18,Ppbench,Ueda16}. $\mu$Suite for example quantifies the system call, context switch, 
and other OS overheads in microservices~\cite{Sriraman18}, while Ueda et al.~\cite{Ueda16} show the impact of compute resource allocation, application framework, and container configuration on the performance 
and scalability of several microservices. DeathstarBench differentiates from these studies by focusing on large-scale applications with tens of unique microservices, allowing us to study effects 
that only emerge at large scale, such as network contention and cascading QoS violations due to dependencies between tiers, as well as by including diverse applications 
that span social networks, media and e-commerce services, and applications running on swarms of edge devices. 

\begin{figure*}
	\centering
\begin{minipage}{0.485\textwidth}
	\includegraphics[scale=0.35, viewport=10 0 825 440]{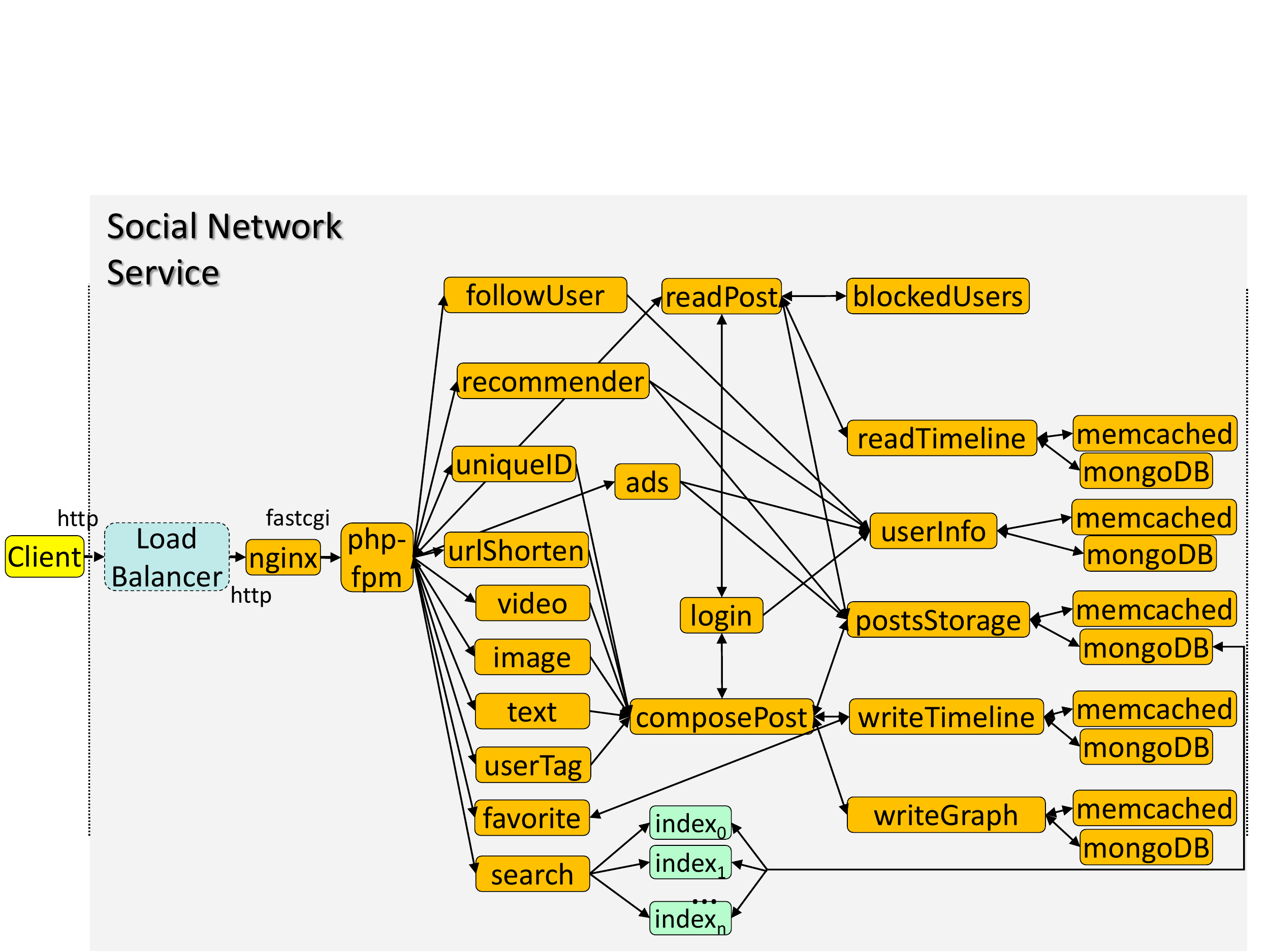}
	\caption{\label{fig:social} The architecture (microservices dependency graph) of \textit{Social Network}. }
	\end{minipage}
	\hspace{0.2cm}
	\begin{minipage}{0.485\textwidth}
	\includegraphics[scale=0.348, viewport = 10 0 825 440]{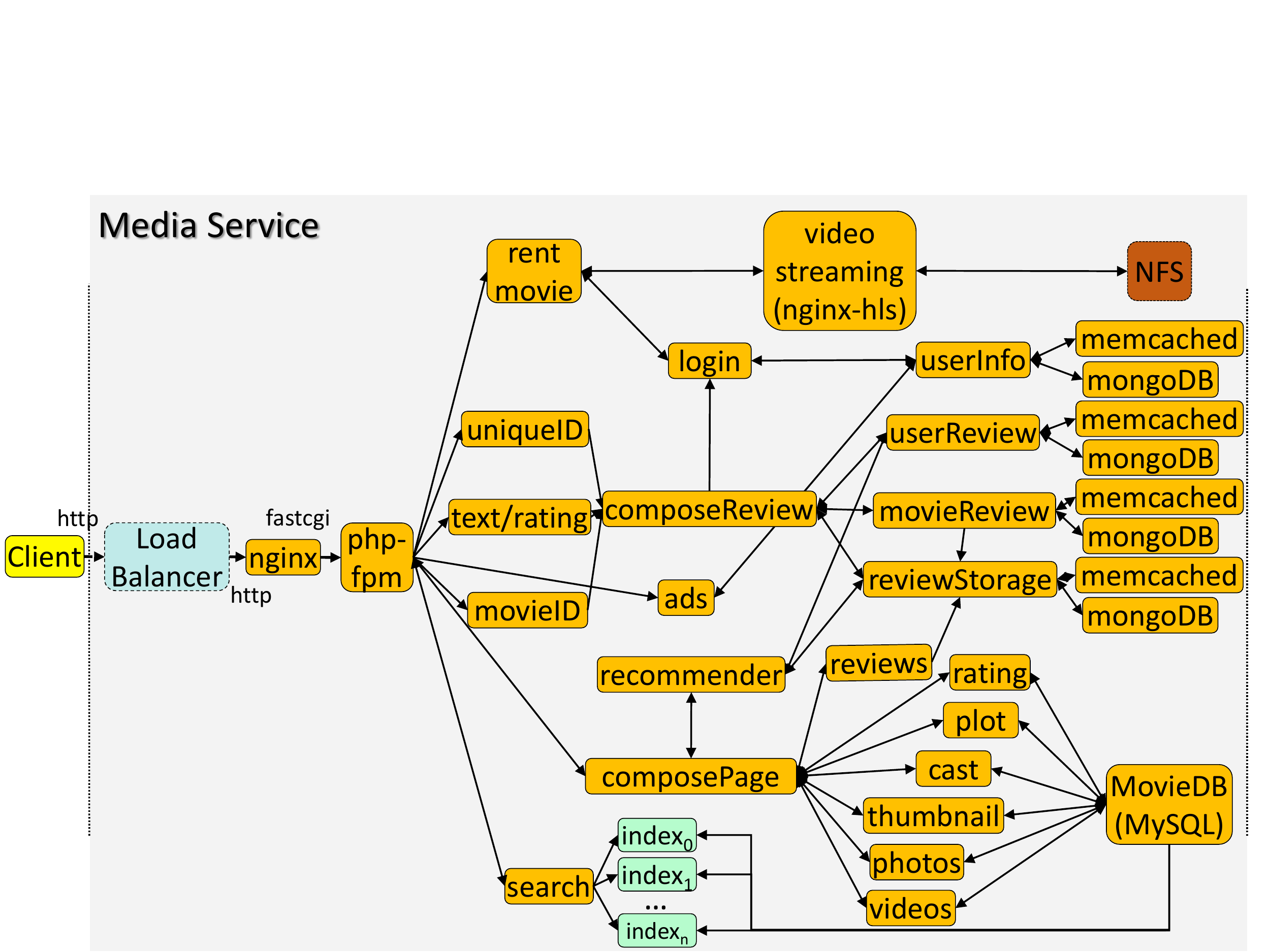}
	\caption{\label{fig:movie} The architecture of the \textit{Media Service} for reviewing, renting, and streaming movies. }
	\end{minipage}
\end{figure*}

\section{The DeathStarBench Suite}
\label{sec:overview}

We first describe the suite's design principles, and then 
present the architecture and functionality of each end-to-end service. 





\vspace{-0.05in}

\subsection{Design Principles}
\vspace{-0.02in}

\noindent DeathStarBench adheres to the following design principles: 
\vspace{-0.02in}
\begin{itemize}\setlength\itemsep{0pt}
	\item {\bf{Representativeness: }}The suite is built using popular open-source applications deployed 
		by cloud providers, such as \textit{NGINX}~\cite{nginx}, \textit{memcached}~\cite{memcached}, \textit{MongoDB}~\cite{mongodb}, \textit{RabbitMQ}~\cite{rabbitmq}, \textit{MySQL}, 
		\textit{Apache http server}, \textit{ardrone-autonomy}~\cite{autonomy,cylon}, and the Sockshop microservices
		by Weave~\cite{sockshop}. Most new code corresponds to interfaces between the services, using Apache Thrift~\cite{thrift}, gRPC~\cite{grpc}, or http requests. 
	\item {\bf{End-to-end operation: }}Open-source cloud services, such as \texttt{memcached}, 
		can function as components of a larger service, but do not capture the impact of inter-service dependencies on end-to-end performance. 
		DeathStarBench instead implements the full functionality of a service from the moment a request is generated at the client until it 
		reaches the service's backend and/or returns to the client. 
	\item {\bf{Heterogeneity: }} The software heterogeneity is both a challenge and opportunity with microservices, as different languages mean different 
		bottlenecks, synchronization primitives, levels of indirection, and development effort. 
		The suite uses applications in low- and high-level, managed and unmanaged languages including 
		C/C++, Java, Javascript, node.js, Python, html, Ruby, Go, and Scala. 
	\item {\bf{Modularity: }}
		We follow Conway's Law~\cite{conwayslaw}, i.e., the fact that the software architecture of a service follows the architecture of the company that built it 
		in the design of the end-to-end applications, to avoid excessive two-way communication between any two dependent microservices, and to ensure they are single-concerned and loosely-coupled. 
	\item {\bf{Reconfigurability: }}Easily updating components of a larger service is one of the main advantages of microservices. 
		Our RPC/HTTP API allows swapping out microservices for alternate versions, 
		with small changes to existing components. 
\end{itemize}

Table~\ref{loc_stats} shows the developed LoCs per service, and the LoCs for the communication protocol; hand-written, and auto-generated by Thrift, where applicable. The majority of 
new code for the \textit{Social Network}, \textit{Media}, \textit{E-commerce}, and \textit{Banking} services goes towards the cross-microservice API, as well as a few microservices for which no open-source 
framework existed, e.g., assigning ratings to movies. For the \textit{Swarm} application, we show code breakdown for two versions; one where the majority of computation happens in a backend cloud (\textit{Swarm Cloud}), 
and one where it happens locally on the edge devices (\textit{Swarm Edge}). 
We also show the number of unique microservices for each application, and the breakdown per programming language. 
Unless otherwise noted, all microservices run in Docker containers. 

\vspace{-0.05in}
\subsection{Social Network}
\vspace{-0.05in}

\noindent{\bf{Scope: }} The end-to-end service implements a broadcast-style social network with uni-directional follow relationships. 

\noindent{\bf{Functionality: }}Fig.~\ref{fig:social} shows the architecture of the end-to-end service. 
Users (\texttt{client})
send requests over \texttt{http}, which first reach a load balancer, implemented with \texttt{nginx}. Once a specific webserver is selected, also in 
\texttt{nginx}, the latter uses a \texttt{php-fpm} module to talk to the microservices responsible for composing and displaying posts, as well 
as microservices for advertisements, search engines, etc. All messages downstream of \texttt{php-fpm} are Apache Thrift RPCs~\cite{thrift}. 
Users can create posts embedded with text, media, links, and tags to other users. Their posts are then broadcasted to all their followers. 
Users can also read, favorite, and repost posts, as well as reply publicly, or send a direct message to another user. The application also includes 
machine learning plugins, such as ads and user recommender engines~\cite{Bottou,Netflix03,Witten,Kiwiel}, a search service using \texttt{Xapian}~\cite{Kasture16}, 
and microservices to record and display user statistics, e.g., number of followers, and to allow users to follow, unfollow, or block other accounts. 
The service's backend uses \texttt{memcached} for caching, and \texttt{MongoDB} for persistent storage for posts, profiles, 
media, and recommendations. Finally, the service is instrumented with a distributed tracing system (Sec.~\ref{sec:tracing}), 
which records the latency of each network request and per-microservice processing; 
traces are recorded in a centralized database. 
The service is broadly deployed at our institution, currently servicing several hundred users. We use this deployment 
to quantify the tail at scale effects of microservices in Section~\ref{sec:tail_at_scale}. 

\vspace{-0.05in}
\subsection{Media Service}
\vspace{-0.05in}

\noindent{\bf{Scope: }} The application implements an end-to-end service for browsing movie information, as well as reviewing, rating, renting, and streaming movies~\cite{Cockroft15,Cockroft16}. 

\noindent{\bf{Functionality: }} Fig.~\ref{fig:movie} shows the architecture of the end-to-end service. 
As with the social network, a client request hits the load balancer, which distributes requests among multiple \texttt{nginx} webservers. 
Users can search and browse information about movies, including their plot, photos, videos, cast, and review information, as well as 
insert new reviews in the system for a specific movie by logging into their account. 
Users can also select to rent a movie, which involves a payment authentication module to verify that the user has enough funds, and 
a video streaming module using \texttt{nginx-hls}, a production \texttt{nginx} module for {\smallcapital HTTP} live streaming. 
The actual movie files are stored in {\smallcapital NFS}, to avoid the latency and complexity of accessing chunked records from non-relational databases, 
while movie reviews are kept in \texttt{memcached} and \texttt{MongoDB} instances. 
Movie information is maintained in a sharded and replicated MySQL database. 
The application also includes movie and advertisement recommenders, 
as well as a couple auxiliary services for maintenance and service discovery, which are not shown in the figure. 
We are similarly deploying \textit{Media Service} as a hosting site for project demos at Cornell, 
which members of the community can browse and review.

\begin{figure*}
	\centering
	\begin{minipage}{0.485\textwidth}
	\includegraphics[scale=0.35, viewport=10 0 825 440]{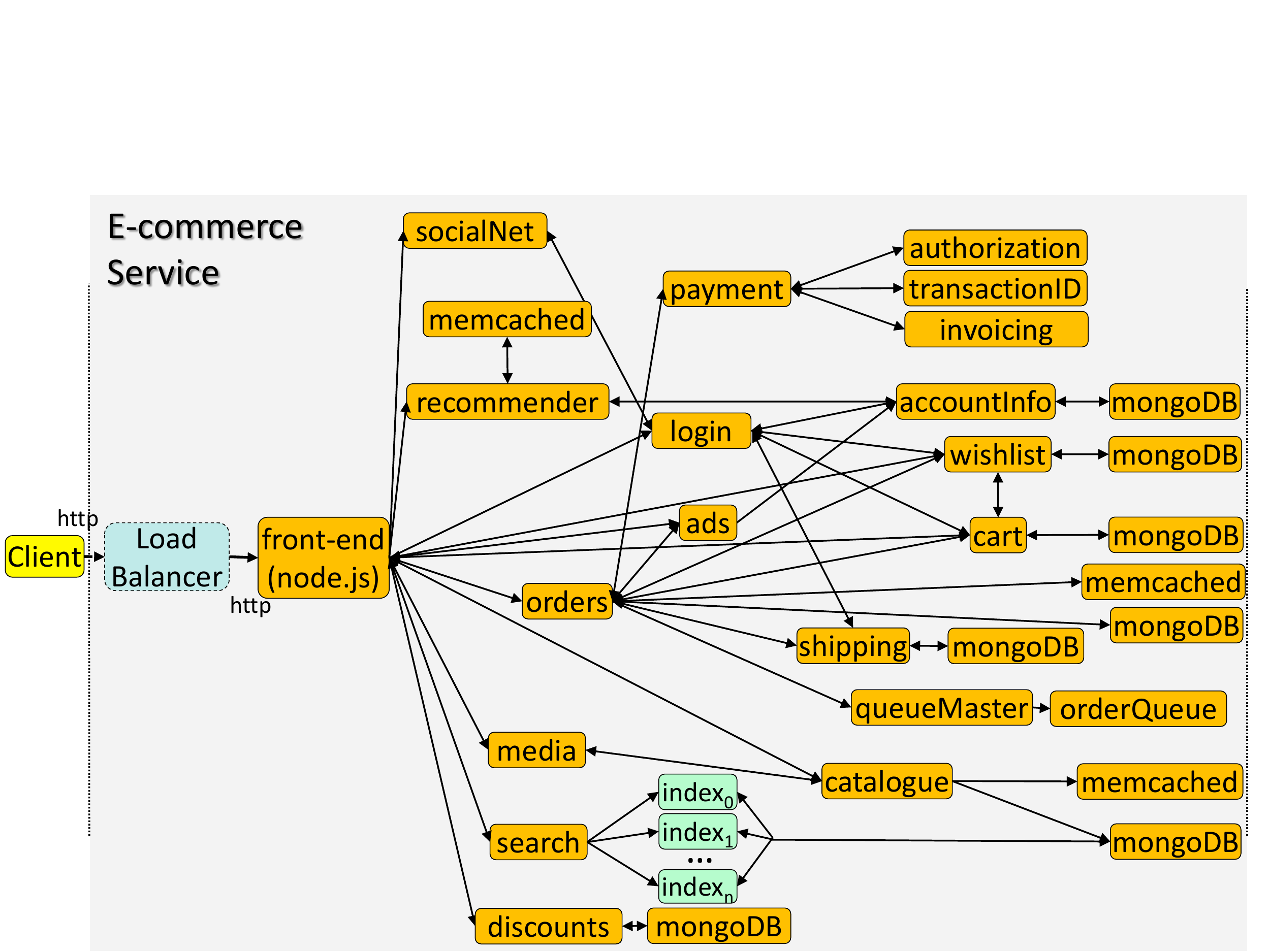}
	\caption{\label{fig:eshop} The architecture of the \textit{E-commerce} service. }
	\end{minipage}
	\hspace{0.2cm}
	\begin{minipage}{0.485\textwidth}
	\includegraphics[scale=0.348, viewport = 10 0 825 440]{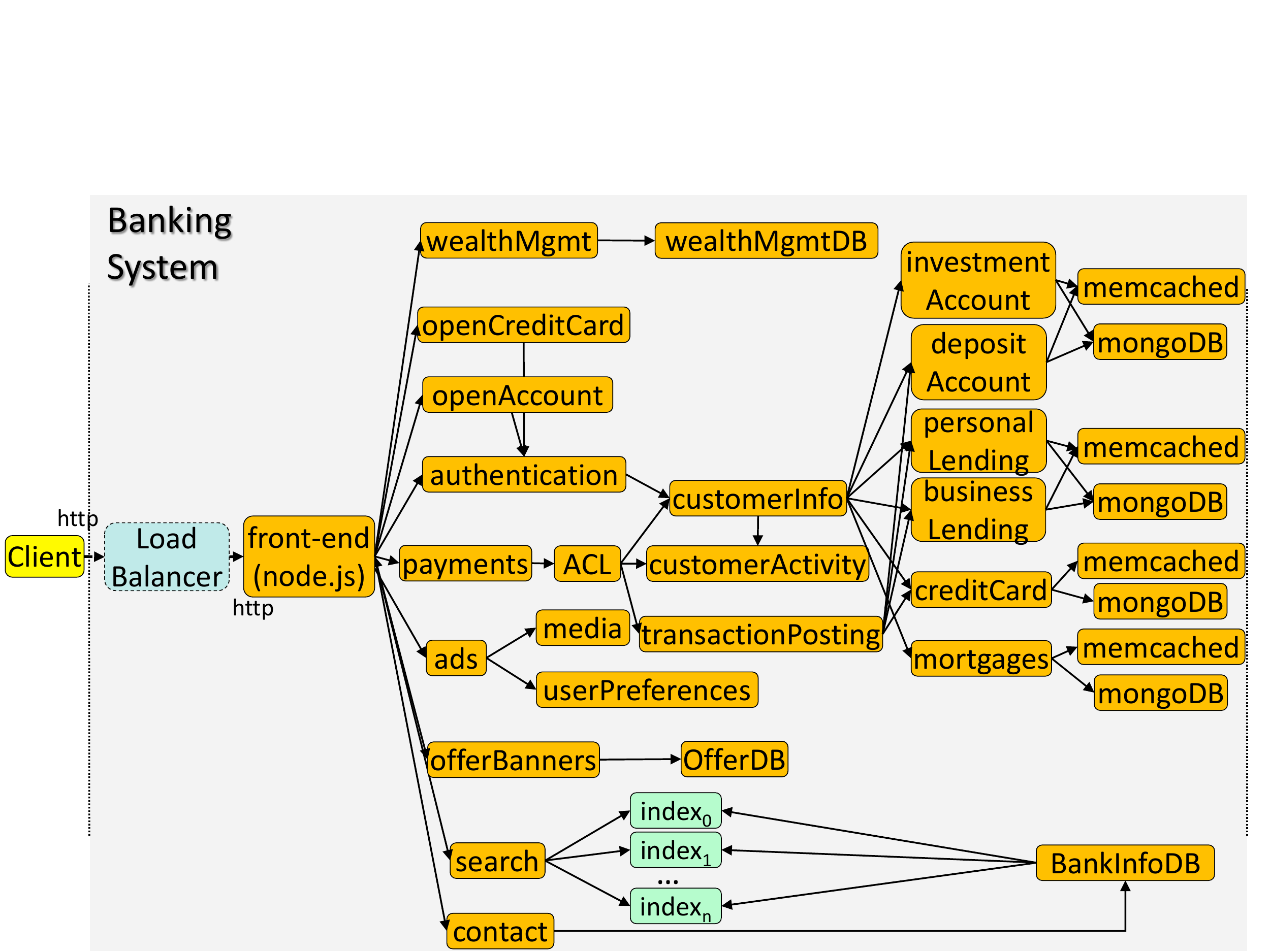}
	\caption{\label{fig:banking} The architecture of the \textit{Banking} end-to-end service. }
	\end{minipage}
\end{figure*}
\begin{figure*}
	\centering
	\begin{tabular}{ccc}
		\includegraphics[scale=0.26, viewport = 5 0 665 520]{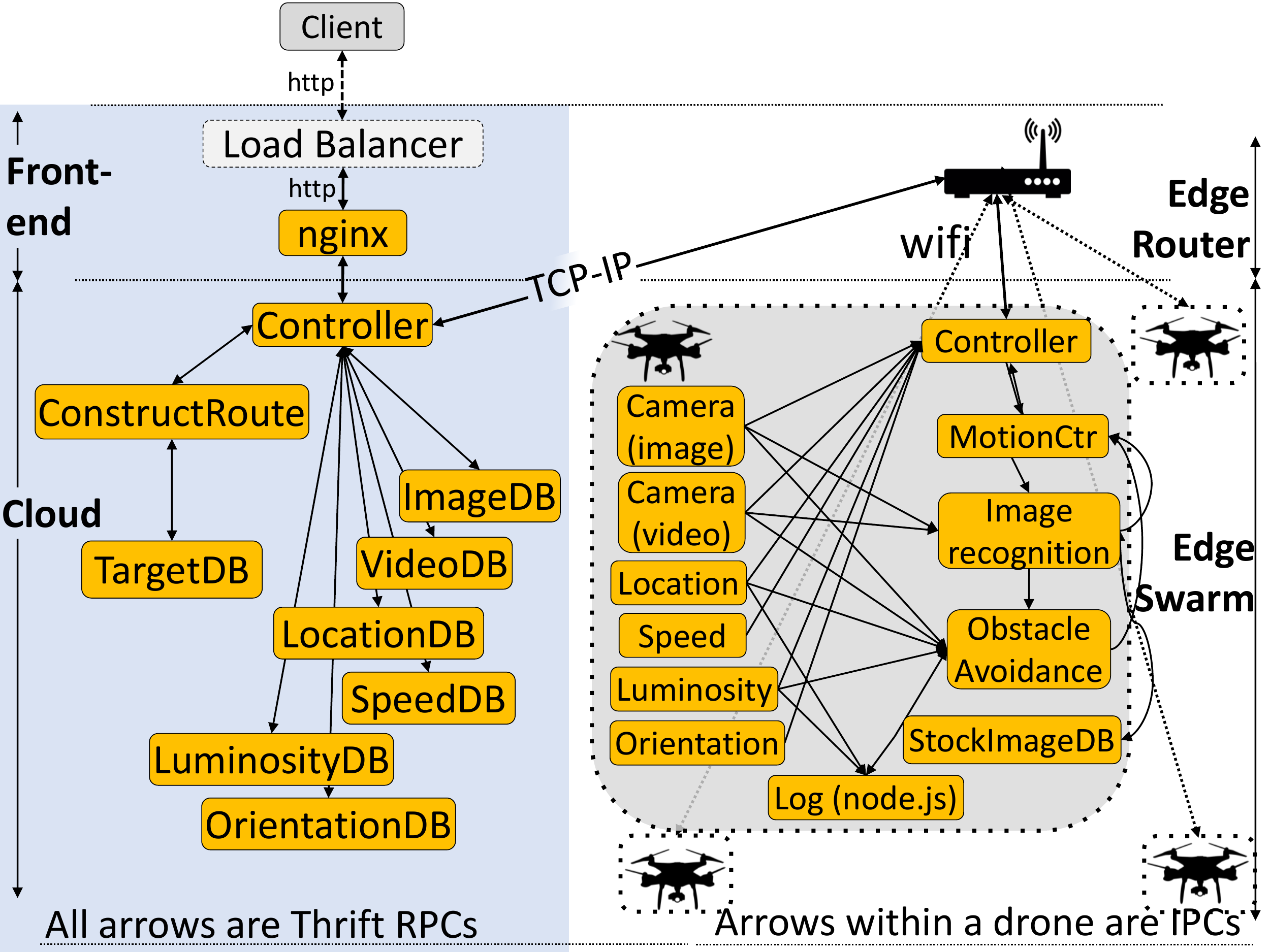} & 
		\includegraphics[scale=0.26, viewport = -35 0 675 520]{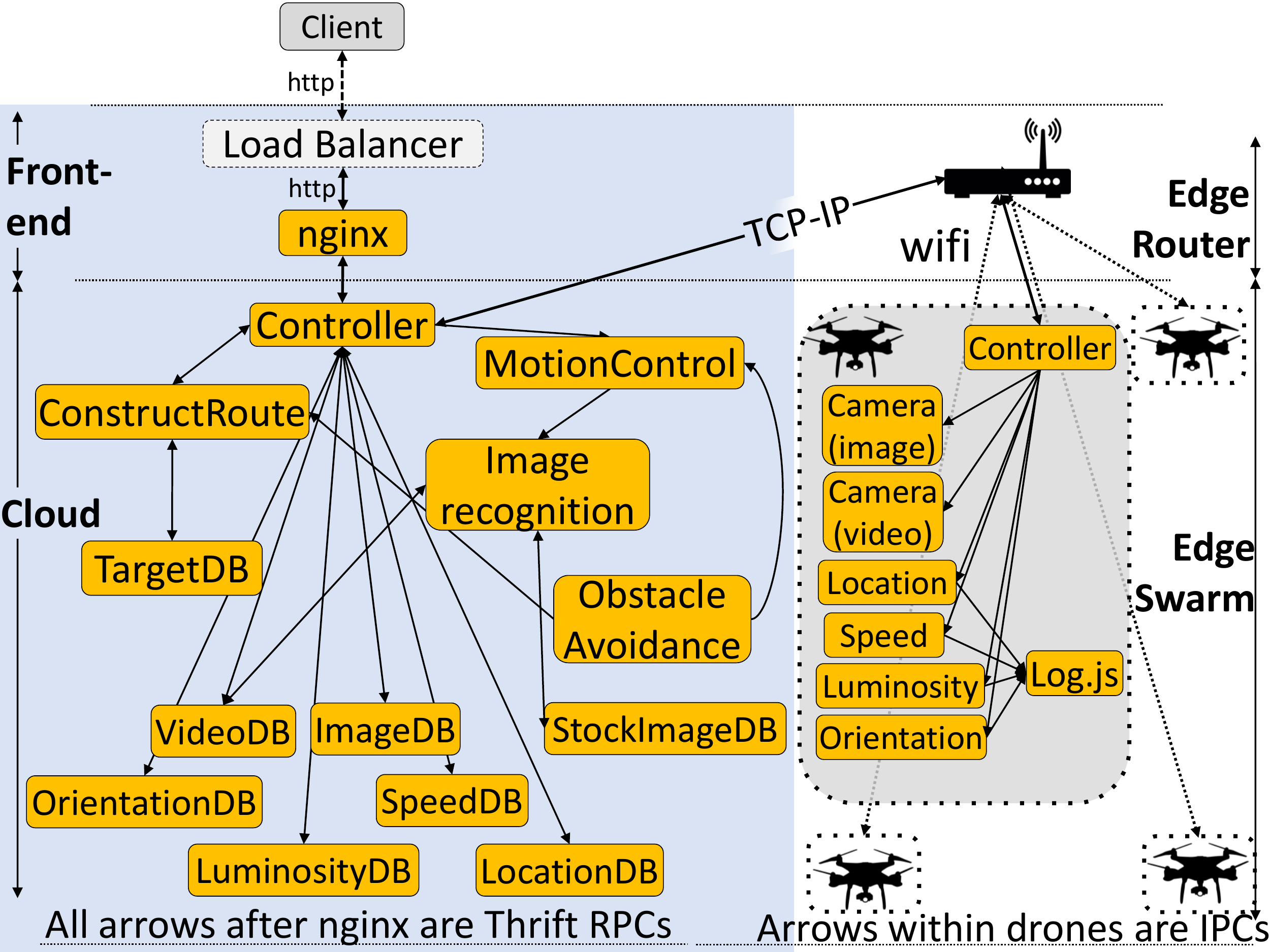} & 
		\includegraphics[scale=0.098, viewport=-50 0 580 1320,clip=true]{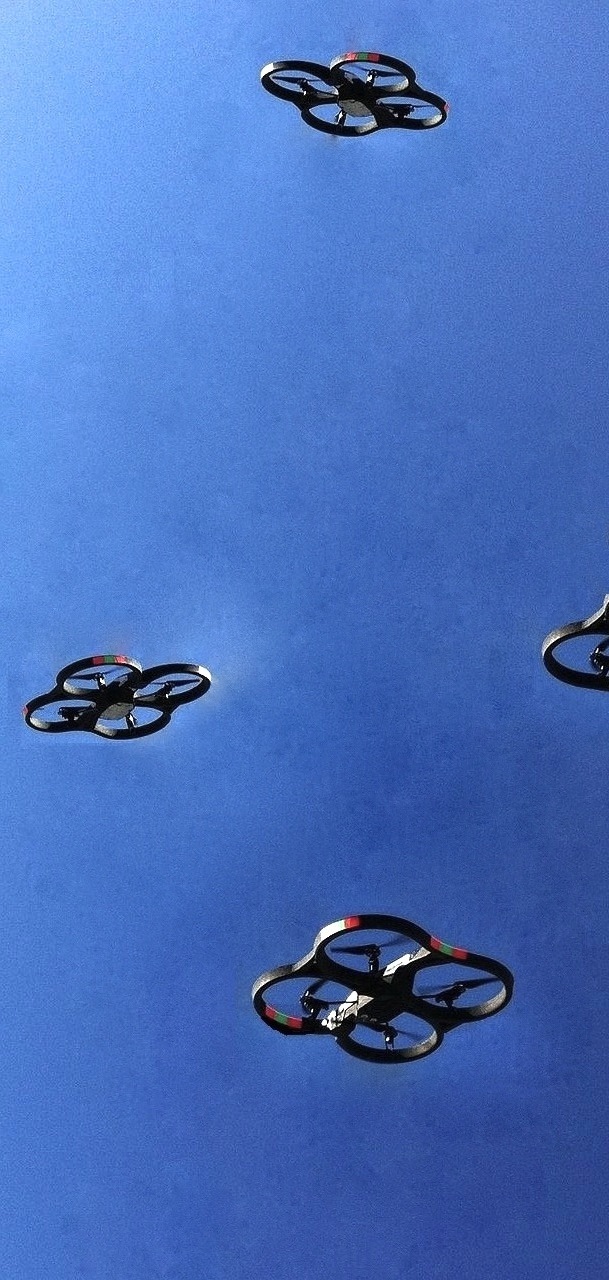} \\
\end{tabular}
\caption{\label{fig:iot} The Swam service running (a) on edge devices, and (b) on the cloud. (c) Local drone swarm executing the service. }
\end{figure*}

\subsection{E-Commerce Service}
\vspace{-0.02in}

\noindent{\bf{Scope: }}The service implements an e-commerce site for clothing. 
The design draws inspiration, and uses several components of the open-source Sockshop application~\cite{sockshop}.  

\noindent{\bf{Functionality: }}Fig.~\ref{fig:eshop} shows the architecture of the end-to-end service. 
The application front-end in this case is a \texttt{node.js} service. 
Clients can use the service to browse the inventory using \texttt{catalogue}, a Go microservice that mines the back-end 
\texttt{memcached} and \texttt{MongoDB} instances holding information about products. 
Users can also place \texttt{orders} (Go) by adding items to their \texttt{cart} (Java). 
After they \texttt{log in} (Go) to their account, they can select \texttt{shipping} options (Java), process their \texttt{payment} (Go), 
and obtain an \texttt{invoice} (Java) for their order. Orders are serialized and committed 
using \texttt{QueueMaster} (Go). Finally, the service includes 
a \texttt{recommender} engine for suggested products, 
and microservices for creating an item {\texttt{wishlist}} (Java), and displaying current discounts. 

\subsection{Banking System}

\noindent{\bf{Scope: }}The service implements a secure banking system, which users leverage to process payments, request loans, or balance their credit card. 

\noindent{\bf{Functionality: }}Users interface with a \texttt{node.js} front-end, similar to the one in the \textit{E-commerce} service to login to their account, search 
information about the bank, or contact a representative. Once logged in, a user can process a payment from their account, pay their credit card or request a new one, 
browse information about loans or request one, and obtain information about wealth management options. Most microservices are written in Java and Javascript. The back-end databases 
consist of in-memory \texttt{memcached}, and persistent \texttt{MongoDB} instances. The service also has a relational database (\texttt{BankInfoDB}) that includes information about 
the bank, its services, and representatives. 

\subsection{Swarm Coordination}

\noindent{\bf{Scope: }} Finally, we explore a different execution environment for microservices, where applications run both 
on the cloud and on edge devices. The service coordinates the routing of a swarm of programmable drones, which perform image recognition and 
obstacle avoidance. 

\noindent{\bf{Functionality: }}We explore two version of this service. In the first (Fig.~\ref{fig:iot}a), the majority of the computation happens on the drones, 
including the motion planning, image recognition, and obstacle avoidance, with the cloud only constructing the initial route per-drone 
(Java service \texttt{ConstructRoute}), and holding persistent copies of sensor data. This architecture avoids the high network latency 
between cloud and edge, however, it is limited by the on-board resources. The \texttt{Controller} and \texttt{MotionController} 
are implemented in Javascript, while \texttt{ImageRecognition} is using \textit{jimp}, a node.js library for image recognition~\cite{jimp}, 
and \texttt{ObstacleAvoidance} in C++. Services on the drones run natively, 
and communicate with each other over {\smallcapital IPC}, while the cloud and drones communicate over \texttt{http} 
to avoid installing the heavy dependencies of Thrift on the edge devices. 

In the second version (Fig.~\ref{fig:iot}b), the cloud is responsible for most of the computation. 
It performs motion control, image recognition, and obstacle avoidance for all drones, 
using the \texttt{ardrone-autonomy}~\cite{autonomy}, and Cylon~\cite{cylon} libraries, 
in OpenCV and Javascript respectively. The edge devices are only responsible for 
collecting sensor data and transmitting them to the cloud, as well as recording some diagnostics using a local node.js \texttt{logging} service. 
In this case, almost every action suffers the cloud-edge network latency, although services benefit from the additional cloud resources. 
We use 24 programmable Parrot AR2.0 drones (a subset is seen in Fig.~\ref{fig:iot}c), 
together with a backend cluster of 20 two-socket, 40-core servers. 
Drones communicate with each other and the cluster over a wireless router. 


\subsection{Methodological Challenges of Microservices}
\label{sec:tracing}

A major challenge with microservices is that one cannot simply rely on the client to report performance, as with traditional client-server applications. 
Resolving performance issues requires determining which microservice(s) is the culprit of a QoS violation, which typically happens through distributed 
tracing. 
We developed and deployed a distributed tracing system that records per-microservice latencies at {\smallcapital RPC} granularity using the Thrift timing interface. 
{\smallcapital RPC}s or {\smallcapital REST} requests are timestamped upon arrival and departure from each microservice by the tracing module, 
and data is accumulated by the \texttt{Trace Collector}, implemented similarly 
to the Zipkin Collector~\cite{zipkin}, and stored in a centralized Cassandra database. We additionally track the time spent processing network requests, as opposed 
to application computation using a similar methodology to~\cite{Li14}. 
We verify that the overhead from tracing is negligible, less than $0.1\%$ 
on end-to-end latency in all cases, which is tolerable for such systems~\cite{dapper,mysterymachine,gwp}. 






\subsection{Provisioning \& Query Diversity}
\label{sec:provisioning}

Before characterizing the architectural behavior of microservices, we provision the end-to-end applications to ensure that microservices are 
used in a balanced way, and that no single microservice introduces early bottlenecks due to resource saturation. To do so, we start with 
a fair resource allocation for all microservices of an end-to-end workload, and upsize saturated microservices until all tiers saturate at about the same load. The 
ratio of resources between tiers varies significantly across end-to-end services, highlighting the need for application-aware resource management. 

\begin{figure}
	\centering
	\includegraphics[scale=0.21, viewport = 200 0 1100 430]{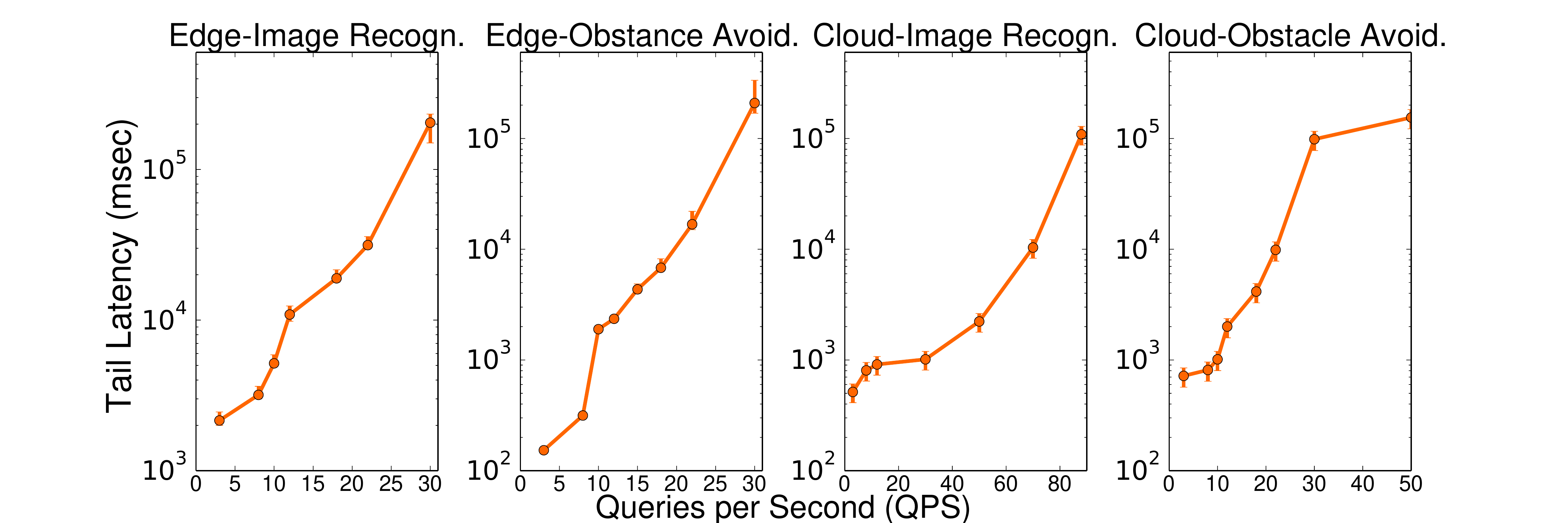} \\
	\caption{\label{fig:load_swarm} Throughput-tail latency for the Swarm service when execution happens at the edge versus the cloud. }
\end{figure}

Different query types also achieve different performance in each service. For example, \texttt{composePost} requests in the \textit{Social Network} 
vary in the media they embed in a message, ranging from text-only messages, to posts including image and video files (we keep videos 
within a few MBs, similar to the allowable video sizes in production social networks like Twitter). 
Reposting a post incurs the longest latency across query types for \textit{Social Network}, as it must first read an existing post, prepend to it, 
and then propagate the message across the user's followers' timelines. 



In \texttt{E-commerce}, on the other hand, placing an order, which includes adding an item to the cart, 
logging in to the account, confirming payment, and selecting shipping, takes 1-2 orders of magnitude longer than browsing the eshop's catalogue. 
In reality, placing an order requires interaction with the end user; in our case we automate the client's decisions so they incur zero delay, 
making latency server-dominated. The trends across query types are similar for the \textit{Media} and \textit{Banking} services, with processing 
payments, either to rent a movie, or to perform a transaction in a bank account, dominating latency and defining each service's saturation point. 

Finally, in Fig.~\ref{fig:load_swarm}, we compare the performance of the IoT application when computation happens at the edge versus the cloud.
Since drones have to communicate with a wireless router
over a distance of several tens of meters, latencies are significantly higher than for the cloud-only services. When processing happens in the cloud, latency at \textit{low load} is higher,
penalized by the long network delay. As load increases however, edge devices quickly become oversubscribed due to the limited on-board resources, with processing on the cloud achieving \textit{7.8x} higher 
throughput for the same tail latency, or \textit{20x} lower latency for the same throughput. 
Obstacle avoidance shows a different trade-off, since it is less compute-intensive, and more latency-critical. Offloading obstacle avoidance to the cloud at low load can have catastrophic
consequences if route adjustment is delayed, which highlights the importance of latency-aware resource management between cloud and edge, especially for safety-critical computation.

\section{Architectural Implications}
\label{sec:architecture}

\noindent{\bf{Methodology: }}We first evaluate the end-to-end services on a local cluster with 20 two-socket 40-core Intel Xeon servers (E2699-v4 and E5-2660 v3) with 
128-256GB memory each, connected to a 10GBps ToR switch with 10Gbe NICs. All servers are running Ubuntu 16.04, and unless otherwise noted power management and turbo 
boosting are turned off. 

\begin{figure}
	\centering
	\begin{tabular}{ccc}
		\multicolumn{2}{c}{\includegraphics[scale=0.202, viewport = 390 0 822 70]{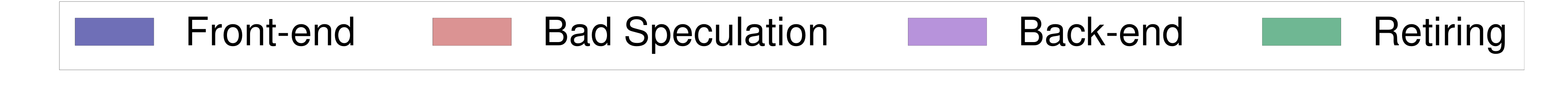}} \\
		\includegraphics[scale=0.202, viewport = 75 20 742 420]{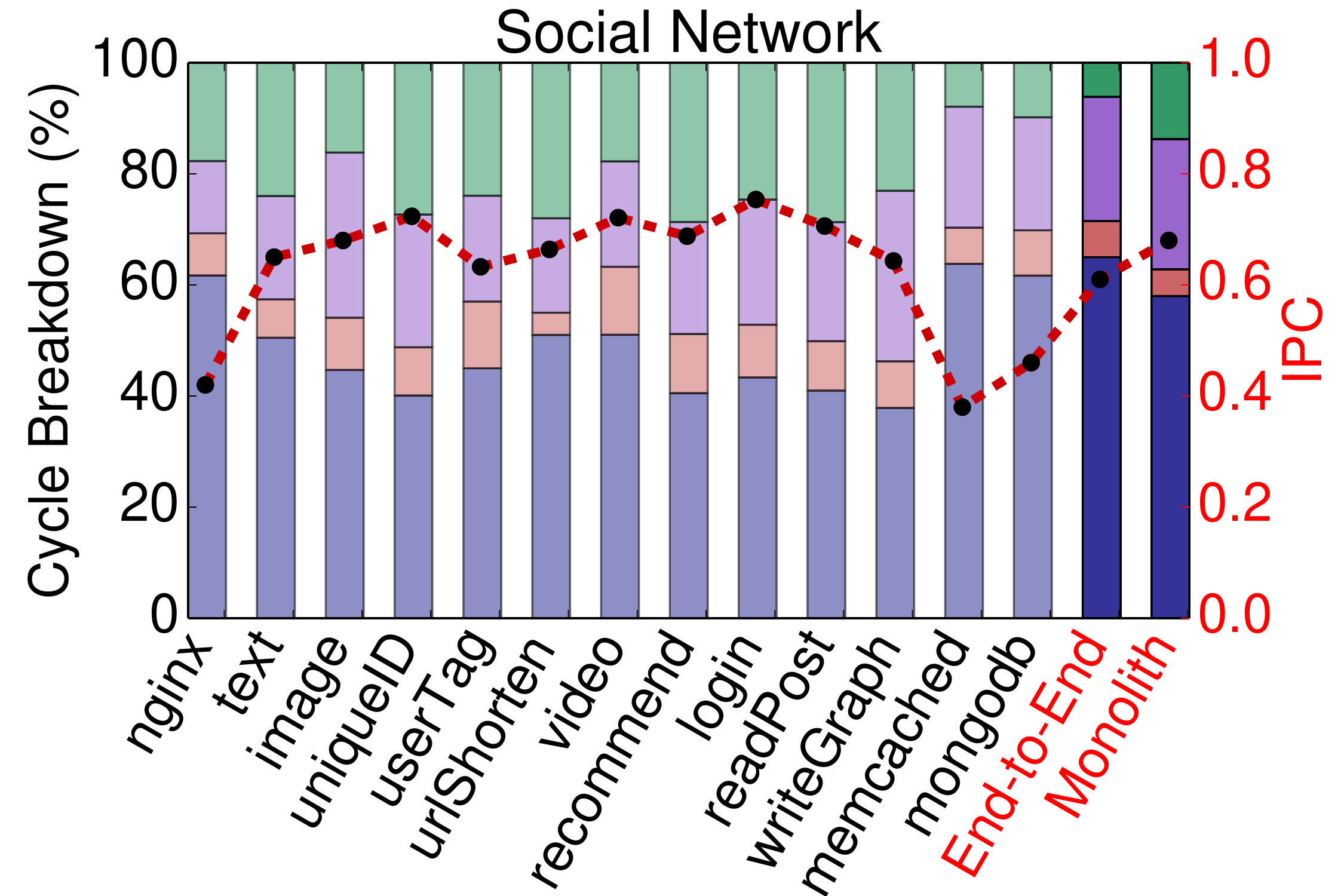} & 
		\includegraphics[scale=0.202, viewport = 150 20 702 420]{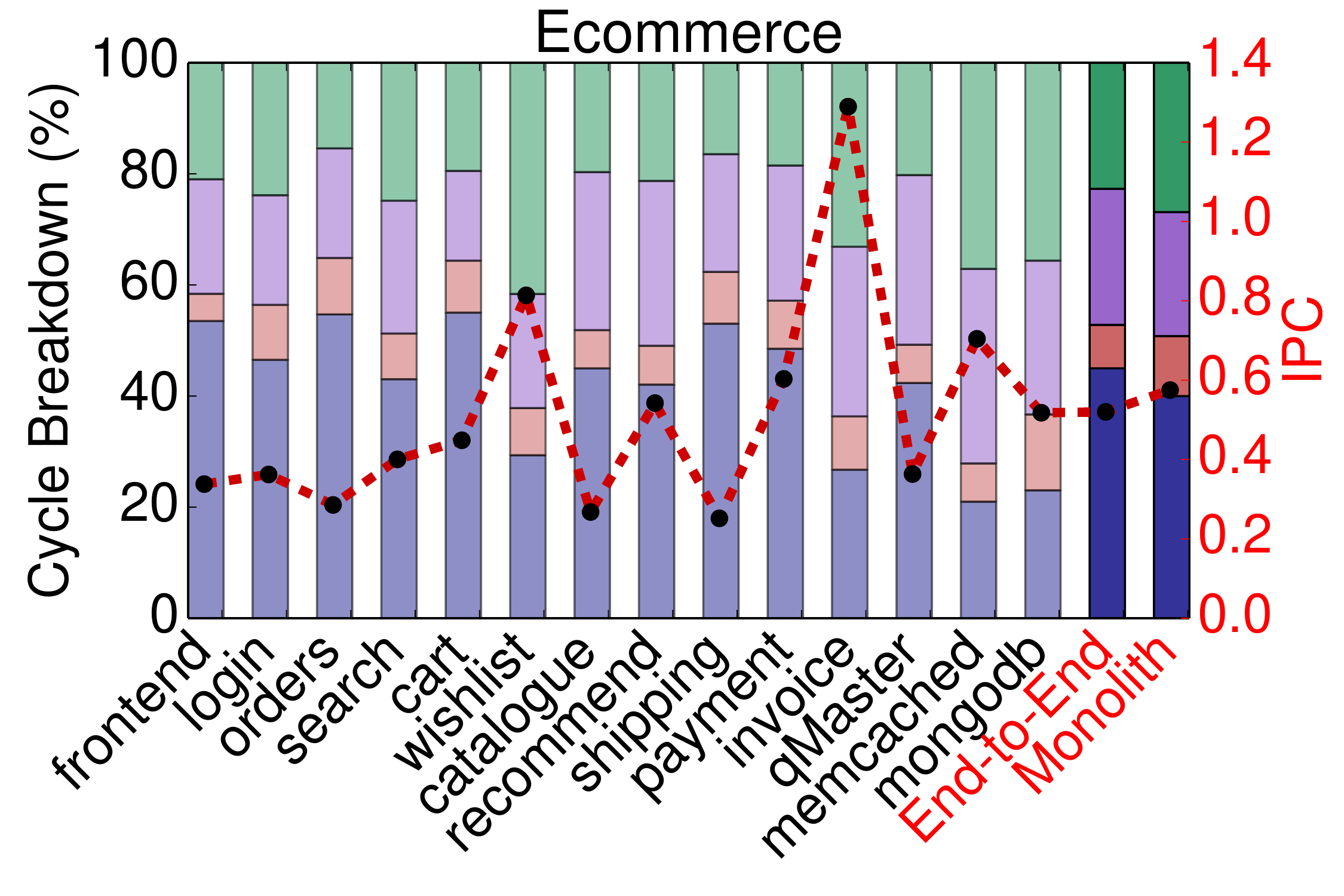} & 
	\end{tabular}
	\caption{\label{fig:cpi} Cycle breakdown and {\smallcapital IPC} for the \textit{Social Network} and \textit{E-commerce} services.}
\end{figure}

\noindent{\bf{Cycles breakdown and IPC: }}
We use Intel vTune~\cite{vtune} to break down the cycles, and identify bottlenecks. 
Fig.~\ref{fig:cpi} shows the {\smallcapital IPC} and cycles for each microservice in the \textit{Social Network} and \textit{E-commerce} services. 
We omit the figures for the other services, however the observations are similar. 

Across all services a large fraction of cycles, often the majority, is spent in the processor front-end. 
Front-end stalls occur for several reasons, including long memory accesses and i-cache misses. 
This is consistent with studies on traditional cloud applications~\cite{Cloudsuite12,Kanev15}, although 
to a lesser extent for microservices than for monolithic services (\textit{memcached}, \textit{mongodb}), given their smaller code footprint. 
The majority of front-end stalls are due to \textit{fetch}, while branch mispredictions account for a smaller fraction of stalls for microservices 
than for other interactive applications, either cloud or IoT~\cite{Cloudsuite12, Reddi15}. 
Only a small fraction of total cycles goes towards committing instructions (21\% on average for \textit{Social Network}), denoting that current systems are poorly 
provisioned for microservices-based applications. 

\textit{E-commerce} includes a few microservices that go against this trend, 
with high {\smallcapital IPC} and high percentage of retired instructions, such as {\smallcapital\texttt{Search}}. Search (xapian~\cite{Kasture16}) is already optimized for memory locality, and has a relatively 
small codebase, which explains the fewer front-end stalls. The same applies for simple microservices, such as the {\smallcapital\texttt{wishlist}} for which i-cache misses are practically 
negligible. \textit{E-commerce} also includes a {\smallcapital\texttt{recommender}} engine, whose {\smallcapital IPC} is extremely low; this is again in agreement with studies on the architectural behavior of 
ML applications~\cite{sirius}. The challenge with microservices is that although individual application components may be well understood, the structure of the 
end-to-end dependency graph defines how individual services affect the overall performance. For both services, we also show the cycles breakdown and {\smallcapital IPC} for corresponding 
applications with the same end-to-end functionality from the user's perspective, but built as monoliths. In both cases, monoliths are developed in Java, and include all 
application functionality, except for the backend databases (in {\smallcapital\texttt{memcached}} and {\smallcapital\texttt{MongoDB}}), in a single binary. The cycles breakdown is not drastically different 
for monoliths compared to microservices, although they experience slightly higher percentages of committed instructions, due to reduced front-end stalls, as they are less likely to 
wait for network requests to complete. {\smallcapital IPC} is also similar to microservices, and consistent with previous studies on cloud services~\cite{Kasture16,Cloudsuite12}. 

\begin{figure}
	\centering
	\begin{tabular}{cc}
		\includegraphics[scale=0.21, viewport = 150 20 552 420]{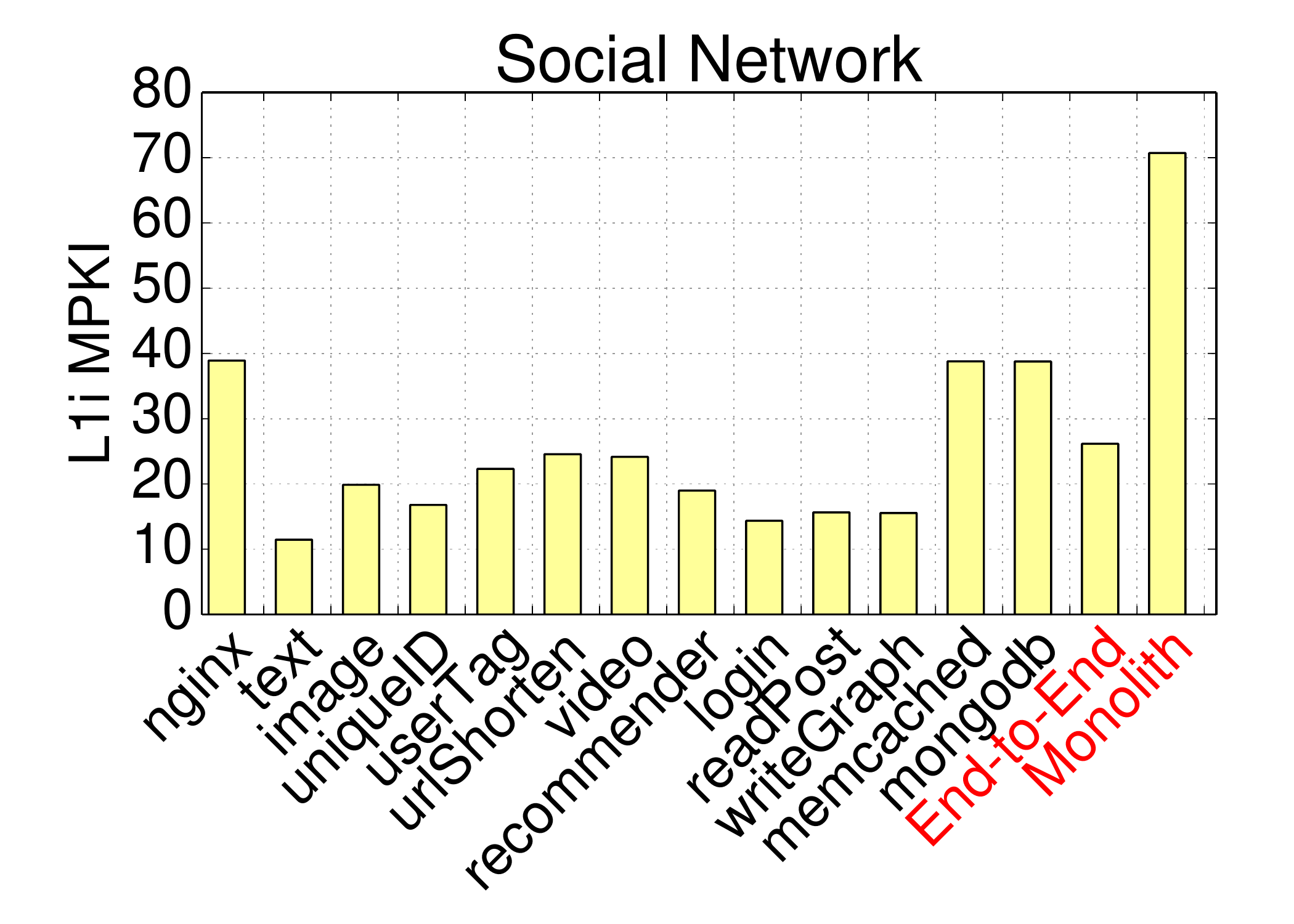} & 
		\includegraphics[scale=0.21, viewport = 0 20 452 420]{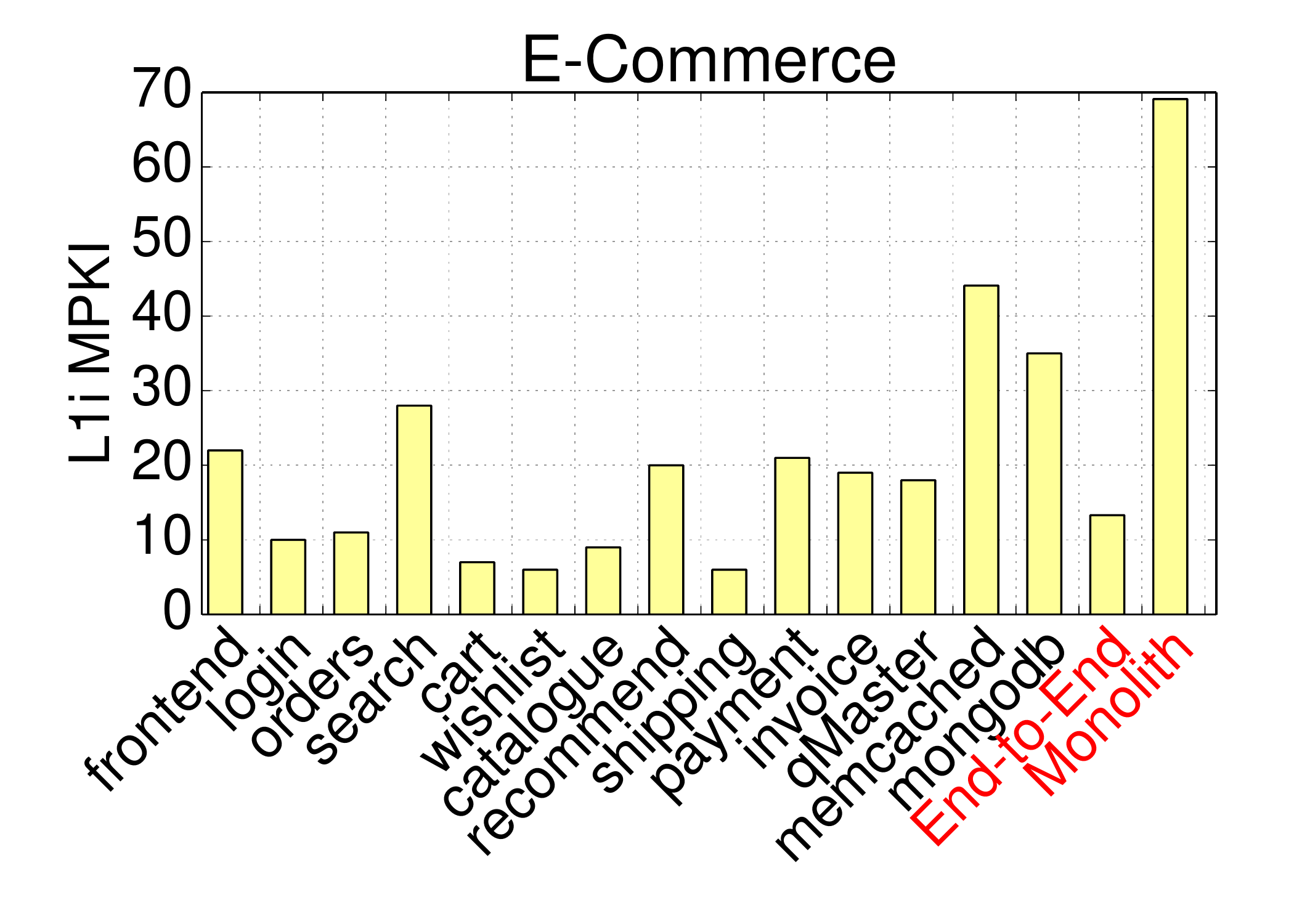} 
	\end{tabular}
	\caption{\label{fig:l1i} L1-i misses in \textit{Social Network} and \textit{E-commerce}.}
\end{figure}


\noindent{\bf{I-cache pressure: }}Prior work has characterized the high pressure cloud applications put on the instruction caches~\cite{Cloudsuite12, Kaynak13}. 
Since microservices decompose what would be one large binary to many small, loosely-connected services, we examine whether previous results on i-cache pressure still hold. 
Fig.~\ref{fig:l1i} shows the {\smallcapital MPKI} of each microservice for the \textit{Social Network} and \textit{E-commerce} applications. 
We also include the back-end caching and database layers, as well as 
the corresponding L1i {\smallcapital MPKI} for the monolithic implementations. 

First, the i-cache pressure of {\smallcapital\texttt{nginx}}, {\smallcapital\texttt{memcached}}, {\smallcapital\texttt{MongoDB}}, and especially the monoliths 
remains high, consistent with prior work~\cite{Cloudsuite12,Kaynak13,Reddi15}. The i-cache pressure of the 
remaining microservices though is considerably lower, especially for \textit{E-commerce}, an expected observation given the microservices' small code footprints. 
Since {\smallcapital\texttt{node.js}} applications outside the context of microservices do not have low i-cache miss rates~\cite{Reddi15}, we conclude that it is the simplicity 
of microservices which results in better i-cache locality. 
Most L1i misses, especially in the \textit{Social Network} happen in the kernel, and are caused by Thrift. 
We also examined the {\smallcapital LLC} and {\smallcapital D-TLB} misses, and found them considerably lower 
than for traditional cloud applications, which is consistent 
with the push for microservices to be mostly stateless. 




\begin{figure}
	\centering
	\begin{tabular}{p{1.20cm}p{1.35cm}p{1.32cm}p{1.32cm}p{1.28cm}}
		\includegraphics[scale=0.14, viewport=100 0 300 340]{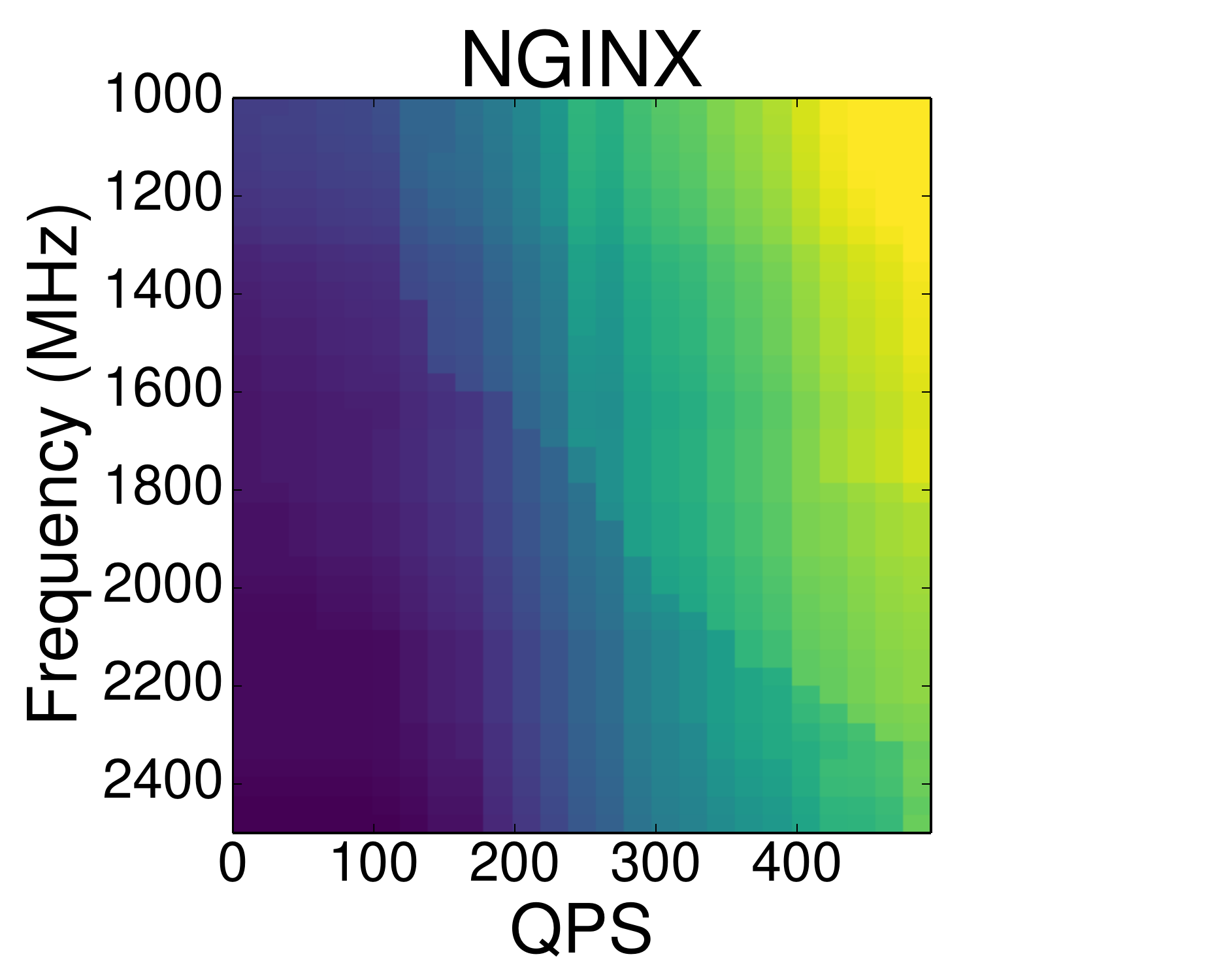} &
		\includegraphics[scale=0.14, trim=2.4cm 0 4.2cm 0, clip=true]{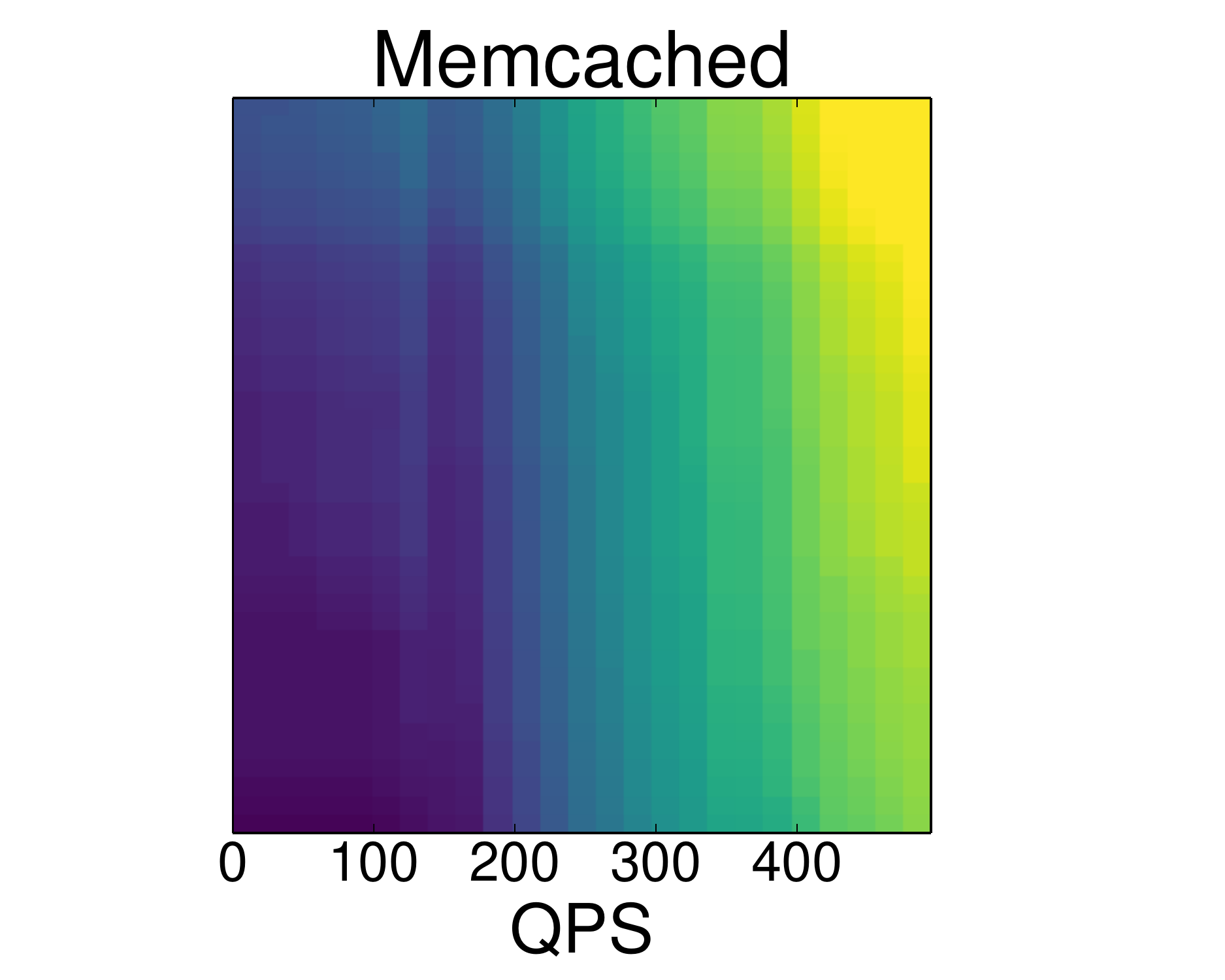} &
		\includegraphics[scale=0.14, trim=2.3cm 0 4.2cm 0, clip=true]{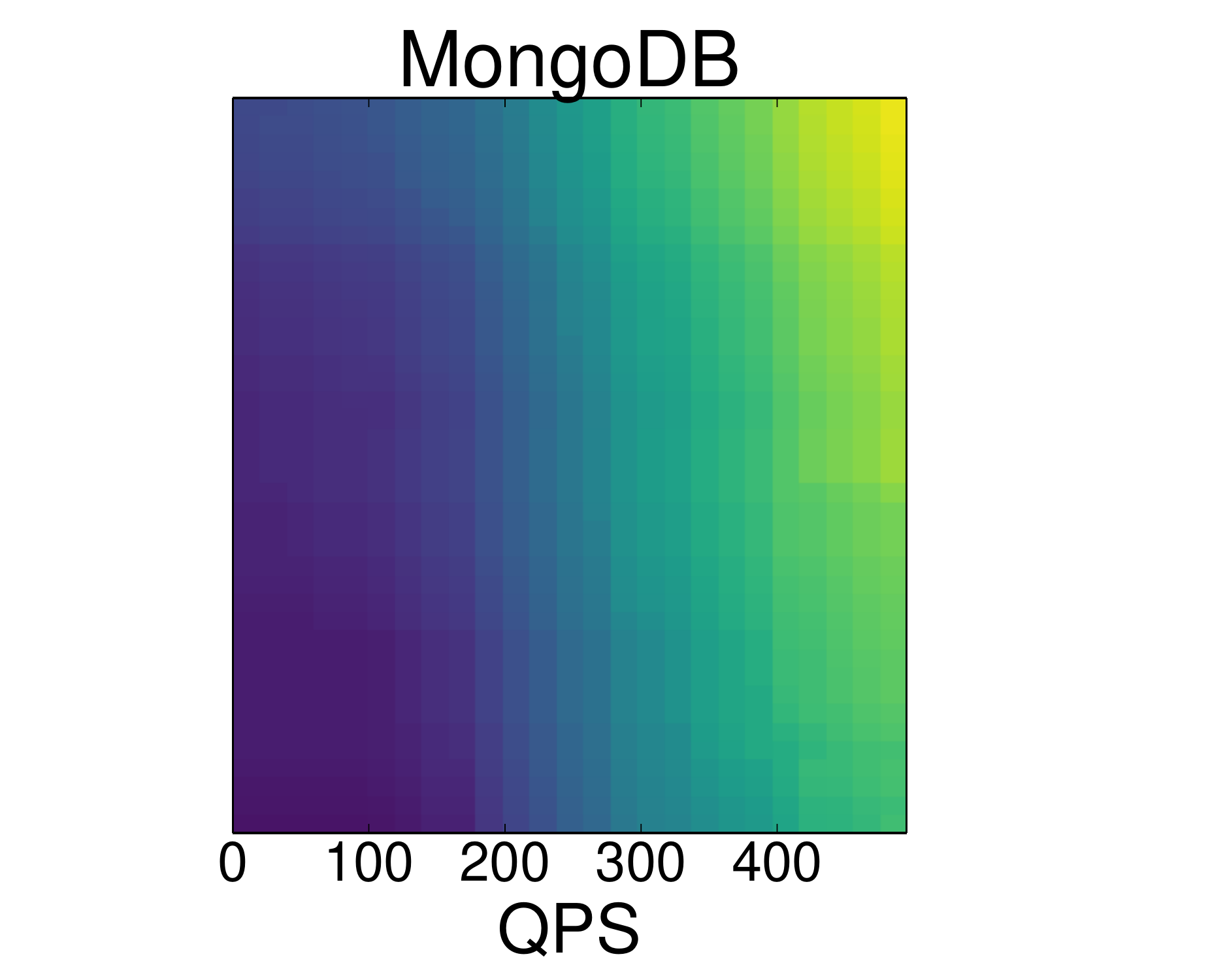} &
		\includegraphics[scale=0.14, trim=2.5cm 0 4.2cm 0, clip=true]{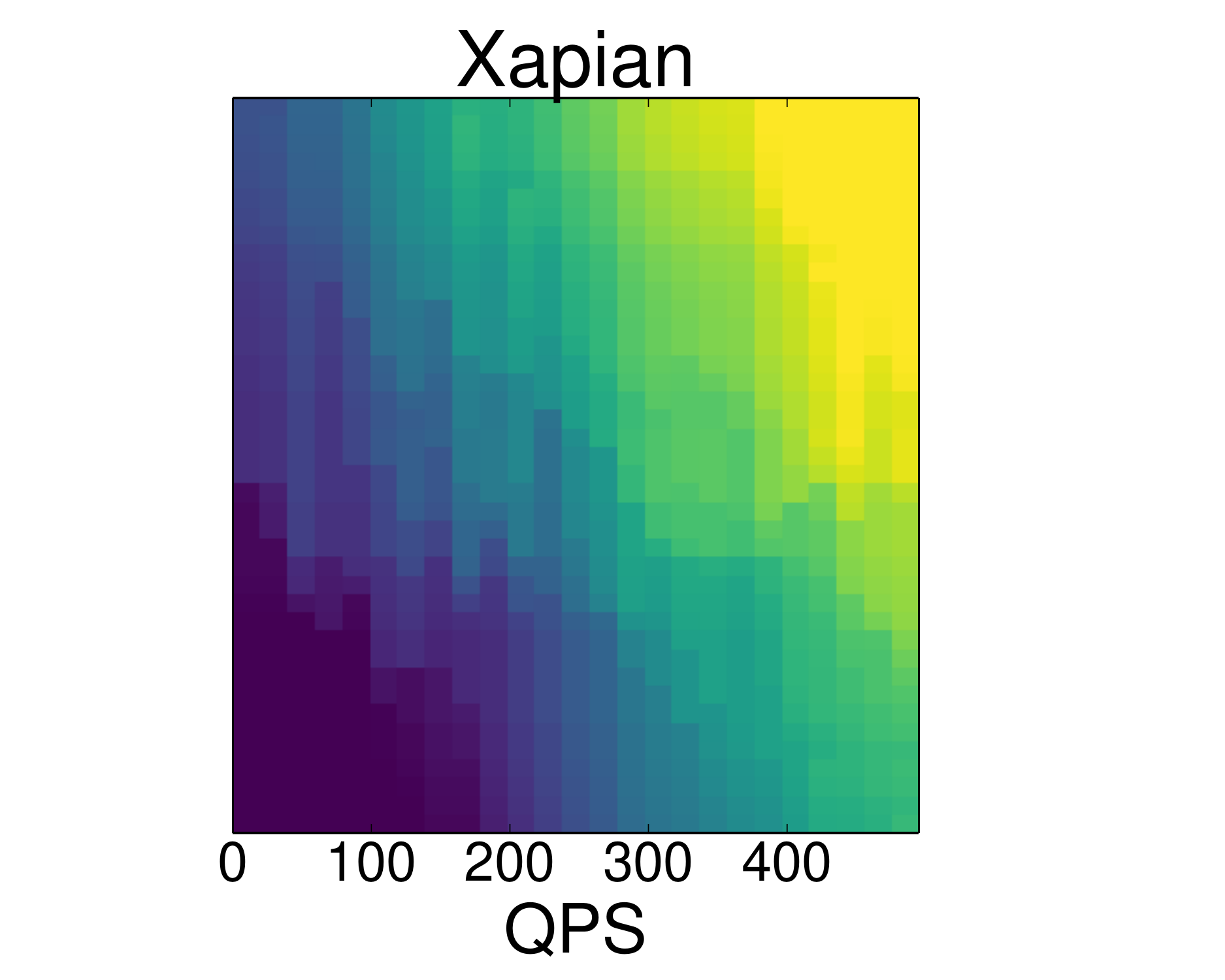} &
		\includegraphics[scale=0.14, trim=2.3cm 0 4.5cm 0, clip=true]{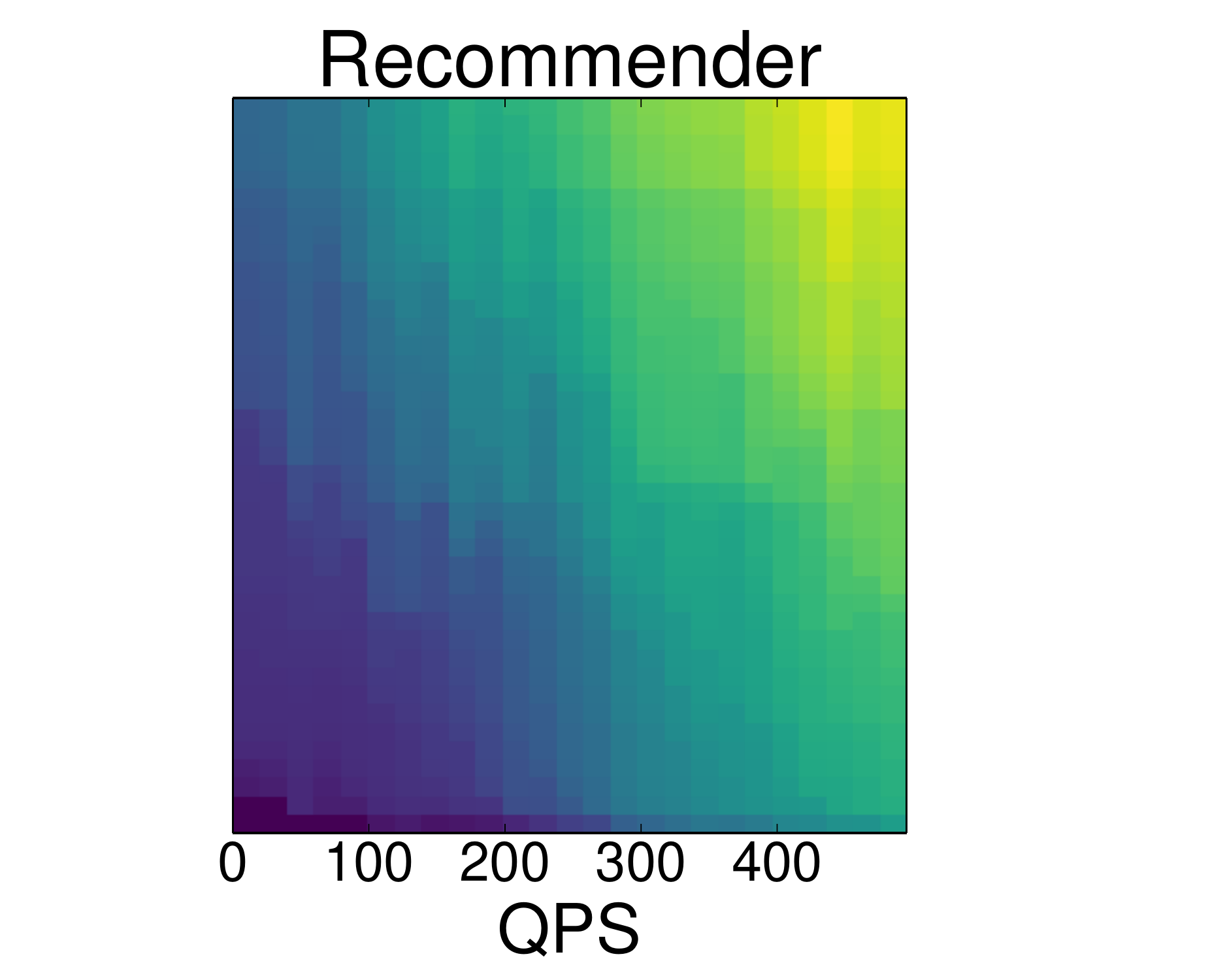} \\
		\includegraphics[scale=0.14, viewport=100 0 300 340]{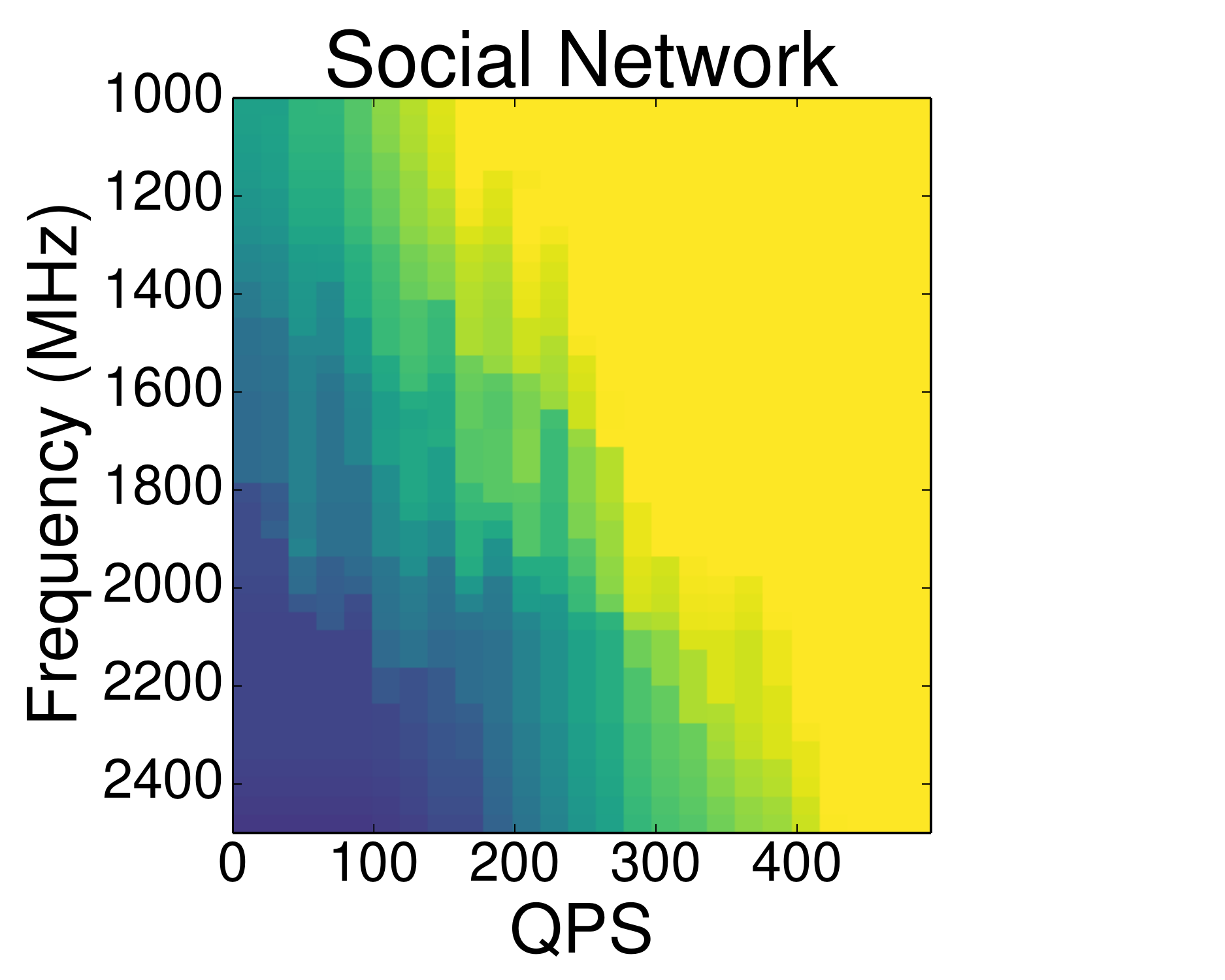} &
		\includegraphics[scale=0.14, trim=2.4cm 0 4.2cm 0, clip=true]{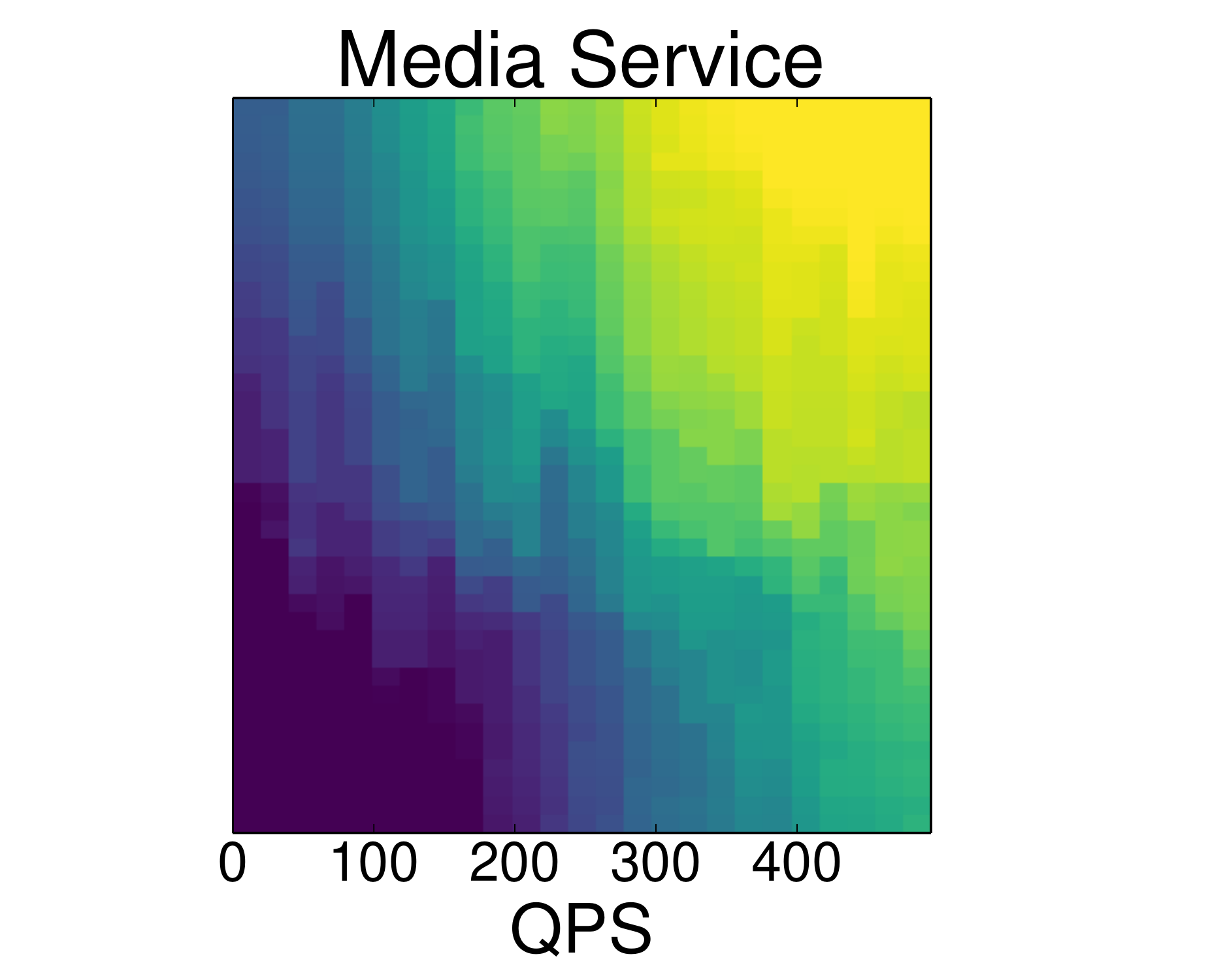} &
		\includegraphics[scale=0.14, trim=2.3cm 0 4.2cm 0, clip=true]{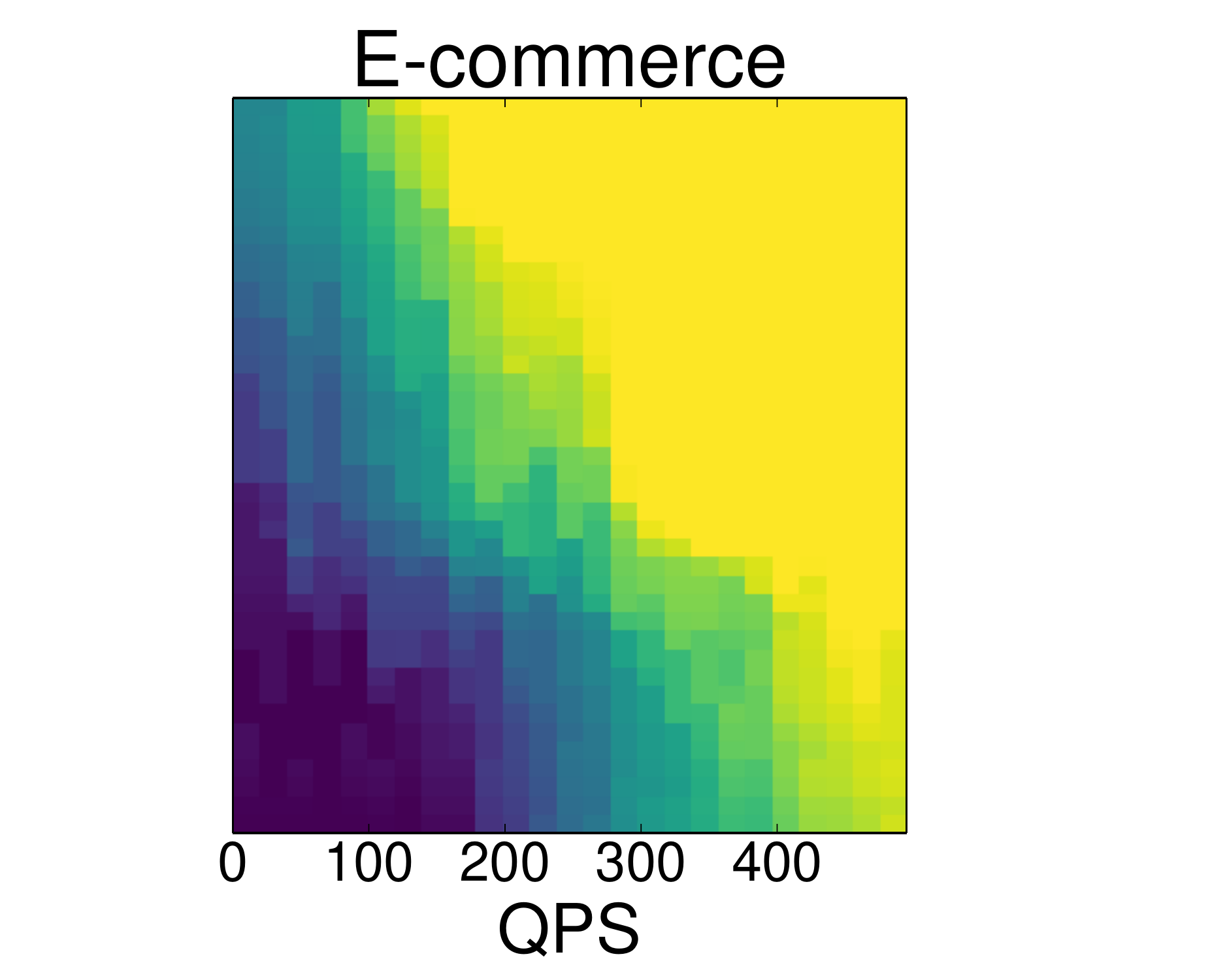} &
		\includegraphics[scale=0.14, trim=2.5cm 0 4.2cm 0, clip=true]{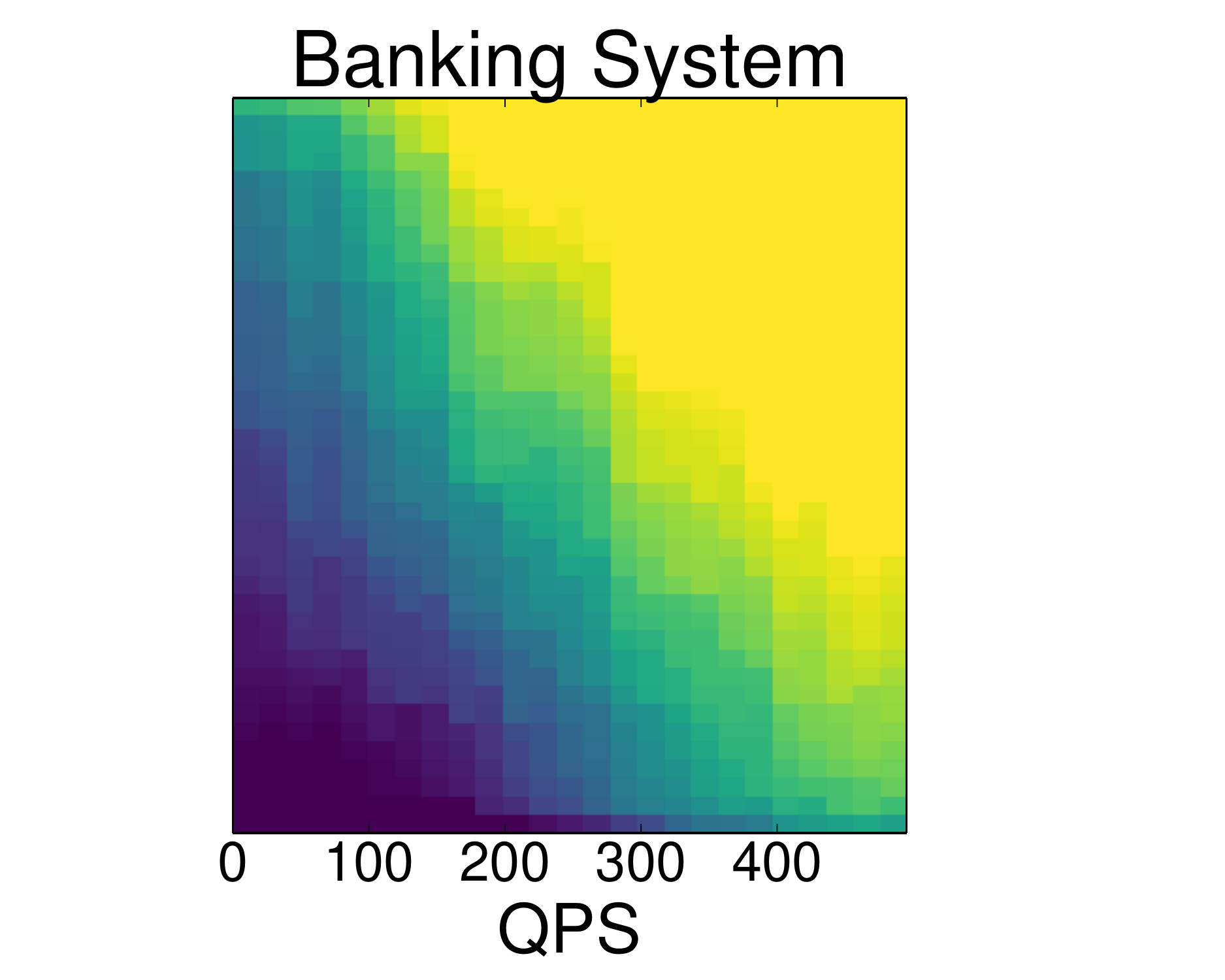} &
		\includegraphics[scale=0.14, trim=2.3cm 0 4.5cm 0, clip=true]{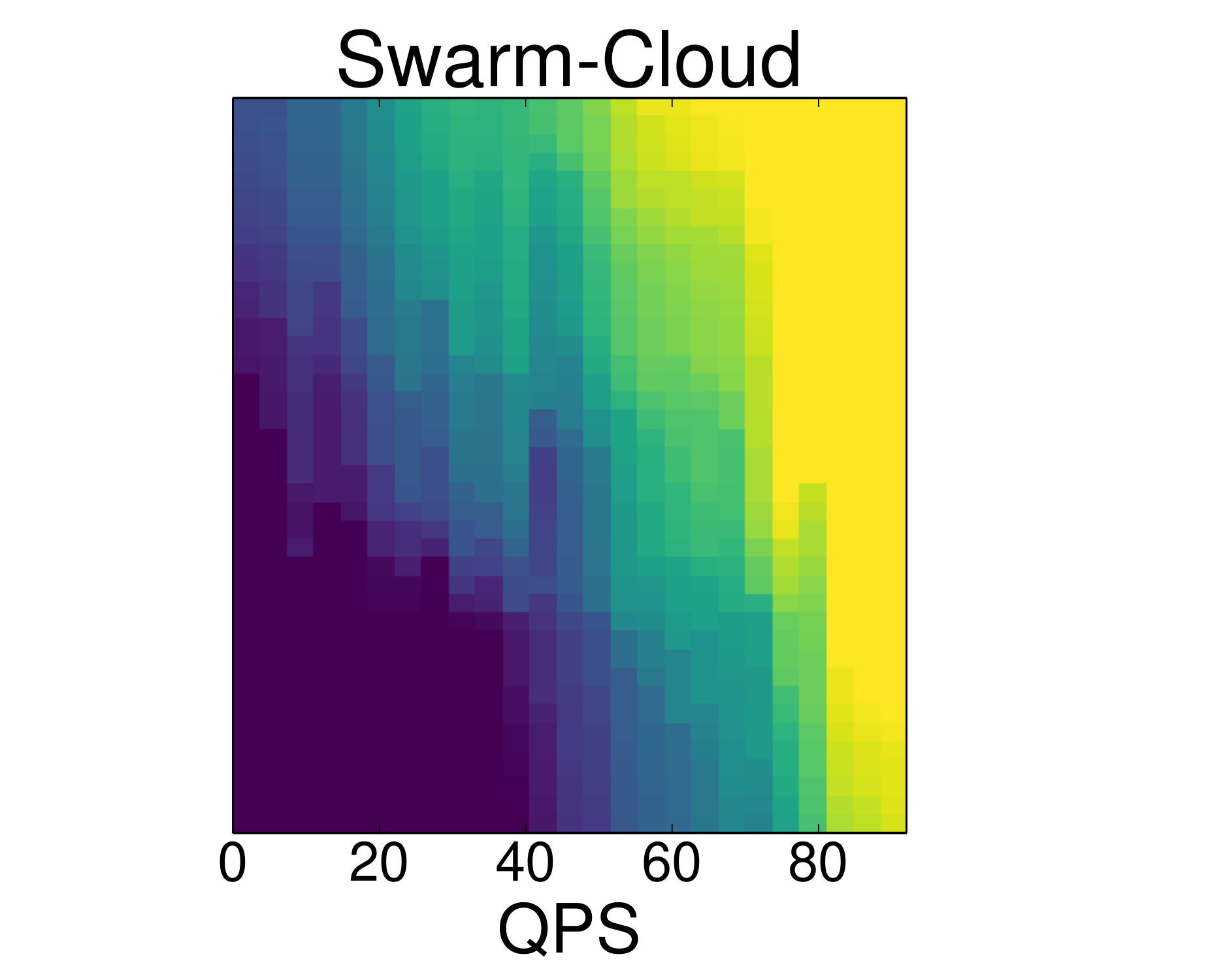} \\
		\vspace{0.2in}
		& & \multicolumn{3}{c}{\includegraphics[scale=0.202, trim=-0.4cm 0.2cm 2cm 4.4cm, clip=true]{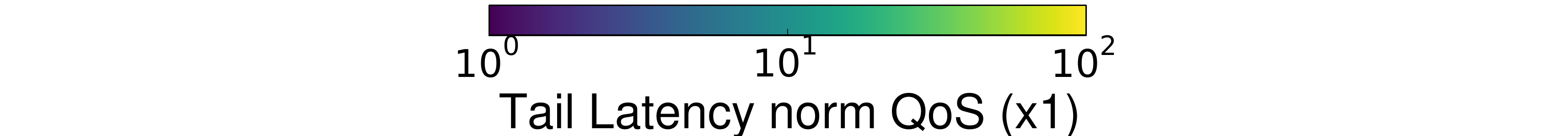}} \\
	\end{tabular}
	\vspace{-0.52in}
	\caption{\label{fig:dvfs} Tail latency with increa-\\ sing load and decreasing frequency\\ (RAPL) for traditional monolithic cloud applications, and the five end-to-end DeathStarBench services. Lighter colors (yellow) 
	denote QoS violations. }
\end{figure}

\vspace{0.06in}
\noindent{\bf{Brawny vs. wimpy cores: }} There has been a lot of work on whether small servers can replace high-end platforms in the cloud~\cite{Reddi10,Reddi11,hoelzle,Chen17}. Despite the 
power benefits of simple cores, interactive services still achieve better latency in servers that optimize for single-thread performance. Microservices offer an appealing target 
for simple cores, given the small amount of computation per microservice. We evaluate low-power machines in two ways. First, we use {\smallcapital RAPL} on our local cluster to reduce the frequency 
at which all microservices run. Fig.~\ref{fig:dvfs} (top row) shows the change in tail latency as load increases, and as the operating frequency decreases for five popular, open-source single-tier 
interactive services: {\smallcapital\texttt{nginx}}, {\smallcapital\texttt{memcached}}, {\smallcapital\texttt{MongoDB}}, {\smallcapital\texttt{Xapian}}, and {\smallcapital\texttt{Recommender}}. 
We compare these against the five end-to-end services (bottom row). 

As expected, most interactive services are sensitive to frequency scaling. Among the monolithic workloads, {\smallcapital\texttt{MongoDB}} is the only one that can tolerate almost minimum frequency at maximum load, due to 
it being I/O-bound. The other four single-tier services experience increased latency as frequency drops, with {\smallcapital\texttt{Xapian}} being the most sensitive~\cite{Kasture16}, followed by {\smallcapital\texttt{nginx}}, 
and {\smallcapital\textit{memcached}}. However, looking at the same study for the microservices reveals that, despite the higher tail latency of the end-to-end service, 
microservices are much more sensitive to poor single-thread performance than traditional cloud applications. Although initially counterintuitive, this result is not surprising, 
given the fact that each individual microservice must meet much stricter tail latency constraints 
compared to an end-to-end monolith, putting more pressure on performance predictability. Out of the 
five end-to-end services (we omit Swarm-Edge, since compute happens on the edge devices), 
the \textit{Social Network} and \textit{E-commerce} are most sensitive to low frequency, while the \textit{Swarm} 
service is the least sensitive, primarily because it is bound by the cloud-edge communication latency, 
as opposed to compute speed. 

\begin{wrapfigure}[14]{r}{0.208\textwidth}
	\vspace{-0.1in}
		{\includegraphics[scale=0.235, viewport = 40 0 332 80]{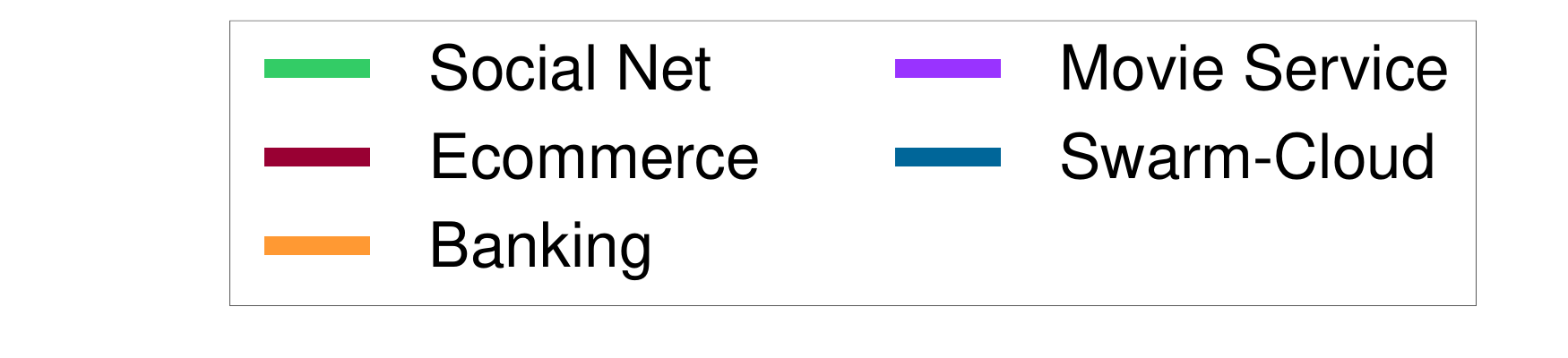}} \\ 
		{\includegraphics[scale=0.24, viewport = -30 20 362 320]{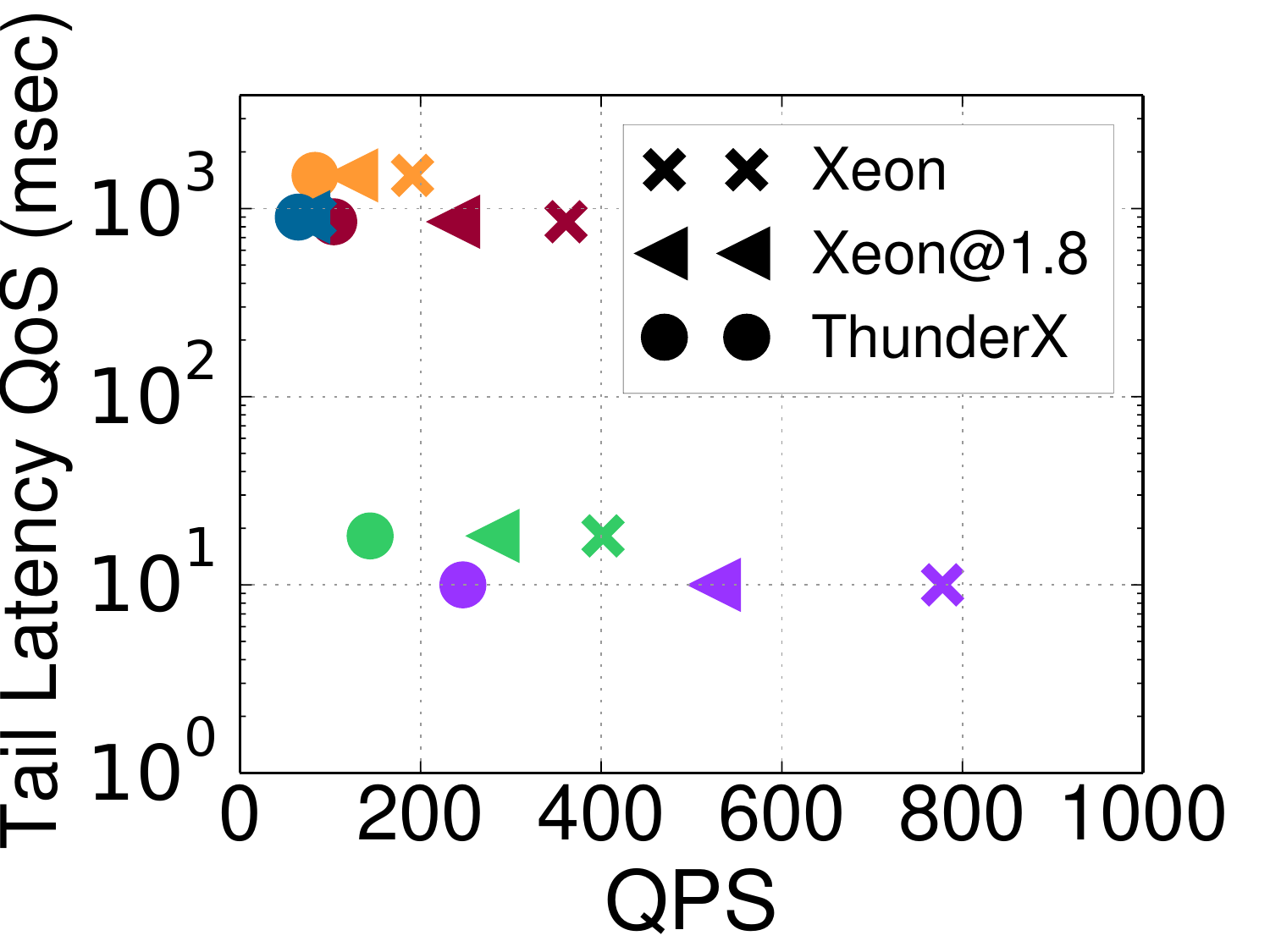}}
		\vspace{-0.08in}
	\caption{\label{fig:cavium} Throughput-tail latency on an Intel Xeon and a Cavium ThunderX server for all end-to-end services.}
\end{wrapfigure}

Apart from frequency scaling, there are platforms designed with low-power cores to begin with. We also evaluate the end-to-end services on two Cavium ThunderX boards (2 sockets, 48 in-order cores per socket, 
1.8GHz each, and a 16-way shared 16MB {\smallcapital LLC})~\cite{Chen17}. The boards are connected on the same ToR switch as the rest of our cluster, and their memory and network subsystems are the same as the other servers. 
Fig.~\ref{fig:cavium} shows the throughput at the saturation point for each application on the two platforms. 
We also show the performance of the Xeon server when equalizing its frequency to the Cavium board. 
Although ThunderX is able to meet the end-to-end QoS target at low load, all five applications 
saturate much earlier than on the high-end server. This is especially the case in \textit{Social Network}, 
and \textit{Media Service} because of their stricter latency requirements, and \textit{E-commerce}, 
because it is more compute intensive. As with power management, \textit{Swarm} does not suffer as much, because it is network-bound. 
Running the Xeon server at 1.8GHz, although worse than its performance at the nominal frequency, 
still outperforms the Cavium SoC considerably. 
Even though low power machines degrade performance in this case, they can still be used for microservices off 
the critical path, or those insensitive to frequency scaling. 








\section{OS \& Networking Implications}
\label{sec:os_network}


We now examine the role of operating systems and networking under the new microservices model. 

\begin{wrapfigure}[11]{l}{0.21\textwidth}
	\centering
		\includegraphics[scale=0.19, viewport = 100 20 522 380]{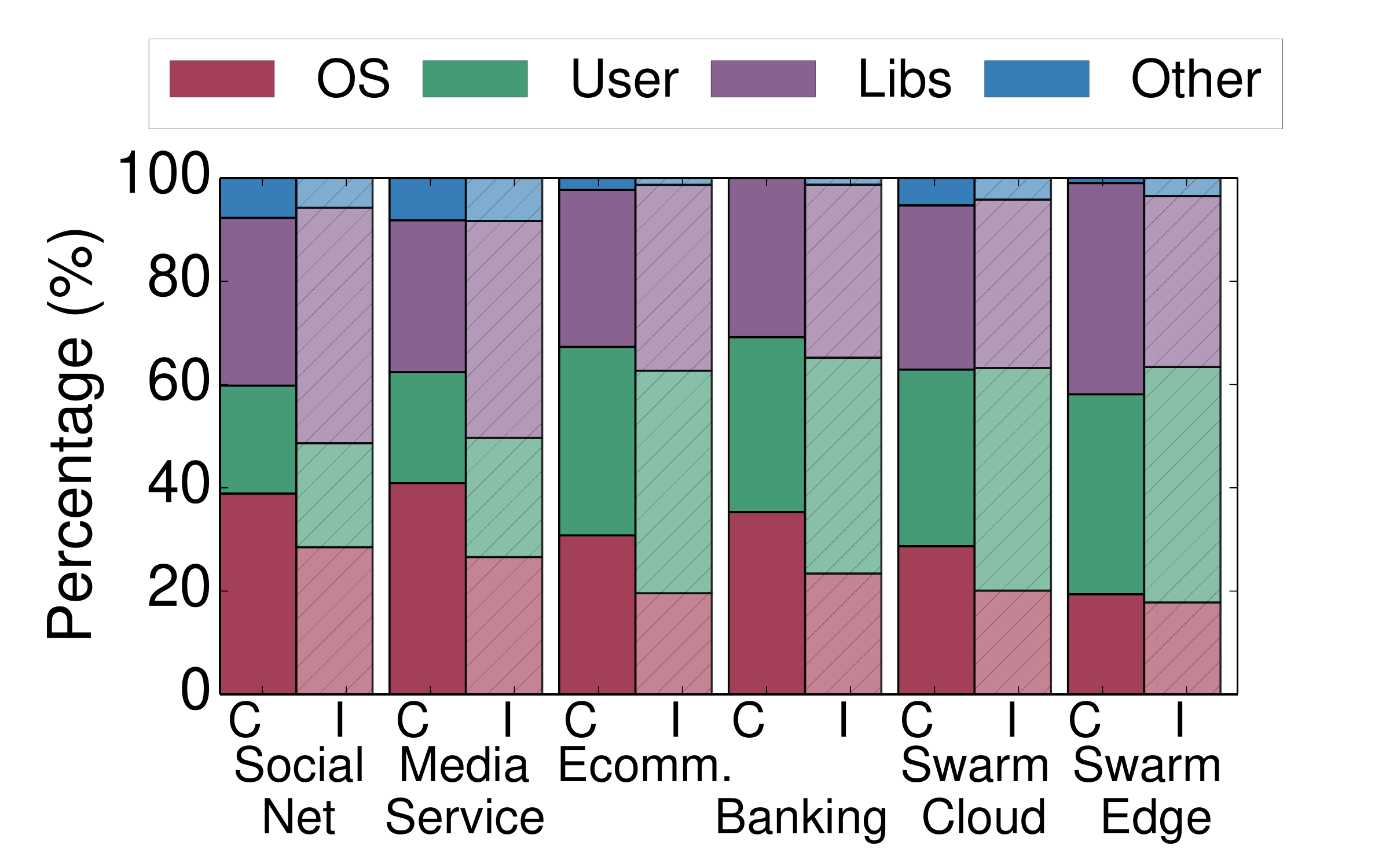} 
		\caption{\label{fig:os} Time in kernel mode, user mode, and libraries for each service. }
\end{wrapfigure}

\noindent{\bf{OS vs. user-level breakdown: }}Fig.~\ref{fig:os} shows the breakdown of cycles (C) and instructions (I) to \textit{kernel}, \textit{user}, 
and \textit{libraries} for each of the end-to-end services. 
For all applications, and especially \textit{Social Network} and \textit{Media Service}, a large fraction of 
execution is at kernel mode, skewed by the use of {\smallcapital\texttt{memcached}} for in-memory caching~\cite{Leverich14}, and the high network traffic, 
with an almost equal fraction going towards libraries like \textit{libc}, \textit{libgcc}, \textit{libstdc}, and \textit{libpthread}. 
The breakdown is less skewed for \textit{E-commerce} and \textit{Banking}, whose microservices are more computationally-intensive, and spend more time in user mode, 
while \textit{Swarm}, both in its cloud and especially edge configurations, spends almost half of the time in libraries. 

The large number of cycles in the kernel is not surprising, 
given that applications like {\smallcapital\texttt{memcached}} and {\smallcapital\texttt{MongoDB}}
spend most of their execution time in the kernel to handle interrupts, 
process {\smallcapital TCP} packets, and activate and schedule 
idling interactive services~\cite{Leverich14}. The large number of 
library cycles is also intuitive, given that microservices 
optimize for speed of development, and hence leverage a lot of 
existing libraries, as opposed to reimplementing the functionality from scratch. 
The overhead of general-purpose Linux has motivated a lot of simpler specialized kernels, 
such as Unikernel~\cite{unikernel}, which trade off compatibility for improved performance. 
Similar OS designs are also applicable to single-concerned microservices. 

\begin{figure}
	\begin{minipage}{0.48\textwidth}
	\begin{tabular}{cc}
		\includegraphics[scale=0.20, viewport= 40 20 580 410]{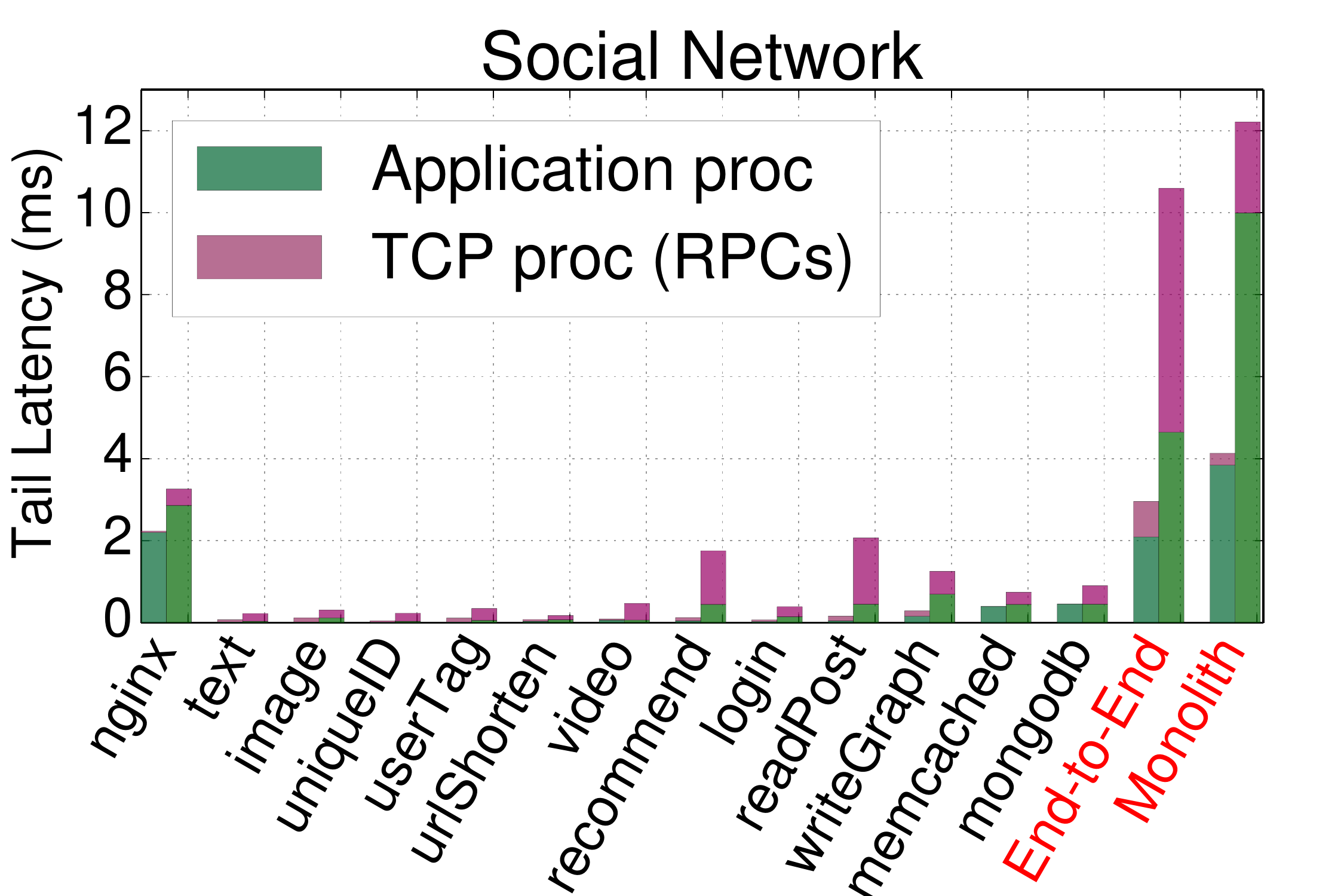} & 
		\includegraphics[scale=0.20, viewport= -40 20 490 410]{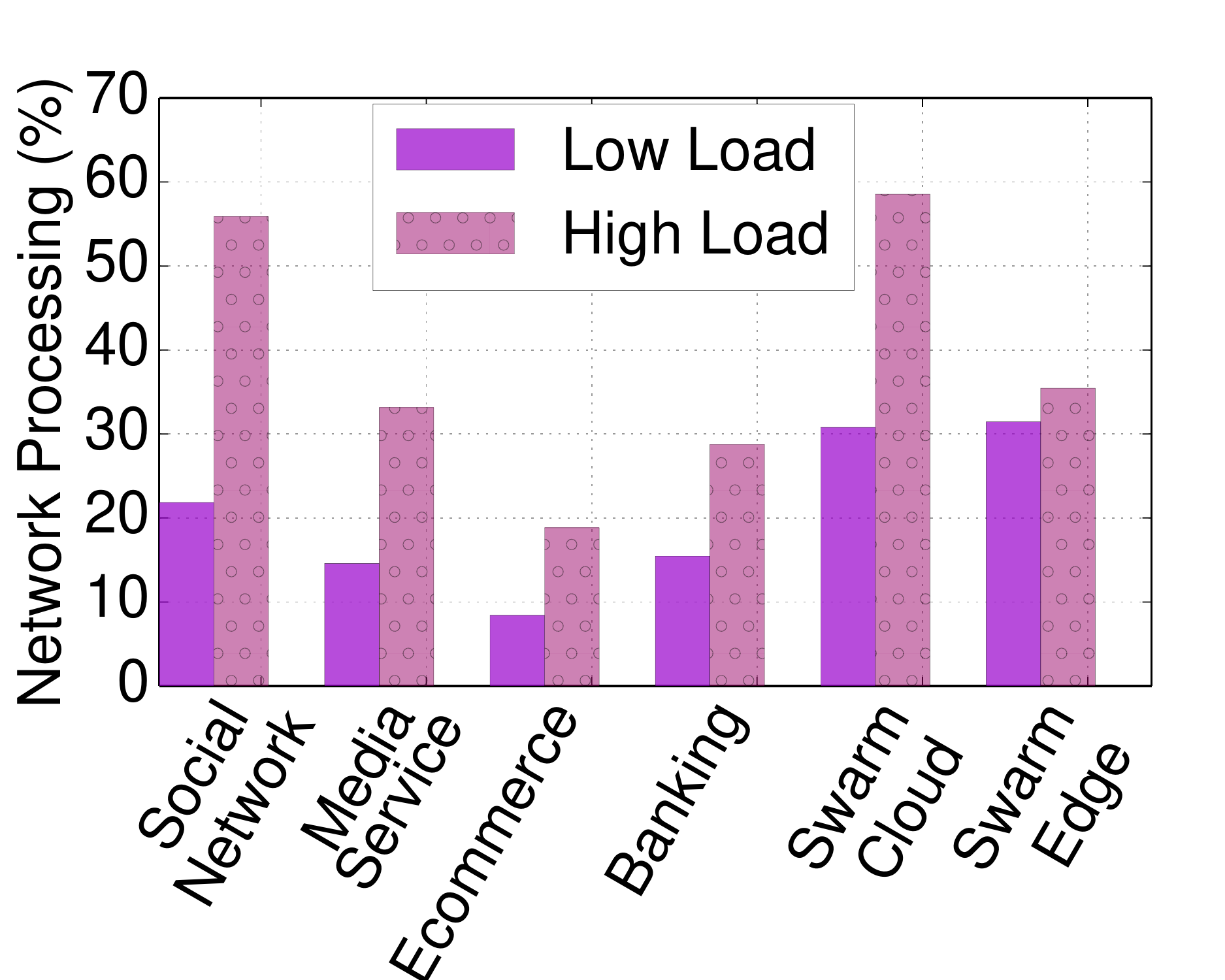} 
	\end{tabular}
	\caption{\label{fig:tcp} Time in application vs network processing for (a) microservices in \textit{Social Network}, and (b) the other services.}
	\end{minipage}
	\vspace{0.08in}
	\begin{minipage}{0.48\textwidth}
	\begin{tabular}{cc}
		\includegraphics[scale=0.384, trim=1.4cm 0.8cm 14.4cm 13.6cm, clip=true]{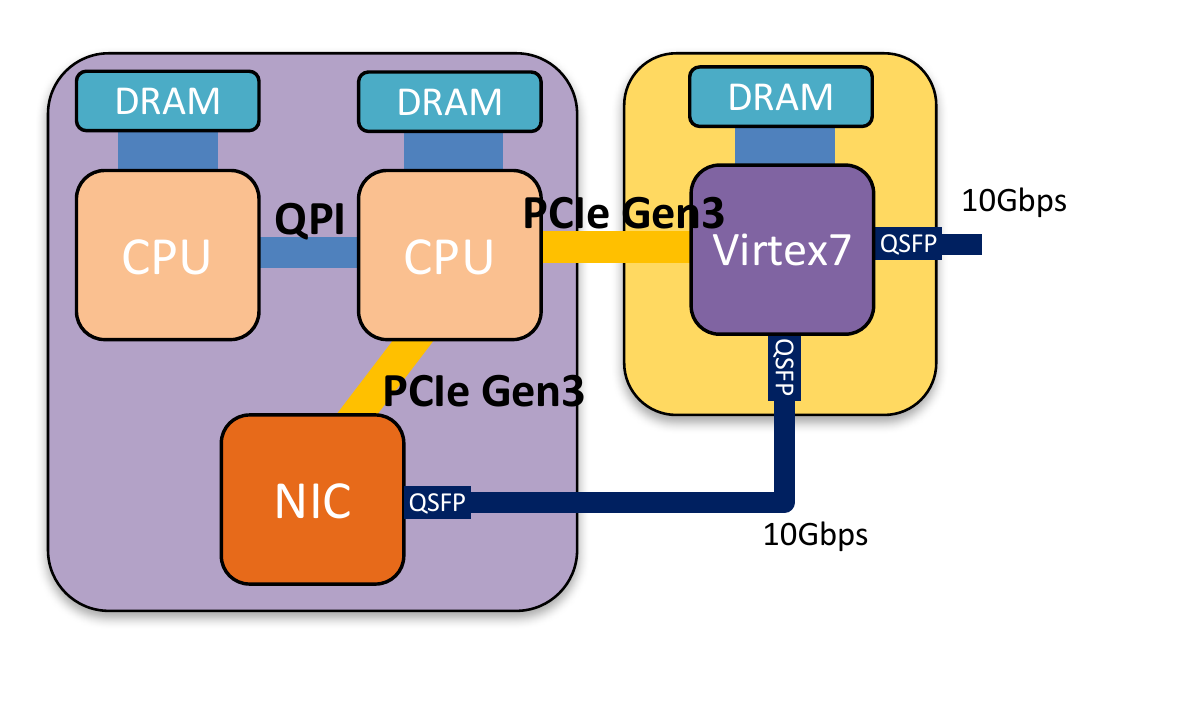} & 
		\includegraphics[scale=0.194, viewport= 120 30 450 410]{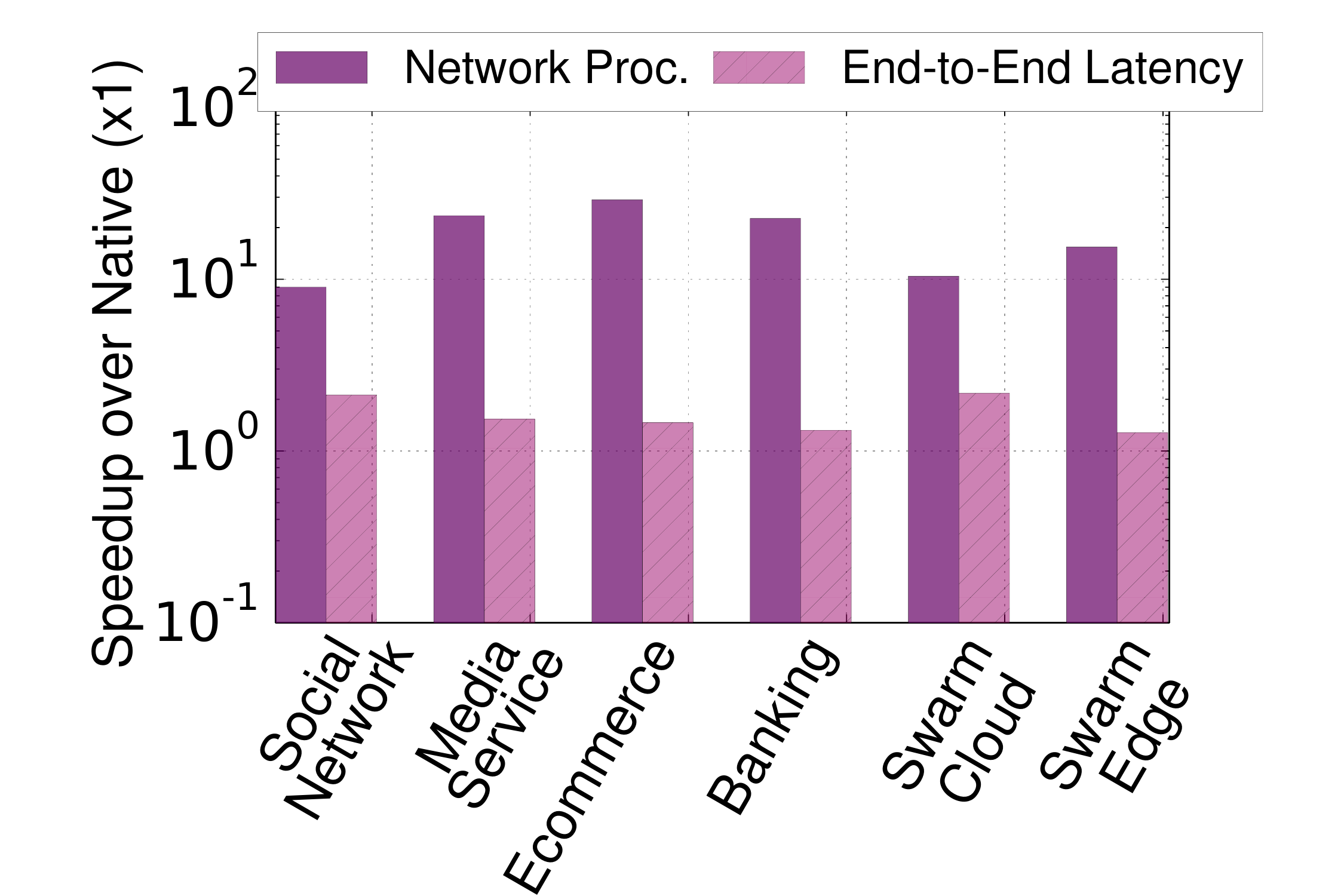} \\
	\end{tabular}
	\caption{\label{fig:acceleration} (a) Overview of the {\smallcapital FPGA} configuration for {\smallcapital RPC} acceleration, and (b) the performance benefits of acceleration in terms of network and end-to-end tail latency. }
\end{minipage}
\end{figure}

\noindent{\bf{Computation:communication ratio: }}Fig.~\ref{fig:tcp}a shows the time spent processing network requests compared to application computation at low 
and high load for the microservices in \textit{Social Network}. Fig.~\ref{fig:tcp}b shows the fraction of tail latency spent processing {\smallcapital RPC} requests for the remaining end-to-end services. 
At low load, {\smallcapital RPC} processing corresponds to 5-75\% of execution time across the \textit{Social Network}'s microservices, and 18\% of end-to-end tail latency. 
This is caused by several microservices being too simple to involve considerable processing. 
In comparison, network processing accounts for a lower fraction of latency in 
\textit{E-commerce} and \textit{Banking}, primarily because their microservices are more computationally intensive. 
Finally, network processing accounts for over 30\% of tail latency in both 
\textit{Swarm} settings, even at low load. 

At high load, network processing becomes a much more pronounced factor of tail latency for all end-to-end services, except for \textit{E-commerce}, and \textit{Banking}, as long queues 
build up in the NICs. This has a significant impact on tail latency, 
with the \textit{Social Network} experiencing a $3.2\times$ increase in end-to-end tail latency. 
The large impact of network processing occurs regardless of whether microservices communicate 
over {\smallcapital RPC}s (\textit{Social Network}, \textit{Media Service}, \textit{Banking}), 
or over \texttt{HTTP} (\textit{E-commerce}, \textit{Swarm-Edge}), although {\smallcapital RPC}s 
introduce considerably lower latencies at low load than {\smallcapital HTTP}. 
Finally, Fig.~\ref{fig:tcp}a also shows the time the monolithic \textit{Social Network} application 
spends processing network requests. Both at low, and especially at high load the difference is 
dramatic, albeit justified, since monoliths are deployed as single binaries, 
with the majority of the network traffic corresponding to client-server communication. 

Given the prominent role network processing has on tail latency, we now examine its potential for acceleration. 

We use a bump-in-the-wire setup, seen in Fig.~\ref{fig:acceleration}a, and similar to the one in~\cite{firestone18} to offload the entire {\smallcapital TCP} stack~\cite{fpga_tcp,sidler16,dhdl,prabhakar16,prabhakar17} 
on a Virtex 7 {\smallcapital FPGA} using Vivado {\smallcapital HLS}. The {\smallcapital FPGA} is placed between the {\smallcapital NIC} and the top of rack switch (ToR), and is connected to both with matching transceivers, 
acting as a filter on the network. We maintain the {\smallcapital PCI}e connection 
between the host and the {\smallcapital FPGA} for accelerating other services, 
such as the machine learning models in the recommender engines, during periods of low network load. 
Fig.~\ref{fig:acceleration}b shows the speedup from acceleration on network processing latency alone, 
and on the end-to-end latency of each of the services. 
Network processing latency improves by $10-68x$ over native {\smallcapital TCP}, while end-to-end tail latency 
improves by 43\% and up to $2.2x$. For interactive, latency-critical services, where even a small improvement 
in tail latency is significant, network acceleration provides a major boost in performance.

\vspace{-0.1in}
\section{Cluster Management Implications}
\label{sec:management}


\begin{figure}
	        \vspace{-0.15in}
		\centering
		\begin{tabular}{cc}
			\hspace*{-0.1in}\includegraphics[scale=0.26, trim=2.0cm 6.65cm 8.4cm 8.6cm, clip=true]{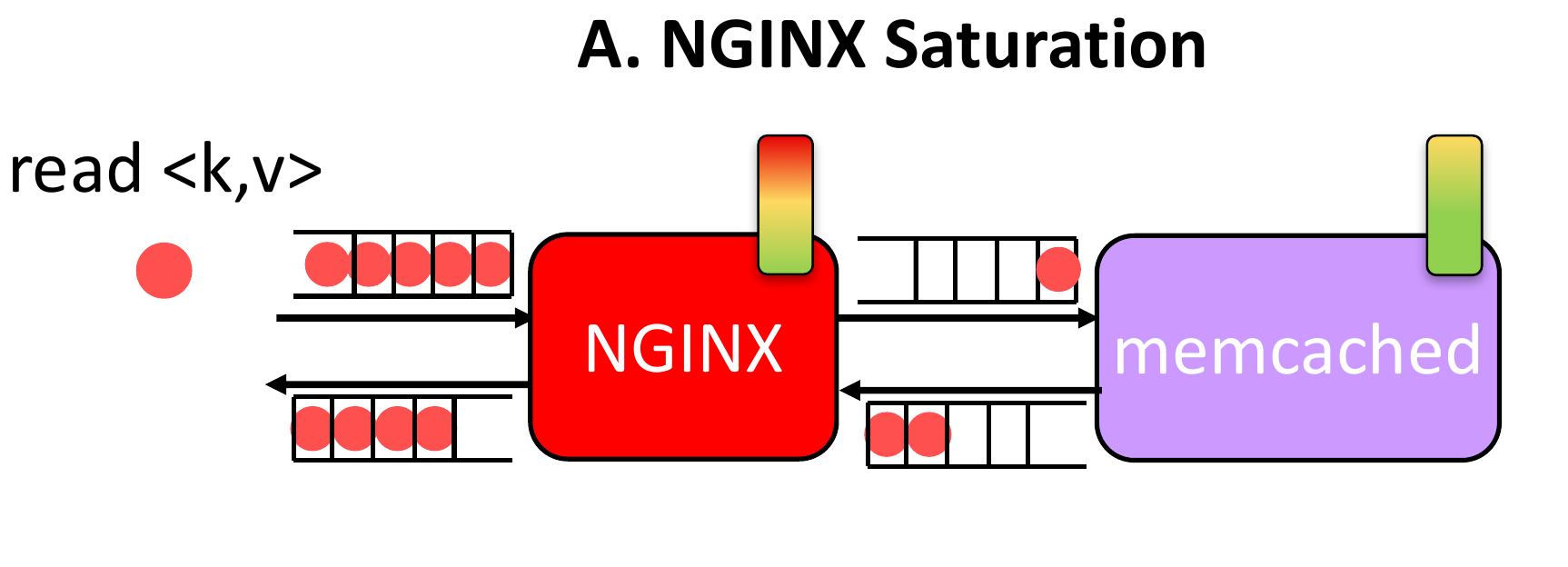} & 
		\includegraphics[scale=0.17, viewport=70 0 700 330]{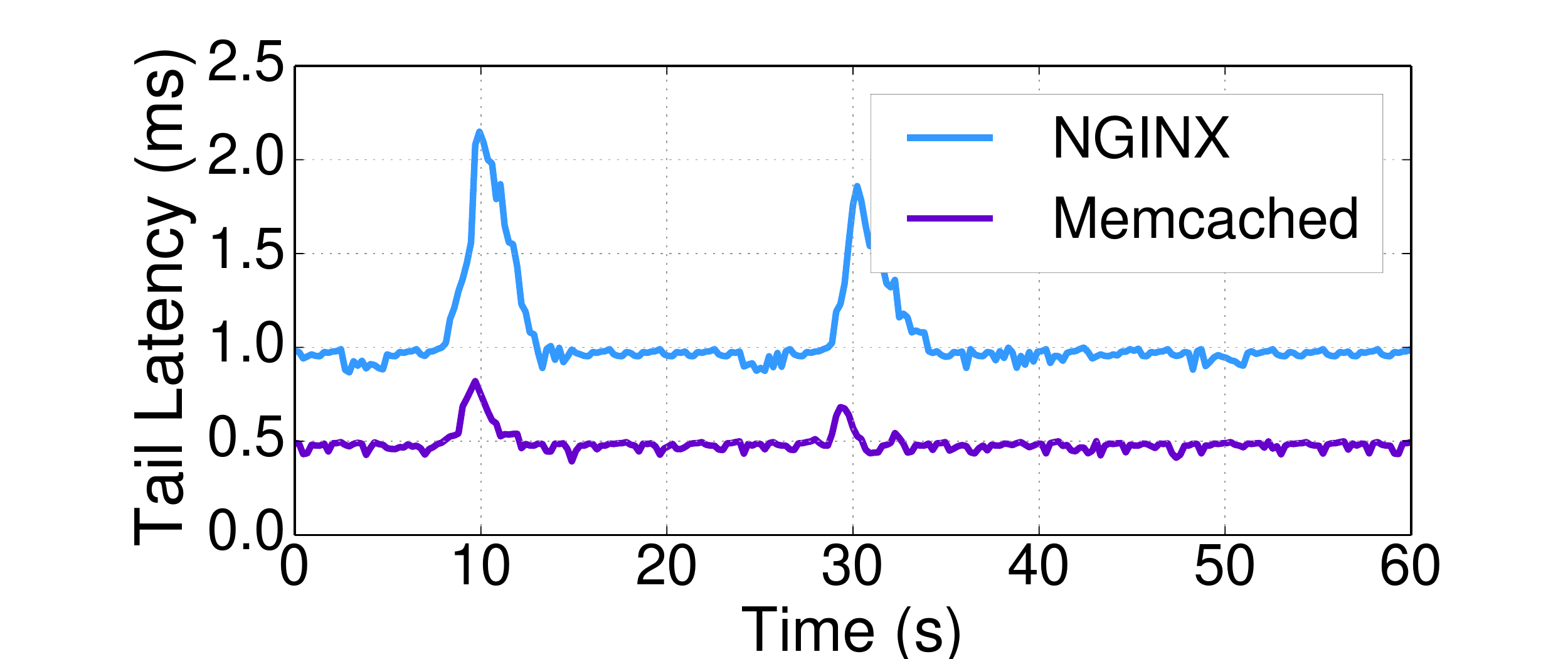} \\ 
		\hspace*{-0.1in}\includegraphics[scale=0.26, trim=2.0cm 0.6cm 8.4cm 15.0cm, clip=true]{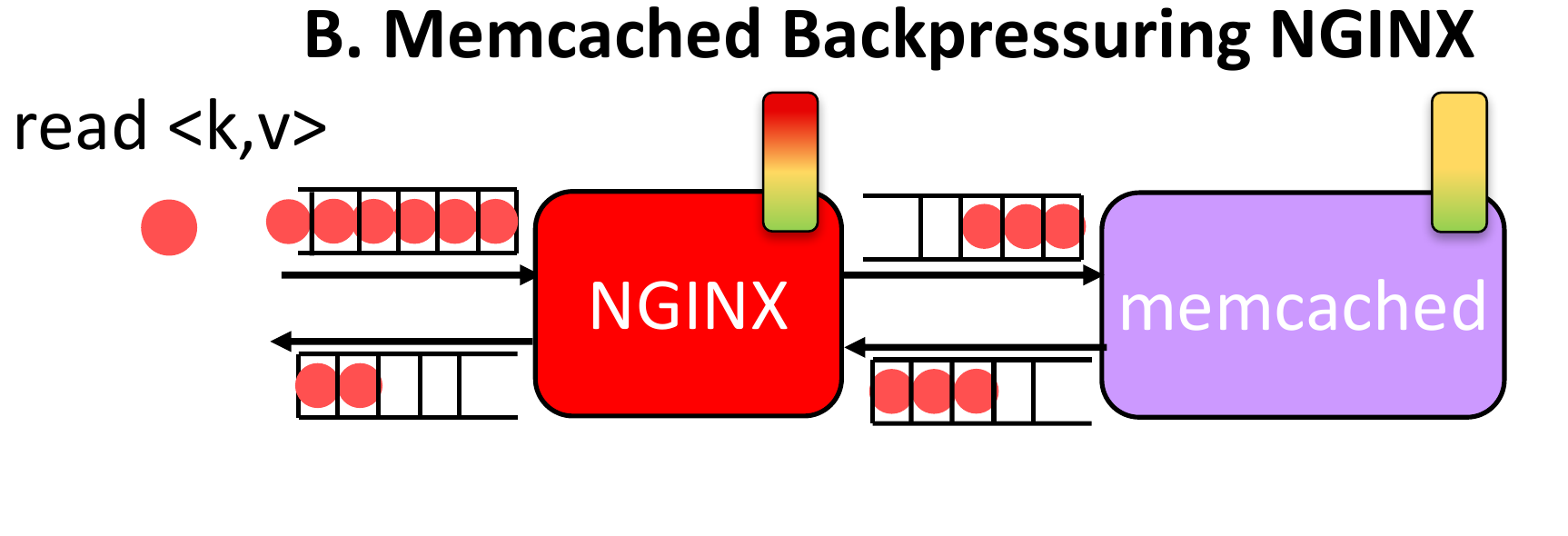} & 
		\includegraphics[scale=0.17, viewport=70 40 700 320]{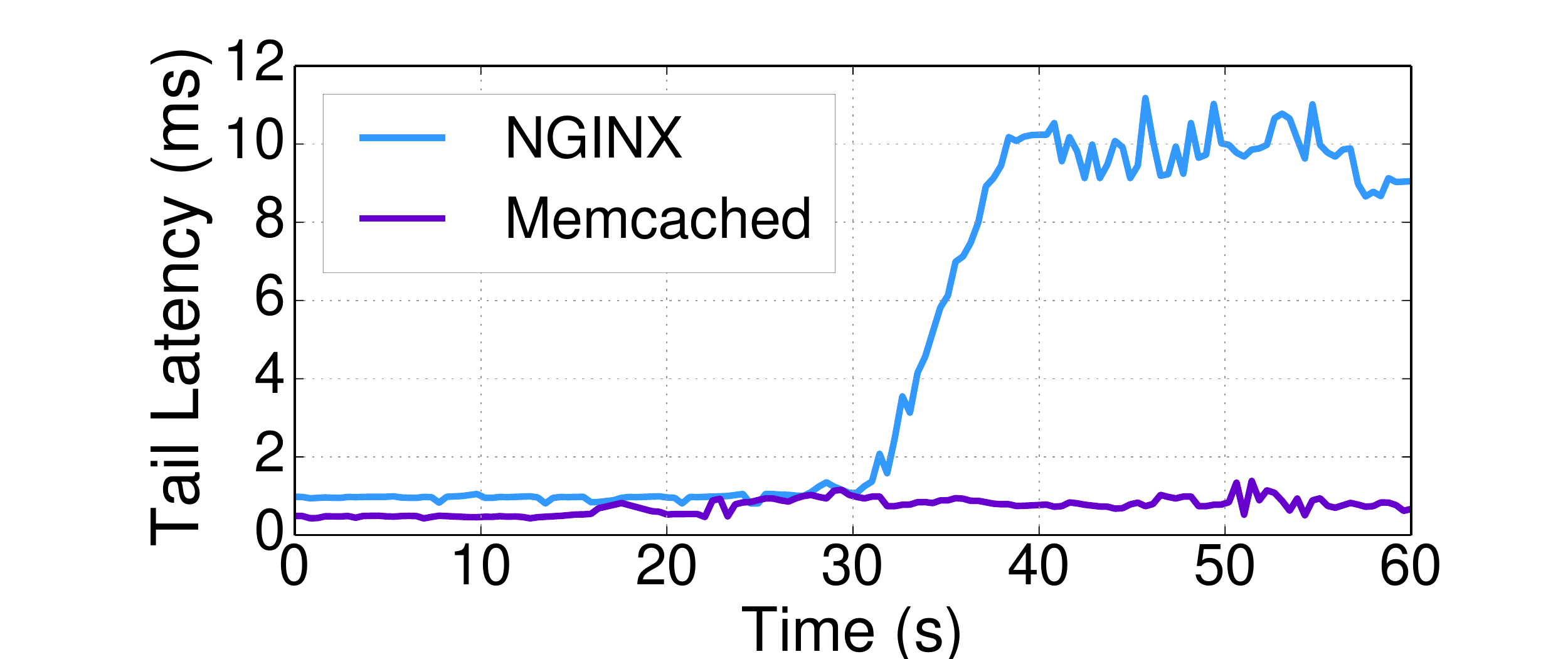} \\
	\end{tabular}
	\caption{{\bf{\label{fig:backpressure}}} {Example of backpressure between microservices in a simple, two-tier application. Case {\smallcapital A} shows a typical hotspot that 
	autoscalers can easily address, while Case {\smallcapital B} shows that a seemingly negligible bottleneck in memcached can cause the front-end {\smallcapital NGINX} service to saturate. }}
\end{figure}

Microservices complicate cluster management, because dependencies between tiers can introduce backpessure effects, leading to system-wide hotspots~\cite{Yang17,Wang17a,Wang17b,Wang17c,Wang17d}. 
Backpressure can additionally trick the cluster manager into penalizing or upsizing a saturated microservice, 
even though its saturation is the result of backpressure from 
another, potentially not-saturated service.
Fig.~\ref{fig:backpressure} highlights this issue for a simplified two-tier application consisting of a webserver
(\texttt{\smallcapital{nginx}}), and an in-memory caching key-value store (\texttt{\smallcapital{memcached}}). In case {\smallcapital{A}}, as the client issues read requests,
\texttt{\smallcapital{nginx}} reaches saturation, causing its latency to increase rapidly, and long queues to form in its input. This is a straightforward case, which autoscaling
systems can easily tackle by scaling out \texttt{\smallcapital{nginx}}, as seen in the figure at $t=14s$ and $t=35s$. 

\begin{wrapfigure}[17]{r}{0.25\textwidth}
	        \vspace{-0.14in}
		\centering
		\includegraphics[scale=0.334, trim=1.2cm 1.8cm 5.6cm 5.8cm, clip=true]{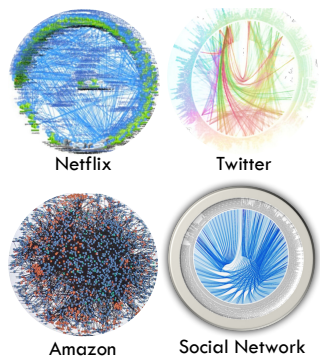}
	        \vspace{-0.25in}
		\caption{{\bf{\label{fig:deathstars}}} {Microservices graphs for three real production 
		cloud providers~\cite{Cockroft15,Cockroft16,twitter_decomposing}. We also show these dependencies for \texttt{Social Network}. }}
	\end{wrapfigure}

Case {\smallcapital{B}} on the other hand, highlights the challenges of backpressure.
When using \texttt{\smallcapital{HTTP1}}, requests within a single connection are blocking, i.e., there can only be one outstanding request
per connection across tiers. Therefore, even though \texttt{\smallcapital{memcached}} itself is not saturated, 
it causes long queues of outstanding requests to form ahead of \texttt{\smallcapital{nginx}}, which in turn cause it to
saturate. Current cluster managers cannot easily address this case, as a utilization-based autoscaling scheme would scale out \texttt{\smallcapital{nginx}}, which is budy waiting and appears saturated. 
As seen in the figure, not only does this not solve the problem, but can potentially make it worse, by admitting even more traffic into the system. 
Even without the connection blocking in \texttt{\smallcapital{HTTP1}}, backpressure still occurs, 
as multi-tier applications are not perfect pipelines where tiers operate 
entirely independently. 

Unfortunately real-world cloud applications are much more complex than this simple example suggests. Fig.~\ref{fig:deathstars} shows
the microservices dependency graphs for three major cloud service providers, and for one of our applications (\textit{Social Network}). The perimeter of the circle (or sphere surface) shows
the different microservices, and edges show dependencies between them. Such dependencies are difficult for developers or users to describe, and furthermore, 
they change frequently, as old microservices are swapped out and replaced by newer services. 

\begin{figure}
	\centering
	\begin{tabular}{cc}
		\includegraphics[scale=0.21, viewport = -50 20 565 410]{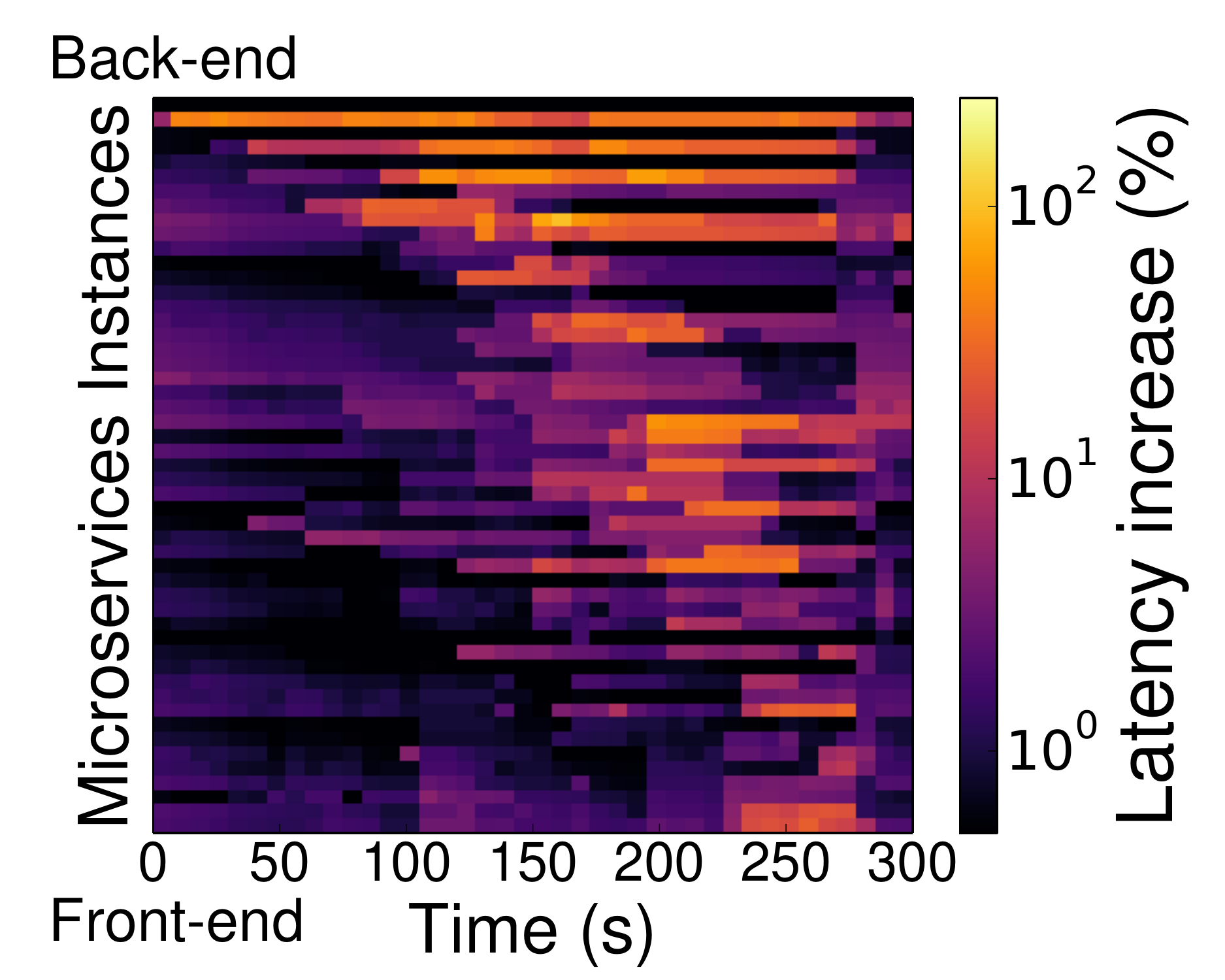} & 
		\includegraphics[scale=0.21, viewport = 80 20 565 410]{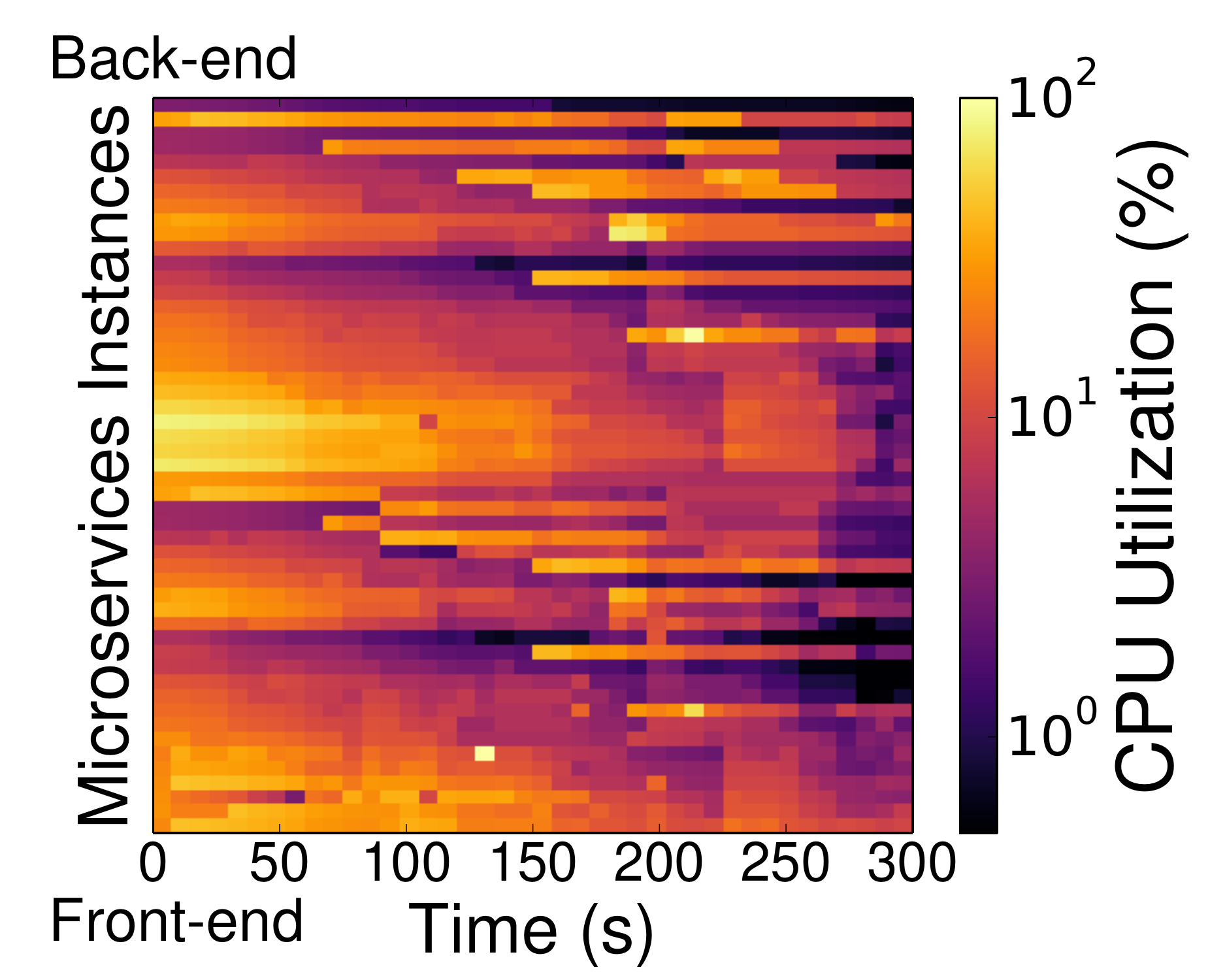} 
	\end{tabular}
	\caption{\label{fig:cascading_small} Cascading QoS violations in \textit{Social Network} compared to per-microservice {\smallcapital CPU} utilization. }
\end{figure}

Fig.~\ref{fig:cascading_small} shows the impact of cascading QoS violations in the \textit{Social Network} service. Darker colors show 
tail latency closer to nominal operation for a given microservice in Fig.~\ref{fig:cascading_small}a, and low utilization in Fig.~\ref{fig:cascading_small}b. 
Brighter colors signify high per-microservice tail latency and high {\smallcapital CPU} utilization. 
Microservices are ordered based on the service architecture, from the back-end services at the top, to the front-end at the bottom. 
Fig.~\ref{fig:cascading_small}a shows that once the back-end service at the top experiences high tail latency, the hotspot propagates to its upstream services, and all the way 
to the front-end. Utilization in this case can be misleading. Even though the saturated back-end services have high utilization in Fig.~\ref{fig:cascading_small}b, microservices 
in the middle of the figure also have even higher utilization, without this translating to QoS violations. 

Conversely, there are microservices with relatively low utilization and 
degraded performance, 
for example, due to waiting on a blocking/synchronous request from another, saturated tier. 
This highlights the need for cluster managers that account for the impact dependencies between microservices have on end-to-end performance 
when allocating resources. 


\begin{figure}
	\centering
	\begin{tabular}{cc}
		\includegraphics[scale=0.285, viewport=10 20 460 308]{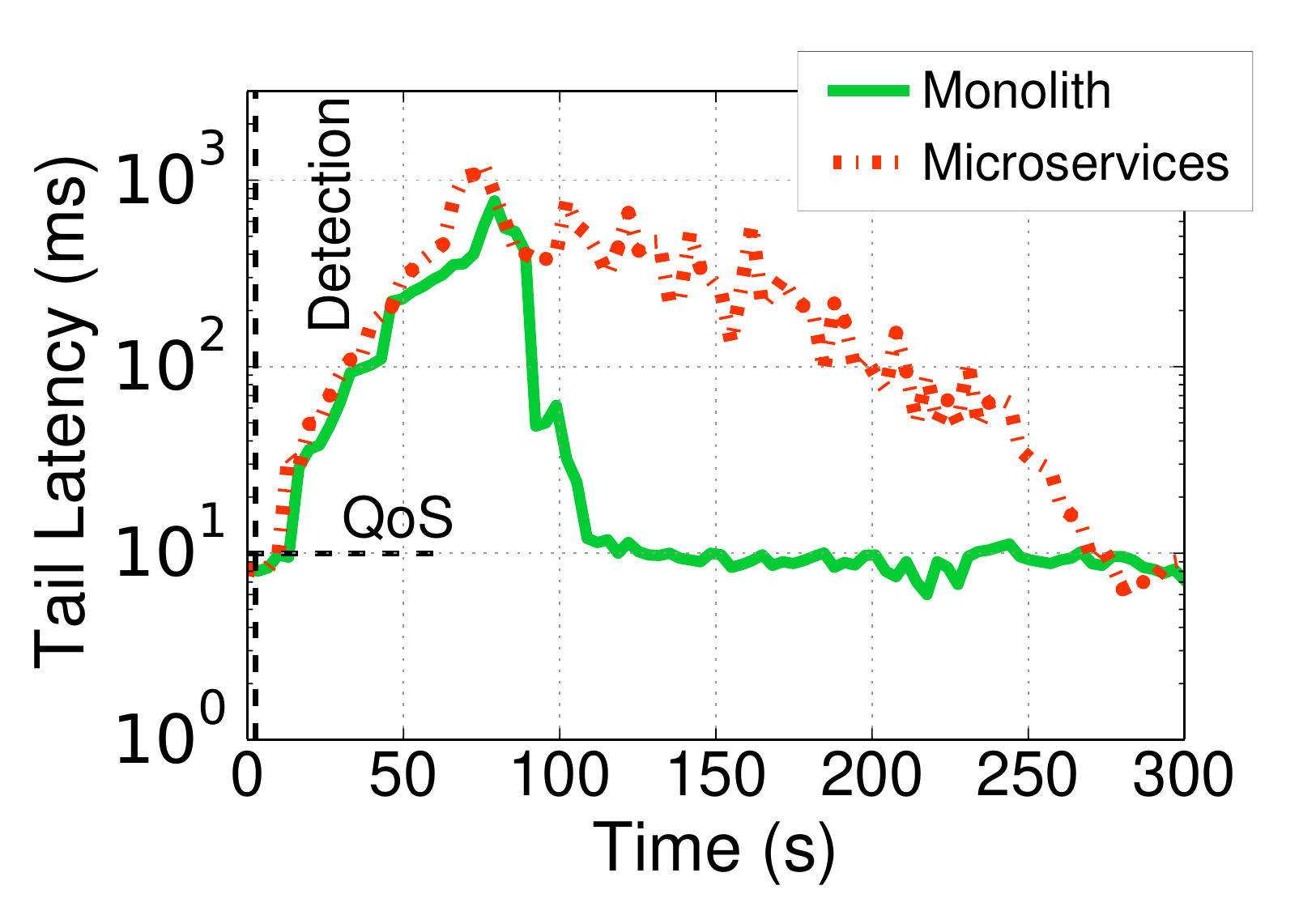} & 
		\includegraphics[scale=0.21, viewport = 60 20 565 240]{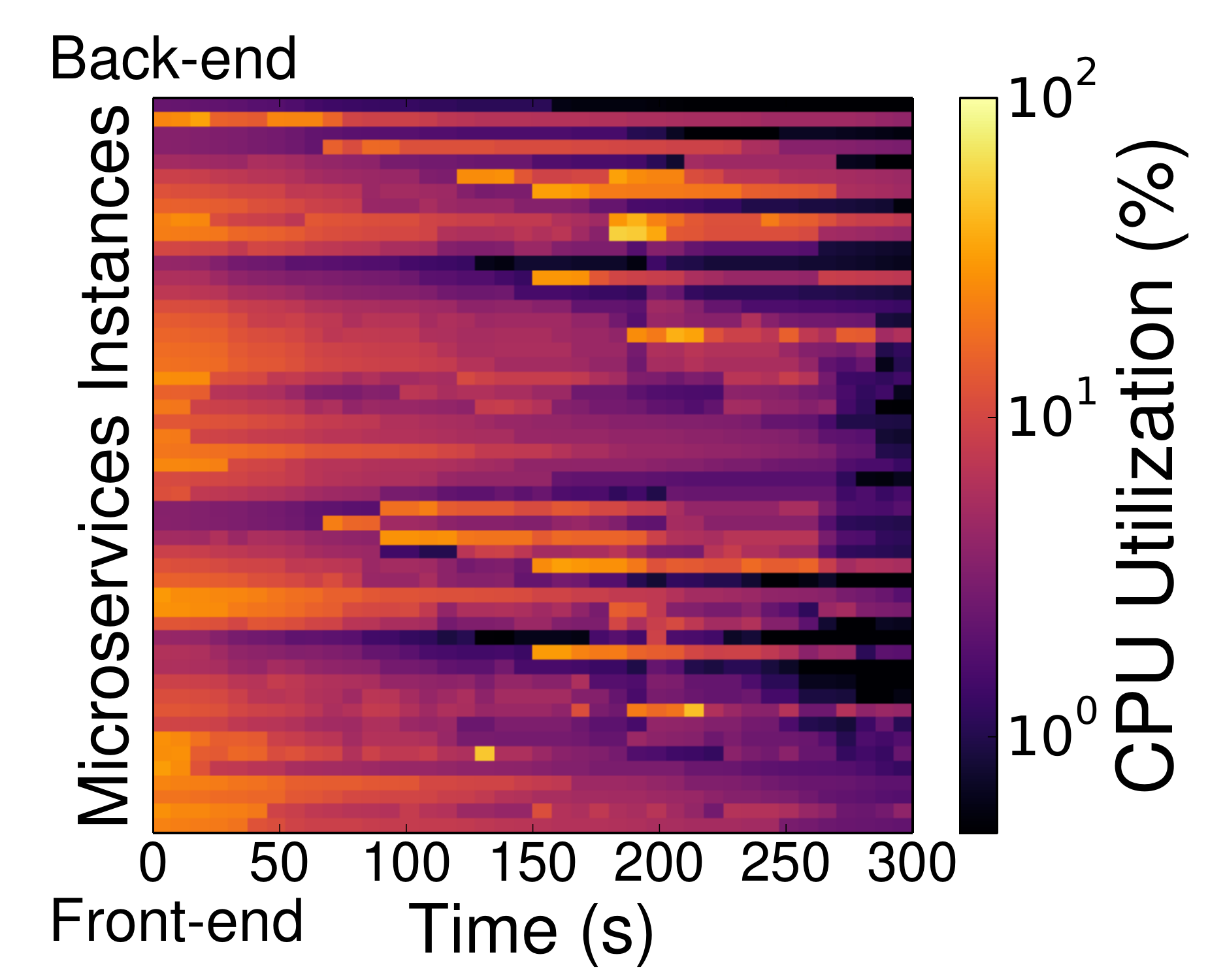} 
	\end{tabular}
	\caption{\label{fig:debugging_motivation} (a) Microservices taking longer than monoliths to recover from a QoS violation, even (b) in the presence of autoscaling mechanisms. }
\end{figure}


Finally, the fact that hotspots propagate between tiers means that once microservices experience a QoS violation, they need longer to recover than traditional monolithic applications, even in the presence of 
autoscaling mechanisms, which most cloud providers employ. 
Fig.~\ref{fig:debugging_motivation} shows such a case for \textit{Social Network} implemented with microservices, and as a monolith in Java. In both cases the QoS violation is detected at 
the same time. However, while the cluster manager can simply instantiate new copies of the monolith and rebalance the load, autoscaling takes longer to improve performance. This is because, 
as shown in Fig.~\ref{fig:debugging_motivation}b, the autoscaler simply upsizes the resources of saturated services - seen by the progressively darker colors of highly-utilized microservices. 
However, services with the highest utilization are not necessarily the culprits of a QoS violation~\cite{Lo14}, taking the system much longer to 
identify the correct source behind the degraded performance and upsizing it. 
As a result, by the time the culprit is identified, long queues have already built up which take considerable time to drain. 


\section{Application \& Programming Framework Implications}
\label{sec:application}






\noindent{\bf{Latency breakdown per microservice: }}We first examine whether the end-to-end services experience imbalance across tiers, 
with some microservices being responsible for a disproportionate amount of 
computation or end-to-end latency, or being prone to creating hotspots. 
We examine each service at low and high load and obtain the per-microservice
latency using our distributed tracing framework, and 
confirm it with Intel's vTune. Both for the \textit{Social Network} and \textit{Media Service} latency at low load is dominated by the front-end ({\smallcapital\texttt{nginx}}), 
while the rest of the microservices are almost evenly distributed. MongoDB is the only exception, accounting for {\smallcapital 8.5\%} and {\smallcapital 10.3\%} 
	of end-to-end latency respectively. 
	
	This picture changes at high load. While the front-end still contributes considerably to latency, overall performance is now limited by the back-end
	databases, and the microservices that manage them, e.g., {\smallcapital\texttt{writeGraph}}. The \textit{Ecommerce} and \textit{Banking} services experience similar fluctuations across load levels, and are additionally 
	impacted by the fact that several of their services are compute intensive, and written in high-level languages, like node.js and Go. This affects execution time, 
	with {\smallcapital\texttt{orders}}, {\smallcapital\texttt{catalogue}}, and {\smallcapital\texttt{payment}} accounting for the majority of end-to-end latency 
	for \textit{Ecommerce}, and {\smallcapital\texttt{payments}} and {\smallcapital\texttt{authentication}} for \textit{Banking}. 
	The back-end databases in this case contribute less to execution time, 
	showing that the choice of programming language affects how hotspots evolve in the system. 
	{\smallcapital\texttt{queueMaster}} also experiences high latency in \textit{E-commerce}, as it uses synchronization 
	to ensure that orders are serialized, processed, and committed in order, which constrains its scalability at high load. 
	
	Finally, the \textit{Swarm coordination} service experiences different trade-offs when running on the cloud compared to the edge devices. While {\smallcapital\texttt{imageRecognition}} 
	dominates latency regardless of where the microservice is running, its impact on tail latency 
	is more severe when running at the resource-limited edge, to the point of preventing 
	the motion controller from engaging, due to insufficient resources. 
	
	This shows that not only bottlenecks vary across end-to-end services, 
	despite individual microservices being same/sim-ilar, but that these 
	bottlenecks additionally change with load, 
	putting more pressure on dynamic and agile management. 

\vspace{0.06in}
\noindent{\bf{Serverless frameworks: }}Microservices are often used in the context of serverless programming frameworks, i.e., frameworks where the application and data are managed by the 
cloud provider, and the user simply launches short-lived ``functions'', and is charged on a per-request basis~\cite{lambda}. Serverless is well-suited for applications with intermittent activity, where 
maintaining long-running instances is cost inefficient. Serverless additionally targets embarrassingly parallel services, which benefit from a massive amount of resources for a brief period of time. 
At the same time, serverless adds an extra level of indirection, as applications have to be instrumented (or re-written) to interface with the serverless framework~\cite{fission,openlambda}. 
Additionally, since serverless functions are ephemeral, data has to be stored in persistent storage for subsequent functions to operate on it. 
On AWS Lambda the output of functions is stored in {\smallcapital S3}, which can introduce significant overheads compared to in-memory computation. 

\begin{figure}
	\centering
		\includegraphics[scale=0.21, viewport = 240 0 1145 340]{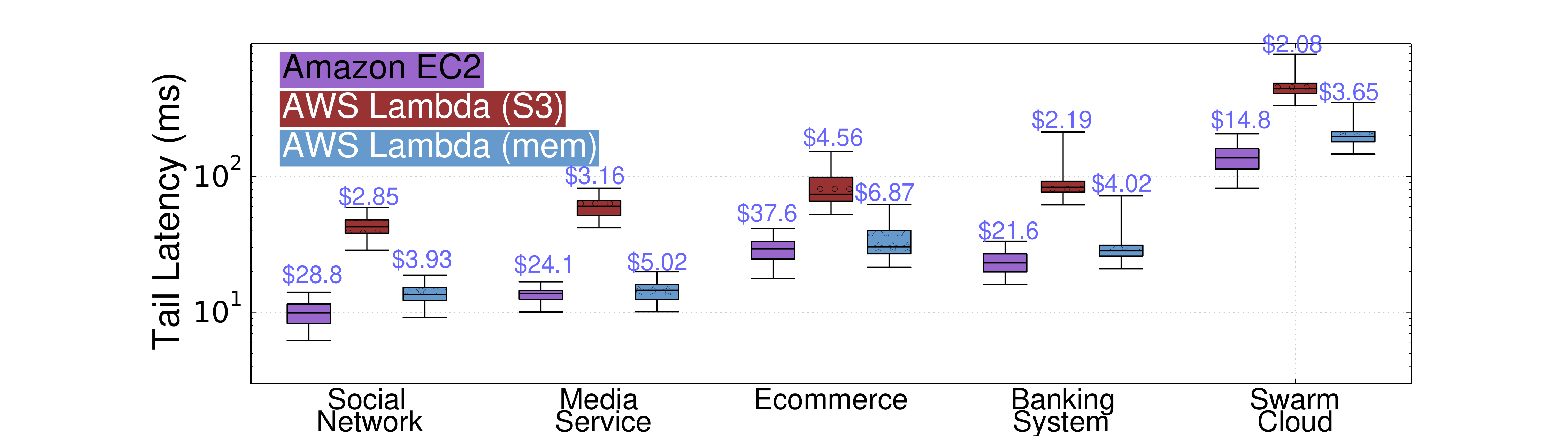} \\
		\includegraphics[scale=0.21, viewport = 266 0 1045 360]{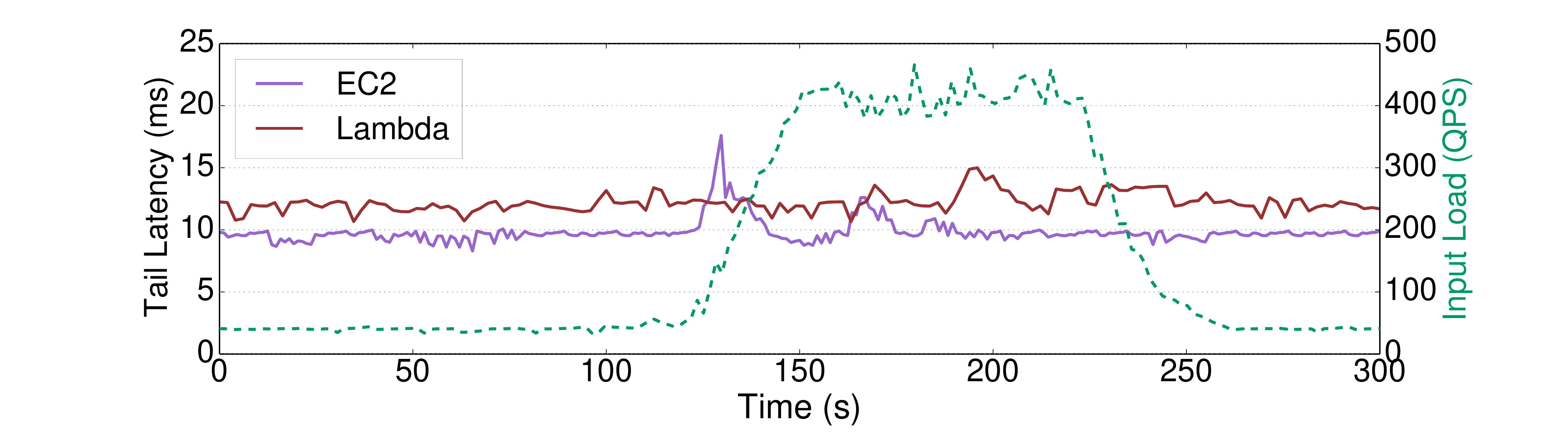} 
		\vspace{-0.1in}
		\caption{\label{fig:serverless} Performance and cost for the five services on Amazon EC2 and AWS Lambda (top). Tail latency for \textit{Social Network} under a diurnal load pattern (bottom). }
\end{figure}

Fig.~\ref{fig:serverless} (top) shows the performance and cost of each end-to-end service on traditional containers on Amazon EC2 versus AWS Lambda functions. Each microservice is instrumented to interface with Lambda's API. 
For a number of microservices written in languages that are not currently supported by Lambda, we also had to reimplement the microservice logic. 
In the case of EC2, each service uses between 20-64 \texttt{m5.12xlarge} instances. We run each service for 10 minutes. 
The margins of box plots show the 25$^{th}$ and 75$^{th}$ latency percentiles, while the whiskers show the 5$^{th}$ and 95$^{th}$. 
In Lambda, we show performance and cost both for the default persistent storage ({\smallcapital S3}), 
and for a configuration that uses the memory of four additional EC2 instances to maintain intermediate state passed through 
dependent microservices. 

Latency is considerably higher for Lambda when using {\smallcapital S3}, primarily due to the overhead and rate limiting of the remote persistent storage. This occurs 
even though the amount of data transfered between microservices is small, to adhere to the design principle that microservices should be mostly stateless~\cite{Cockroft16}. 
The majority of this overhead disappears when using remote memory to pass state between dependent serverless functions. 
Even in this case though, performance variability is higher in Lambda, as functions can be placed anywhere in the datacenter, 
incurring variable network latencies, and suffering interference from external functions co-scheduled on the same physical machines (EC2 instances are dedicated to our services). 
Note that even in the EC2 scenario, dependent microservices are placed on different physical machines to ensure a fair comparison in terms of network traffic. 
On the other hand, cost is almost an order of magnitude lower for Lambda, especially when using {\smallcapital S3}, as resources are only charged on 
a per-request basis. 


The bottom of Fig.~\ref{fig:serverless} highlights the ability of serverless to elastically 
scale resources on demand. The input load is real user traffic in \textit{Social Network}, which follows a diurnal pattern. 
In the interest of cost, we have compressed the load pattern to a shorter period of time and replayed it using our open-loop workload generator. 
Even though EC2 experiences lower tail latency than Lambda during low load periods, consistent with the findings above, 
when load increases, Lambda adjusts resources to user demand faster than EC2. 
This is because the increased number of requests translates to more Lambda functions without requiring the user to intervene. 
In comparison, in EC2, we use an autoscaling mechanism that examines utilization, 
and scales allocations by requesting extra instances, when it exceeds a pre-determined threshold (70\% in this case, consistent 
with EC2 default autoscaler~\cite{AutoscaleLimit}). 
This has a negative impact on latency, since the system waits for load to increase substantially before employing additional resources, and initializing new resources is not instantaneous. 
For microservices to reach the potential serverless offeres, they need to remain mostly stateless, 
and leverage in-memory primitives to pass data between dependent functions. 

\section{Tail At Scale Implications}
\label{sec:tail_at_scale}


We now focus on the \textit{Social Network} service to study the tail at scale effects of microservices, i.e., 
effects that occur because of the large-scale of systems and applications~\cite{tailatscale}. We 
The \textit{Social Network} has several hundred registered users, and 165 active daily users on average. 
The input load for this study is real user-generated traffic. 
To scale to larger clusters than our local infrastructure allows, 
we deploy the service on a dedicated EC2 cluster with 40 up to 
200 \texttt{c5.18xlarge} instances (72 vCPUs, 144GB RAM each).


\begin{figure}
	\centering
		\includegraphics[scale=0.184, viewport = 80 0 525 420]{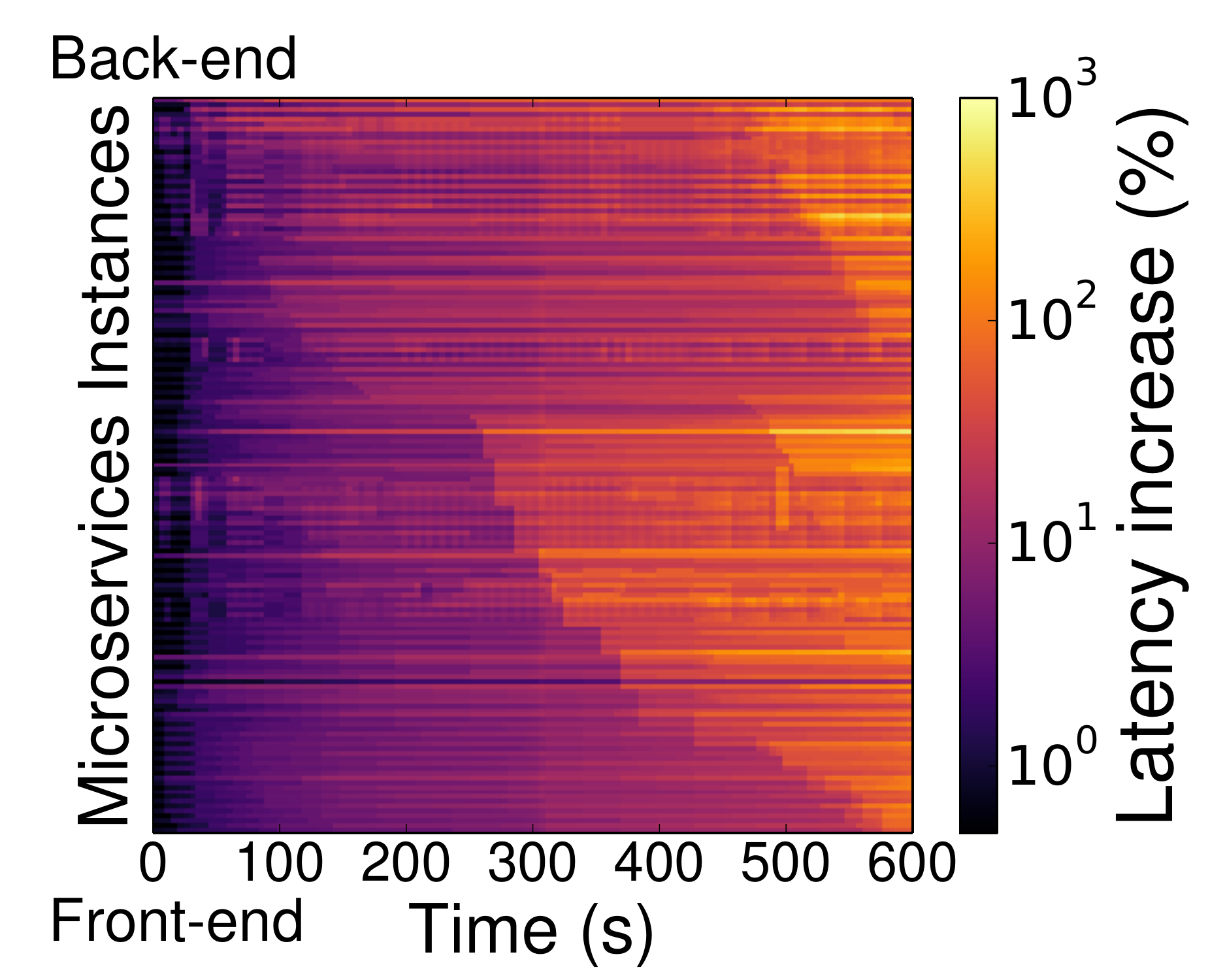}
		\includegraphics[scale=0.186, viewport = 0 0 280 420]{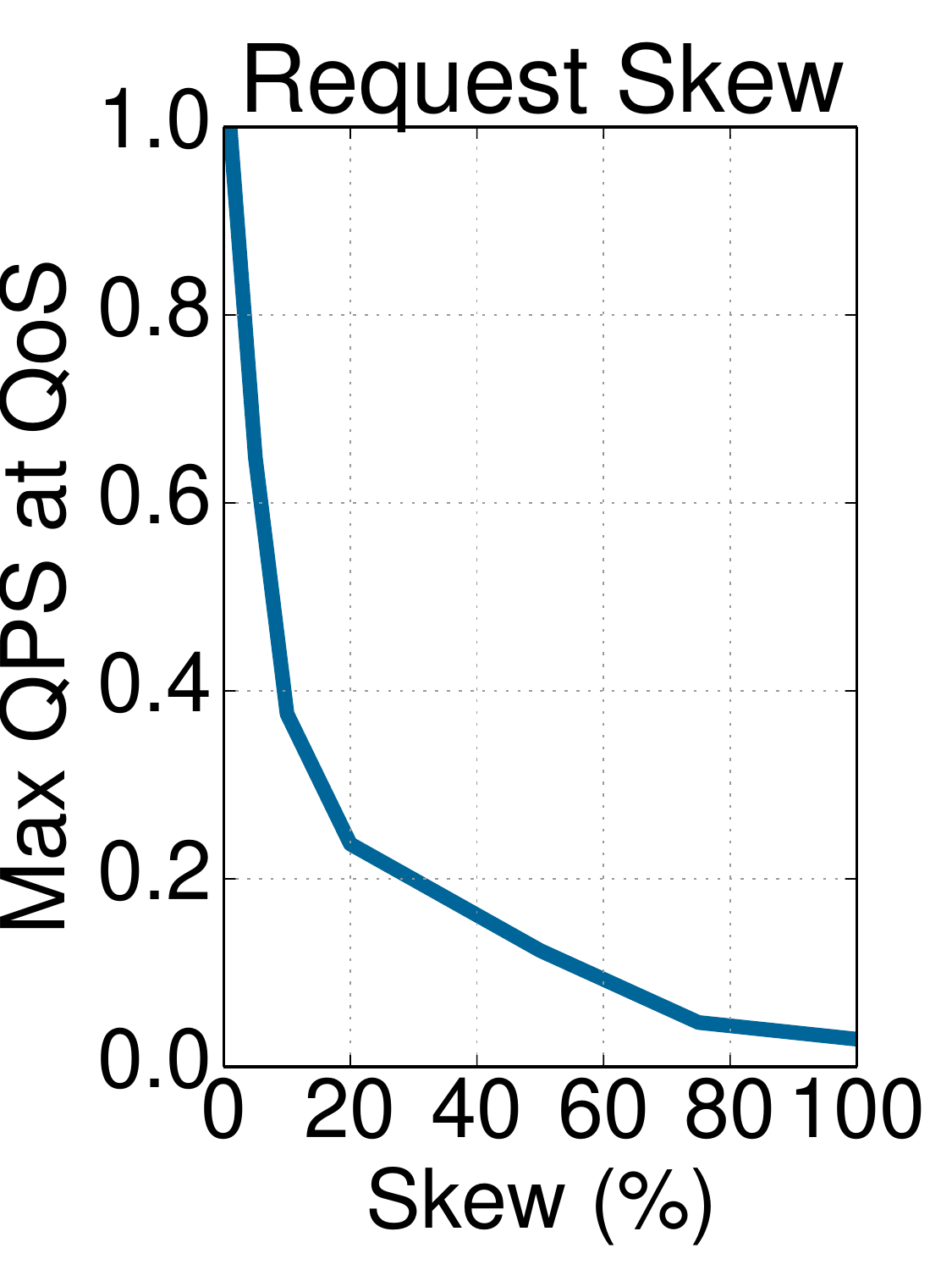}
		\includegraphics[scale=0.186, viewport = -5 0 450 420]{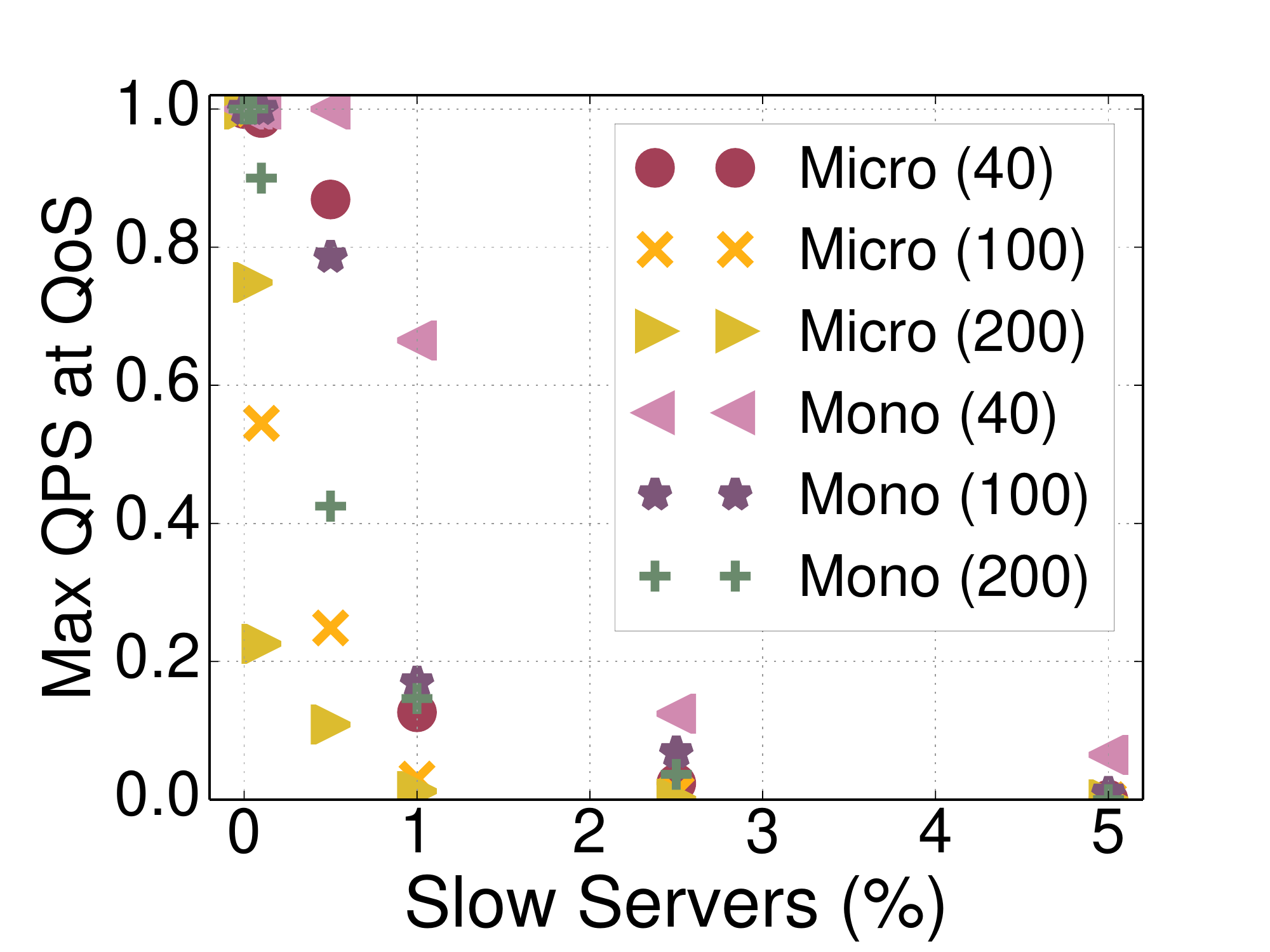}
		\caption{\label{fig:tailatscale} (a) Cascading hotspots in the large-scale \textit{Social Network} deployment, and tail at scale effects from (b) request skew, and (c) slow servers. }
\end{figure}

\noindent{\bf{Large-scale cascading hotspots: }}Fig.~\ref{fig:tailatscale}a shows 
the performance impact of dependencies between microservices on 100 {\smallcapital EC2} instances. 
Microservices on the y-axis are again ordered from the back-end in the top to the front-end in the bottom. 
While initially all microservices are behaving nominally, at $t=260s$ the middle tiers, 
and specifically \texttt{composePost}, and \texttt{readPost} become saturated due to a switch routing misconfiguration 
that overloaded one instance of each microservice, instead of load balancing requests across different instances. 
This in turn causes their downstream services to saturate, causing a similar waterfall pattern in per-tier latency 
to the one in Fig.~\ref{fig:cascading_small}. Towards the end of the sampled time ($t>500s$) 
the back-end services also become saturated for a similar reason, causing microservices 
earlier in the critical path to saturate. This is especially evident for microservices 
in the middle of the y-axis (bright yellow), whose performance was already degraded from the previous QoS violation. 
To allow the system to recover in this case we employed rate limiting, which constrains 
the admitted user traffic until current hotspots dissipate. Even though rate limiting is effective, 
it affects user experience by dropping a fraction of requests. 

\noindent{\bf{Request skew: }} Load is rarely uniform in user-facing cloud services, 
with some users being responsible for a disproportionate amount of generated load. 
Real traffic in the \textit{Social Network} usually adheres to this principle, 
with a small fraction of users, around 5\% being responsible for more than 30\% of the requests. 
To study request skew to its extreme we additionally inject synthetic users that generate 
a much larger number of requests than typical users. Specifically, we vary \textit{skew} 
from 0 to 99\%, where skew is defined as $[100-u]$, with $u$ the fraction of users 
initiating 90\% of total requests. Skew of 0\% means uniform request distribution. 
Fig.~\ref{fig:tailatscale}b shows the impact of skew on the max sustained load for 
which QoS is met. When skew=0\%, the service achieves its max QPS under QoS 
for that cluster size (100 instances). As skew increases, goodput (throughput under QoS) 
quickly drops, and when less than 20\% of users are responsible for the majority of requests, 
\textit{goodput} is almost zero. 

\noindent{\bf{Impact of slow servers: }}Fig.~\ref{fig:tailatscale}c shows the impact 
a small number of slow servers has on overall QoS as cluster size increases. 
We purposely slow down a small fraction of servers by enabling aggressive power management, 
which we already saw is detrimental to performance (Sec.~\ref{sec:architecture}). 
For large clusters (>100 instances), when 1\% or more of servers behave poorly, 
the goodput is almost zero, as these servers host at least one microservice 
on the critical path, degrading QoS. Even for small clusters (40 instances), 
a single slow server is the most the service can sustain and still achieve some QPS under QoS. 
Finally, we compare the impact of slow servers in clusters of equal size 
for the monolithic design of \textit{Social Network}. In this case goodput is higher, even as 
cluster sizes grow, since a single slow server only affects the instance of the monolith 
hosted on it, while the other instances operate independently. The only exception are back-end databases, 
which even for the monolith are shared across application instances, 
and sharded across machines. If one of the slow servers is hosting a database shard, 
all requests directed to that instance are degraded. 
In general, the more complex an application's microservices graph, 
the more impactful slow servers are, as the probability that a service on the critical path will be 
degraded increases. 

\vspace{-0.08in}

\section{Conclusions}
\label{sec:Conclusions}

We have presented DeathStarBench, an open-source suite for cloud and IoT
microservices. The suite includes representative services, 
such as social networks, video streaming, e-commerce, and swarm control services. 
We use DeathStarBench to study the implications microservices have across the cloud system stack, from datacenter 
server design and hardware acceleration, to OS and networking overheads, and cluster management 
and programming framework design. We also 
quantify the tail-at-scale effects of microservices as clusters grow in size, and services become more complex, and 
show that microservices put increased pressure in low tail latency and performance predictability. 

\section*{DeathStarBench Release}

The applications in DeathStarBench are publicly available at: {\color{blue} \href{http://microservices.ece.cornell.edu}{\url{http://microservices.ece.cornell.edu}}} under a GPL licence. 
We welcome feedback and suggestions, and hope that by releasing the benchmark suite publicly, we can encourage more work in this emerging field. 

\balance
%
\bibliographystyle{plain}
\bibliography{references}

\end{document}